\documentclass[
aps,pra,reprint,
amssymb, amsfonts, amsmath,
eqsecnum,
nofootinbib
]{revtex4-2}
\usepackage[utf8]{inputenc}
\usepackage[T1]{fontenc}
\usepackage{dsfont,bm,wasysym,amsbsy,mathtools}
\usepackage{graphicx}
\usepackage{hyperref}
\usepackage{tcolorbox}
\usepackage{qcircuit,physics}
\usepackage[english]{babel}
\usepackage[hats]{./quantum_hqcs}
\usepackage{soul}
\hypersetup{
    colorlinks=true,
    linkcolor=blue,
    urlcolor=blue,
    citecolor=blue,
}

\newcommand{\q}{\ope{q}}
\newcommand{\p}{\ope{p}}
\newcommand{\uint}{\op{m}}

\renewcommand{\vec}[1]{\mathbf{#1}}

\renewcommand{\l}{\op{\ell}}
\newcommand{\mGop}[1][]{\op{m}_{G #1}}
\newcommand{\uGop}[1][]{\op{u}_{G #1}}
\newcommand{\mG}[1][]{m_{G #1}}
\newcommand{\uG}[1][]{u_{G #1}}

\newcommand{\GKP}{\mathrm{GKP}}
\newcommand{\Hcv}{\mc{H_{\text{CV}}}}
\newcommand{\Hcvp}{\mc{H^{\prime}_{\text{CV}}}}

\newcommand{\threevec}[3]{
    \begin{smallmatrix}
        {#1} \\ {#2} \\{#3}
    \end{smallmatrix}
}
\newcommand{\twovec}[2]{
    \begin{smallmatrix}
        {#1} \\ {#2}
    \end{smallmatrix}
}
\newcommand{\m}{\uint}

\renewcommand{\iint}{\int \! \! \! \! \int}

\newcommand{\envop}{\op{\epsilon}}
\newcommand{\sparam}{\zeta}

\newcommand{\eqr}[1]{Eq.~\eqref{#1}}

\newcommand{\secr}[1]{Sec.~\ref{#1}}

\newcommand{\psitelst}{
    \bar{\psi}^{\text{tel}}(s,t)
}
\newcommand{\psitelstnobar}{
    \psi^{\text{tel}}(s,t)
}
\newcommand{\psitelstx}{
    \bar{\psi}^{\text{tel}}(x;s,t)
}
\newcommand{\psitelstxprime}{
    \bar{\psi}^{\text{tel}}(x';s,t)
}

\newcommand{\psitelGKPjstx}{
    \bar{\psi}^{\text{tel}}_{\GKP,j}(x;s,t)
}

\newcommand{\psitelGKPst}{
    \bar{\psi}^{\text{tel}}_{\GKP}(s,t)
}

\newcommand{\psitelGKPstx}{
    \bar{\psi}^{\text{tel}}_{\GKP}(x;s,t)
}
\newcommand{\psitelGKPstxprime}{
    \bar{\psi}^{\text{tel}}_{\GKP}(x';s,t)
}

\newcommand{\intalpha}{
\int_{-\alpha/2}^{+\alpha/2}\hspace{-4pt}
}

\newcommand{\shrinkeq}{
\medmuskip=2mu
\thinmuskip=0mu
\thickmuskip=0mu
\nulldelimiterspace=0pt
\scriptspace=0pt
}

\newcommand{\shrinkeqless}{
\medmuskip=1mu
\thinmuskip=1mu
\thickmuskip=1mu
\nulldelimiterspace=1pt
\scriptspace=1pt
}

\begin{document}

\title{Subsystem analysis of continuous-variable resource states}

\begin{abstract}
Continuous-variable (CV) cluster states are a universal resource for fault-tolerant quantum computation when supplemented with the Gottesman-Kitaev-Preskill (GKP) bosonic code.  We generalize the recently introduced subsystem decomposition of a bosonic mode \href{https://link.aps.org/doi/10.1103/PhysRevLett.125.040501}{[Phys.\ Rev.\ Lett.\ \textbf{125}, 040501 (2020)]}%
, and we use it to analyze CV cluster-state quantum computing with GKP states.  Specifically, we decompose squeezed vacuum states and approximate GKP states to reveal their encoded logical information, and we decompose several gates crucial to CV cluster-state quantum computing.  Then, we use the subsystem decomposition to quantify damage to the logical information in approximate GKP states teleported through noisy CV cluster states. Each of these studies uses the subsystem decomposition to circumvent complications arising from the full CV nature of the mode in order to focus on the encoded qubit information.

\end{abstract}
\date{\today}
\author{Giacomo Pantaleoni}
\email{gpantaleoni@null.net}
\affiliation{Centre for Quantum Computation \& Communication Technology, School of Science, RMIT University, Melbourne, VIC 3000, Australia}
\author{Ben Q. Baragiola}
\affiliation{Centre for Quantum Computation \& Communication Technology, School of Science, RMIT University, Melbourne, VIC 3000, Australia}
\author{Nicolas C. Menicucci}
\affiliation{Centre for Quantum Computation \& Communication Technology, School of Science, RMIT University, Melbourne, VIC 3000, Australia}
\maketitle

\section{ Introduction }

In the original formulation of measurement-based quantum computing~\cite{raussendorf_one-way_2001}, one first prepares a highly entangled state---the cluster state---and then performs quantum computations by measuring the nodes of the cluster~\cite{raussendorf_computational_2001, nielsen_optical_2004, nielsen_cluster-state_2006}. This paradigm may have advantages over implementations of quantum computations that rely on direct control of the gates \cite{nielsen_optical_2004}.

Continuous-variable (CV) cluster-state quantum computing~\cite{braunstein_error_1998,menicucci_universal_2006} is based on the degrees of freedom of a number of bosonic modes rather than on discrete degrees of freedom in standard qubit cluster states. CV cluster-state quantum computing is particularly suited to optics (arguably its most natural implementation) due the experimental straightforwardness of Gaussian operations and quadrature measurements (both staples of CV quantum computing) as well as room-temperature operation, long decoherence times, and scalability~\cite{takeda_toward_2019}. Indeed, large optical cluster states have been realized in the laboratory~\cite{yoshikawa_invited_2016,asavanant_generation_2017,asavanant_generation_2019,takeda_universal_2017,larsen_deterministic_2019,asavanant2020onehundred,larsen2020deterministic} as have measurement-based Gaussian operations~\cite{asavanant2020onehundred, larsen2020deterministic}.

The ability of CV cluster states to perform universal, measurement-based operations is limited by the fact that noise originating in the CV cluster state itself accumulates during computation. A path forward is to combine CV cluster states with bosonic codes~\cite{menicucci_fault-tolerant_2014}, which encode discrete quantum information into bosonic modes~\cite{cochrane_macroscopically_1999,gottesman_encoding_2001,chuang_bosonic_1997,terhal_encoding_2016,albert_performance_2018,joshi2020quantum, cai_bosonic_2021}.
The bosonic code known to provide both universality and fault tolerance in CV cluster-state quantum computing is the Gottesman-Kitaev-Preskill (GKP) encoding~\cite{gottesman_encoding_2001}.
Together, the GKP code and CV cluster states allow
CV architectures to execute qubit-based quantum algorithms and, for resource states of sufficiently high quality, they also allow for error correction, which underpins fault tolerance~\cite{menicucci_fault-tolerant_2014, wu_quantum_2020,bourassa2021blueprint,larsen2021faulttolerant}.

Such hybrid approaches combine two types of CV state: GKP states that encode qubit information and ``unencoded'' momentum-squeezed states that comprise the CV cluster state. Here, we use the framework of bosonic subsystem codes and the underlying subsystem decomposition (SSD)---introduced in Ref.~\cite{pantaleoni_modular_2020}---to show that both types of state encode compatible logical qubits. We further employ the SSD to assess damage to these encoded qubits when the CV states are only approximations of their ideal counterparts.

The subsystem decomposition is a method for partitioning a CV mode into two virtual subsystems: a discrete quantum system (the \emph{logical subsystem}) and a remainder (the \emph{gauge mode}), which is itself another CV mode. This provides a method to identify a logical component for any object defined on the CV mode, most notably states and operators. Tracing over the gauge-mode component of a CV state yields a discrete, logical reduced state, which itself can be analyzed using any of the standard tools of quantum information without having to worry about its CV origin.

In this work, we first generalize the modular subsystem decomposition in Ref.~\cite{pantaleoni_modular_2020} to qu\emph{dit} logical subsystems. Then, returning to qubits, we employ the SSD to analyze necessary elements of CV cluster-state quantum computation.
We use the SSD to reveal the logical information in single-mode momentum-squeezed and approximate GKP states (discussed above) and also to identify the logical, gauge, and entangled components of several operators useful for CV cluster-state quantum computing: unitary position and momentum shifts as well as the regularizing noise operator that generates approximate GKP states from ideal ones.

The foundation for CV measurement-based quantum computing is the teleportation protocol, which transfers an input state along a one-dimensional CV cluster state by measuring the nodes sequentially.
Ideally, the input state is preserved; however, physical implementations of teleportation cause damage, because the nodes of the cluster are not infinitely squeezed.
In \secr{sec:measurementsCVCS}, we use the SSD to extract the logical information from teleported ideal and approximate GKP states in order to quantify damage to it and to find the logical effects arising from the measurement outcomes.

While a number of the results that we show along the way hold true regardless of the SSD, this work heavily relies on the idea of organizing the presentation of calculations at the logical subsystem level. As such, it is natural to start by discussing the details of the subsystem decomposition---both conceptually and as a procedure that one can simply apply to any CV state in order to get a (useful) qubit density matrix. And that is where we begin the next section.

This article is organized as follows. In Sec.~\ref{sec:ssd}, we introduce the subsystem decomposition, discuss its use, and apply it to position and momentum translation operators. In Sec.~\ref{sec:decsqueezedandGKP}, we use the SSD to study the logical information of finitely momentum-squeezed vacuum and approximate GKP states. In Sec.~\ref{sec:measurementsCVCS}, we analyze the logical effects when approximate GKP states undergo noisy teleportation.

\section{ Subsystem decomposition } \label{sec:ssd}

We first review and extend the modular subsystem decomposition introduced in Ref.~\cite{pantaleoni_modular_2020}. The main idea is to decompose the CV Hilbert space as a tensor product
$\Hcv \simeq \mathbb{C}^d \otimes \Hcvp$%
, where $\mathbb{C}^d$ is the Hilbert space of a qudit, and $\Hcvp$ the Hilbert space of a virtual mode (referred to as the \emph{gauge mode}) that is isomorphic to, but distinct from, the original mode.
When decomposed with respect to the \emph{subsystem basis} associated with this tensor-product decomposition, any CV state of the mode is revealed to have an encoded \emph{logical} qudit, although this subsystem qudit is in general entangled with the gauge mode.
There are many ways to perform subsystem decompositions of a mode, each having its own subsystem basis that describes a different tensor-product partitioning of $\mathcal{H}_\text{CV}$. Because of this, different subsystem decompositions can identify different logical information within a fixed CV state.

In this work, we focus on a specific subsystem decomposition based on modular variables~\cite{aharonov_modular_1969} due to its compatibility with the GKP code.
More specifically, our starting point is the modular-variable decomposition of the position quadrature operator, just as was done in Ref.~\cite{pantaleoni_modular_2020}.

We first set our notation and conventions.
In terms of the creation and annihilation operators for the mode, $\op{a}$ and $\op{a}^\dagger$, the position and momentum quadrature operators are
    \begin{subequations}
    \begin{align}
        \op q &\coloneqq \tfrac {1} {\sqrt 2} (\op a + \op a^\dag)
        \, , \\
        \op p &\coloneqq \tfrac {-i} {\sqrt 2} (\op a - \op a^\dag)
        \,.
    \end{align}
    \end{subequations}
They satisfy the canonical commutation relation $[\op q, \op p] = i$, and the measured variance of the vacuum state is~$\tfrac 1 2$ in every quadrature ($\hbar = 1$). The position operator is diagonalized by position eigenstates $\ket{x}[q]$ with spectrum $x \in \reals$. Similarly, $\op{p}$ has eigenstates $\ket{x}[p]$ for $x \in \reals$.  These satisfy the scalar product
\begin{equation}\label{eq:qpinprod}
    \bra{x}\ket{y}[q][p] = \frac {e^{i x y}} {\sqrt{2\pi}}
\end{equation}
for all $x,y \in \reals$.
Henceforth we emphasize the position basis since it is the basis we will use for the subsystem decomposition.

Position eigenstates are orthonormal, $\bra{x}\ket{y}[q][q] = \delta\pqty{x - y}$, and form a basis for $\Hcv$,
\begin{align} \label{eq:positionbasis}
    \id =  \int dx\, \ket{x}[q] \bra{x}[q]
    \, .
\end{align}
A pure, CV state vector $\ket{\psi}$
can then be expanded as
\begin{align}
    \ket{\psi} = \int dx\ \psi(x) \ket{x}[q]
    \, ,
\end{align}
where the position wave function evaluated at $x$ is
    \begin{align} \label{eq:positionwavefunction}
        \psi(x) \coloneqq \bra{x}\ket{\psi}[q][]
        \, .
    \end{align}

Our strategy to introduce the modular-position subsystem decomposition is to decompose a position eigenstate $\ket{x}[q]$. Its associated eigenvalue $x$ is a real number that can be written as a set of three separate numbers: a decimal part, an integer part modulo another integer $d$, and a quotient modulo $d$. These three numbers will then label position eigenstates and, ultimately, define three separate virtual subsystems of different nature.

We illustrate this decomposition of a real number by example, without worrying about which ideas will carry over to the quantum case (where a real number is interpreted as an eigenvalue for a specific state). A real number represented in base $10$, such as
\begin{align}
        177.2453851 \dots \ ,
        \label{eq:base10}
\end{align}
can be specified by a pair of numbers: an integer (177) and fractional part (0.2453851\dots). This means that, effectively, we can slice up the number across the decimal point and think of it as two separate constituents. While this ``decomposition'' is self evident, note that there is no particular reason (other than convenience) to stop the slicing at the decimal point. Let us further separate the real number, by simply electing that the very last digit of the integer is going to be one of the \emph{three} constituents that specify the number. We may, for example, write
\begin{align}
        17
        \mathbf{\underline{7}}
        .2453851\dots
        \label{eq:base10ssd}
\end{align}
and talk about our number as a triplet: an integer (17), a ``digit'' (7, bold underlined above), and a fractional part (0.2453851\dots). As there is no reason to prefer base 10,  we consider the binary representation of the same number, which is
\begin{align}
        1011000
        \mathbf{\underline{1}}
        .0011111 \dots
        \, ,
        \label{eq:base2}
\end{align}
In this case the triplet consists of an integer (1011000), a single bit (1, bold underlined), and the remainder (0.0011111\dots). Note that the elements of the binary triplet are different from those for the base-10 triplet.

Consider a situation where an engineer needs to design a computer, but their architecture at the physical level is ``constrained'' to use real numbers.  One solution they could employ to recover a binary digit-based design is to use the base-2 decomposition of a real number into triplets described above. Specifically, given any real number, they may just use the underlined, bold digit, and throw away the ``leftover information,'' interpreted as useless numbers padding the bit from left and right. Whether the discarded information is actually useless depends on further assumptions on the error model of their all-real-number architecture.

We adopt a different approach: rather than discard the leftover integer and real-number remainder, we stitch them together into a \emph{new} real number. Using this idea, we represent the original number as
 \begin{align}
        1011000
        \mathbf{\underline{1}}
        .0011111...
        =
        \begin{cases}
        1
        \,,
        \\
        1011000.0011111...
        \,,
        \end{cases}
        \label{eq:base2gauge}
\end{align}
where the top row gives the useful binary information, and the bottom row gives the new real number representing the leftover information.  It can be useful to keep this leftover information rather than discarding it. Suppose the same engineer is told that the real numbers from their computer are occasionally shifted by a small amount.  This results in a mismatch between data types: the errors are real numbers, yet the useful information is binary (\emph{i.e.},~bits).  As long as the discussion is limited to classical bits, this problem is simple enough to treat: large, real shifts can flip the useful digit while small shifts leave it alone. Importantly, whether a shift is ``large'' and ``small'' is determined by the leftover information.

The modular-position subsystem decomposition follows in the same spirit, with the logical subsystem and the gauge mode being analogs of the useful and the leftover information, respectively. Although error correction is not a focus in this work, we note that the state of gauge mode keeps track of the leftover information in a modular bosonic subsystem code. Error correction resets the gauge-mode state so that the logical information is most resilient to future shifts.

\subsection{ Partitioned-position basis }
\label{sec:partitionedpositionbasis}
\begin{figure}[t]
    \centering
    \includegraphics[width=\columnwidth]{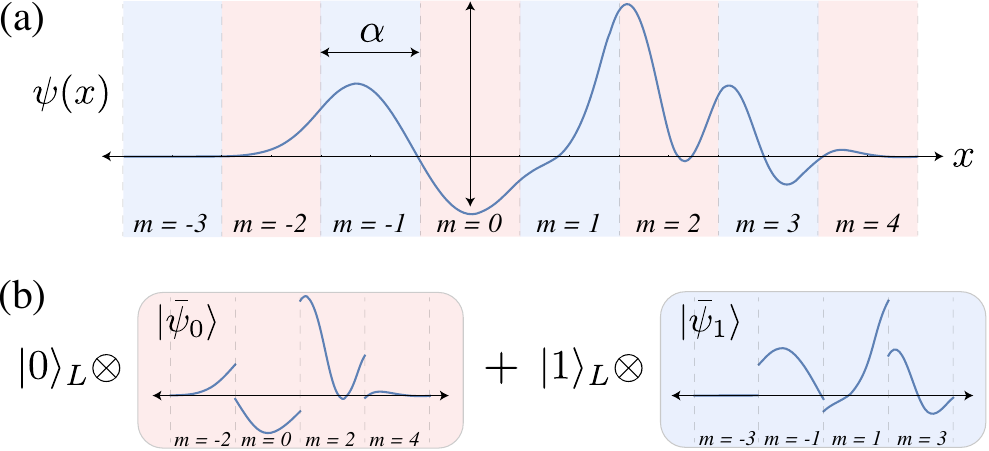}
    \caption{
     Illustration of the modular-position subsystem decomposition of a wave function. (a) Wave function $\psi(x)$ partitioned with respect to bins of size $\alpha$ labeled by bin number $m$. Within each bin is the partitioned-position wave function $\psi(m,u)$. (b) Subsystem decomposition for a logical qubit ($d=2$). The unnormalized wave functions for the gauge states $\ket{\bar{\psi}_0}$ and $\ket{\bar{\psi}_1}$ are obtained by ``stitching'' together $\psi(x)$ from the even- and odd-$m$ bins in (a). Note that here we use the CV-mode bin numbers $m$ and not the gauge bin-number labels $\mG$.
}
    \label{fig:piecewise}
\end{figure}

As the first step towards the modular-position subsystem decomposition, we construct an intermediate basis based on modular position. This basis is founded on the fact that a real number $x \in \mathbb{R}$ can be written as the sum of its quotient and its remainder with respect to some divisor $\alpha \in \mathbb{R}$,
\begin{align} \label{eq:originaldecomp}
    x = \closestint{x}{\alpha} + \fracpart{x}{\alpha}
    = \alpha I_\alpha(x) + \fracpart{x}{\alpha}\, ,
\end{align}
where the integer multiple and fractional parts of $x$,
\begin{subequations}
\begin{align}
    \closestint{x}{\alpha} &\coloneqq \alpha \Big\lfloor \frac{x}{\alpha} + \frac{1}{2} \Big\rfloor \label{eq:closestintegermultiple}
    \, ,\\
    \fracpart{x}{\alpha} &\coloneqq x - \closestint{x}{\alpha}
    \, ,
\end{align}
\end{subequations}
indicate the closest (centered) integer multiple of $\alpha$ and the (centered) remainder, respectively \cite{pantaleoni_modular_2020}.
If $\alpha$ carries units, the fractional part and the closest integer multiple of $\alpha$ are given in the same units of $\alpha$.
We have also defined the related, unitless function,
\begin{align} \label{eq:closestint}
    I_\alpha \pqty{x}
    \coloneqq \frac{ \closestint{x}{\alpha} }{\alpha}
    = \closestint{\frac x \alpha }{}
    \, ,
\end{align}
that gives the closest (centered) integer, as opposed to closest integer multiple of $\alpha$ in \eqr{eq:closestintegermultiple}.
In the rightmost expression, $\closestint{\cdot}{}$ is shorthand for $\closestint{\cdot}{1}$.
The spectrum of a quadrature operator (any linear combination of $\op{q}$ and $\op{p}$) is the set of real numbers, so we can use \eqr{eq:originaldecomp} to decompose its eigenbasis. We focus here to the position operator $\op{q}$, for which
\begin{align}
    \q
    = \closestint{\q}{\alpha} + \fracpart{\q}{\alpha}
    = \alpha \m + \op{u}
    \, ,
    \label{eq:positiondecomposition}
\end{align}
where $\alpha \m \coloneqq \closestint{\q}{\alpha}$ and $\op{u} \coloneqq \fracpart{\q}{\alpha}$ are the integer and modular operator first studied by Aharonov~\cite{aharonov_modular_1969}.  The \emph{bin-number operator} $\op{m}$ and \emph{modular-position operator} $\op{u}$ commute,
$   [\op{m}, \op{u}] = 0 $,
and define a basis of common eigenstates~\cite{aharonov_modular_1969,zak_finite_1967,zak_dynamics_1968,englert_periodic_2006}
\begin{align}
\ket{x}[q] = \ket{\alpha m  + u}[q] \eqqcolon \ket{m, u}
    \, ,
    \label{eq:ppbasis}
\end{align}
with
$m \in \mathbb{Z}$ and $u \in [ -\alpha/2,\alpha/2 )$.
These states are orthonormal, $\bra{m,u}\ket{m',u'} = \delta_{m, m'}\delta\pqty{u' - u}$, and form a basis giving a resolution of the identity
\begin{align}
    \id =
    \sum_{m\in\integers} \intalpha du\
    \ketbra{m,u}{m,u}
    \, .
    \label{eq:identitymu}
\end{align}
Expanding a CV state $\ket{\psi}$ in this basis gives
\begin{align}
    \ket{\psi} = \sum_{m\in\integers} \intalpha
    du\
    \psi\pqty{m,u} \ket{m,u}
    \, ,
\end{align}
where the \emph{partitioned-position wave function} $\psi(m,u)$ is defined as
\begin{align}
 \psi(m,u) \coloneqq \braket{m,u}{\psi}
     \, ,
\end{align}
which can also be thought of as a pointwise mapping
$\psi(m,u) = \psi(\alpha m + u )$, where the right-hand side is the position wave function~$\psi(\cdot)$ [\eqr{eq:positionwavefunction}], evaluated at the partitioned-position decomposition of its argument. The partitioned-position  wave function $\psi(m,u)$ can then be interpreted as a piecewise wave function, parametrized by integer bin numbers $m\in\mathbb{Z}$ and modular position variables defined in the interval $[-\frac \alpha 2, \frac \alpha 2)$. This is akin to the fact that a function of a real variable is defined over a set of intervals
$\bigl[(m - \frac 1 2) \alpha, (m + \frac 1 2) \alpha \bigr)$ parametrized by $m$. This idea is shown in Fig.~\ref{fig:piecewise}(a).

The partitioned-position basis can be interpreted in terms of two separate subsystems, one identified with the bin number $m$ and the other with the fractional position $u$. From this perspective, the basis states above are shorthand for
\begin{align} \label{eq:murotors}
    \ket{m,u} = \ket{m} \otimes \ket{u}
    \, .
\end{align}
The $u$-subsystem is a virtual planar rotor~\cite{albert_general_2017,albert2020robust} whose states can be described by continuous wave functions over basis states $\ket{u}$ in the compact interval $[-\alpha/2,\alpha/2)$ or by their Fourier coefficients, which are indexed by unbounded integers.  It follows from this latter fact that the $m$-subsystem is also a planar rotor, with $\ket{m}$ describing the rotor basis in the Fourier domain.  The continuous degree of freedom of this second planar rotor, modular momentum, lies in the interval $[-\pi/\alpha,\pi/\alpha)$.\footnote{A basis for a CV mode is given by the eigenstates of these two commuting ``modular variables''---modular position and modular momentum~\cite{aharonov_modular_1969}. These basis states can be found by taking the Zak transform of the position eigenstates~\cite{zak_dynamics_1968,janssen_zak_1988}.}
We do not often need to interpret $m$ and $u$ as labeling separate subsystems, although doing so can be useful in some circumstances.

\subsection{ Modular-position subsystem decomposition }
\label{sec:ssd2}

We now isolate a qudit subsystem from the CV mode. This qudit constitutes a logical subsystem, \emph{i.e.},~where discrete-variable information can be stored within the continuous Hilbert space of the mode. Above, we decomposed the position operator $\op{q}$ into integer-valued bin-number operator $\op{m}$ and modular-position operator $\op{u}$, \eqr{eq:positiondecomposition}, which allowed us to define the partitioned-position basis, \eqr{eq:ppbasis}.
By performing a modulo operation on $\op{m}$, we further decompose the partitioned-position basis and introduce the $d$-dimensional qudit subsystem.\footnote{This type of subsystem encoding was first performed in the context of a quantum rotor~\cite{raynal_encoding_2012}.}

An integer $m \in \mathbb{Z}$ can be decomposed with respect to some other positive integer $d$ into a quotient part and a remainder,
$m = d \mG + \ell$, where
\begin{subequations}
\begin{align}
    \label{eq:maps1}
    \mG &= I_d(m)
    \, ,
    \\
    \ell &= \fracpart{m}{d}
    \label{eq:maps2}
    \, ,
\end{align}
\end{subequations}
with $\mG \in \mathbb{Z}$ and $\ell \in \{0,1,\cdots,d-1\}$. With slight abuse of notation, it is understood that whenever $d$ is an integer, $I_d$ is the standard quotient function (\emph{i.e.}, division without remainder), namely $ I_d(m) = \lfloor m / d \rfloor $.

Applying this decomposition to the integer operator $\m$ itself gives
\begin{align}
    \m = \closestint{\m}{d} + \fracpart{\m}{d}
    =
    d \mGop + \l
    \, ,
    \label{eq:mdecomp}
\end{align}
where $\mGop \coloneqq I_d(\m) $ is an operator whose spectrum is $\integers$ and $ \l \coloneqq\fracpart{\m}{d} $. This decomposition procedure is largely analogous to the previous section [compare \eqr{eq:intbasis} to \eqr{eq:ppbasis}] with the difference being the spectrum of the operators being decomposed. Indeed, the operators $\l$ and $\mGop$ commute, hence their simultaneous eigenstates define a new tensor-product basis,
\begin{align}
    \ket{m} &= \ket{d \mG + \ell} \eqqcolon \ket{\ell}[L] \otimes \ket{\mG}[G]
    \label{eq:intbasis}
    \, ,
\end{align}
We refer to $\l$ as a \emph{logical} operator and to $\mGop$ as a \emph{gauge} operator since they act on two different subsystems (labeled $L$ and $G$). Their eigenvalue relations are
\begin{align}
    \mGop\ket{\ell}[L] \otimes \ket{\mG}[G]
    &= \mG \ket{\ell}[L] \otimes \ket{\mG}[G]
    \, ,
    \\
    \l\ket{\ell}[L] \otimes \ket{\mG}[G]
    &= \ell\ket{\ell}[L] \otimes \ket{\mG}[G]
    \, .
\end{align}
The logical operator $\l$ generates logical-$Z$ rotations via exponentiation,
\begin{align} \label{eq:logicalZ}
    \Z_L
    \coloneqq \exp\pqty{ \frac{2 \pi i}{d} \l }
    \, ,
\end{align}
as one can easily verify by acting on the basis for the logical subsystem $\ket{\ell}[L]$.
The qudit $\op X _L$ operator is defined through its action on the logical subsystem as
\begin{align} \label{eq:quditX}
    \X_L \ket{\ell}[L] \coloneqq \ket{\fracpart{\ell+1}{d}}[L]
    \, .
\end{align}

We now use the decomposition of $\op{m}$ given above to construct the change of basis between the position eigenstates of a CV mode and the subsystem-decomposed eigenstates. The position operator in the subsystem basis decomposes as
\begin{align} \label{eq:qsubsystemdecomp}
    \q
    = \alpha \m + \op{u}
    = \alpha \l  + d \alpha \mGop + \op{u}
    \,
\end{align}
and is diagonalized by the simultaneous eigenvalues of $\l$, $\mGop$, and $\op{u}$. Thus, a position eigenstate $\ket{x}_q$ can be decomposed as
\begin{equation}
    \ket{x}[q] = \ket{m,u}
    =  \ket{\ell}[L] \otimes \gket{\mG, \uG}
    \label{eq:ssdbasis}
    \, ,
\end{equation}
and the subsystem basis states inherit orthonormality:
    \begin{align}
        \big(\bra{\ell}[L] \otimes  \gbra{\mG, \uG} \big) &
        \big(\ket{\ell'}[L] \otimes \gket{\mG', \uG'} \big) \nonumber
        \\
        &=
        \delta_{\ell, \ell'}
        \delta_{\mG, \mG'}
        \delta(\uG - \uG')
        \, .
    \end{align}
The maps between the subsystem and the partitioned-position eigenvalues are given in Eqs.~\eqref{eq:maps1} and~\eqref{eq:maps2} along with
\begin{align}
    \uG = u
    \, ;
\end{align}
that is, the gauge modular position is equal to the modular position from the partitioned-position decomposition.
The three quantum numbers in the subsystem decomposition---$\ell$, $\mG$, and $\uG$---are functions solely of the position eigenvalue $x$ through
\begin{subequations} \label{eq:doubledecomp}
\begin{align}
    \ell &=
    \fracpart{
    I_\alpha(x)
    }{d}
    \, ,
    \\
    \mG &= I_d(I_\alpha(x))
    \, ,
    \\
    \uG &= \fracpart{x}{\alpha}
    \, ,
    \label{eq:maps}
\end{align}
\end{subequations}
with the closest integer function given in \eqr{eq:closestint}.  In simpler terms, with respect to expressions like \eqr{eq:base10} and \eqr{eq:base2}, $\ell$ is a base-$d$ digit, whereas $\mG$ and $\uG$ are the leftover integer and fractional part respectively. Since changing between the position basis and the subsystem basis is notationally complex, we often discuss the connection to the subsystem basis in terms of the intermediate partitioned-position basis.

The completeness of the subsystem basis follows from \eqr{eq:ssdbasis} and is given by
{\shrinkeqless
\begin{align}
    \id &=
    \sum_{ \ell = 0 } ^{d-1}
    \ket{\ell}[L] \bra{\ell}[L]
    \hspace{2pt} \otimes \hspace{-4pt}
    \sum_{\mG\in\integers}
    \intalpha d\uG
    \ket{\mG,\uG\!}[G] \bra{\mG,\uG\!}[G] \\
    &= \op{I}_L \otimes \op{I}_G
    \, .
    \label{eq:subsystembasisidentity}
\end{align}%
}
From the partitioned-position decomposition, \eqr{eq:identitymu}, we have shown that the states $\ket{\mG,\uG}[G]$ comprise a basis for $\Hcv$. Since the gauge subsystem is formally the same as $\Hcv$ (their Hilbert spaces are isomorphic), the \emph{modular subsystem decomposition} then splits a full oscillator into a tensor product of a qudit and a new virtual oscillator, $\Hcv = \complex^d \otimes \Hcvp$.

\subsubsection{ Decomposing states in the subsystem basis }
Having the resolution of the identity in the subsystem basis, \eqr{eq:subsystembasisidentity}, it is possible to write any state of the CV mode in the subsystem basis to identify the logical and gauge components. Here we describe this procedure, which can be performed in various ways that will be used throughout the rest of this work.

Applying the logical-subsystem identity to an arbitrary, pure CV state $\ket{\psi}$ gives
\begin{align} \label{eq:babysfirstdecomp}
    \ket{\psi} &= \sum_{\ell=0}^{d-1} \ket{\ell}[L] \otimes \ket{\bar{\psi}_\ell}_G
    \, ,
\end{align}
where the gauge-mode state associated with each logical state, $\ket*{\bar{\psi}_\ell}[G]
 \coloneqq \braket{\ell}{\psi}[L][]$, arises from a partial projection and as such is not normalized (indicated by the overbar). Also, the gauge-mode states associated with different values of $\ell$ are typically not orthogonal.
 \blk
Expanding $\ket*{\bar{\psi}_\ell}[G] $ in the gauge-mode position basis gives the subsystem decomposition,
\begin{align}
    \ket{\psi} &=\sum_{ \ell = 0 }^{d-1}
    \int_{-\infty}^{+\infty} dx_G\
    \psi(\ell,x_G)
    \ket{\ell}[L] \otimes \ket{x_G}[q,G]
    \,,
    \label{eq:babysseconddecomp}
\end{align}
where $\psi(\ell,x_G) \coloneqq \big( \bra{\ell}[L] \otimes \bra{x_G}[q,G]  \big)\ket{\psi}$.
The gauge mode can also be expanded in a partioned-position basis, which will be useful for calculations throughout this work. Doing this yields
{\shrinkeq
\begin{align} \label{eq:babysthirddecomp}
    \ket{\psi} &=\sum_{ \ell = 0 }^{d-1}  \sum_{\mG\in\integers} \intalpha d\uG\
    \psi(\ell,\mG,\uG)
    \ket{\ell}[L] \otimes \ket{\mG,\uG}[G]
    \ ,
\end{align}%
}
with subsystem-basis wave function
\begin{align} \label{eq:subsystemwavefunction}
    \psi(\ell,\mG,\uG) \coloneqq \big( \bra{\ell}[L] \otimes \bra{\mG,\uG}[G]  \big)\ket{\psi}
    \, .
\end{align}
Note that the subsystem wave functions for fixed $\ell$ are indeed the gauge-mode wave functions for each $\ket{\bar{\psi}_\ell}$ in \eqr{eq:babysfirstdecomp}.
If a position wave function is $\psi(x)$, $x \in \reals$, the corresponding subsystem-basis wave function is obtained by setting $\psi(\ell,\mG,\uG) = \psi(\alpha \ell + d \alpha\mG + \uG)$ with $\ell \in \{0,\dotsc,d-1\}$, $\mG \in \integers$, and $\uG \in [-\frac{\alpha}{2},\frac{\alpha}{2})$.

For mixed CV states, the procedure is analogous, with the subsystem decomposition applied to the density matrix. A state $\op{\rho}$ is expressed in the position basis as
\begin{align} \label{eq:densitymatrixpositionbasis}
    \op{\rho} = \iint dx dx' \rho(x,x')  \ketbra{x}{x'}[q][q]
    \; ,
\end{align}
with matrix elements $\rho(x,x') \coloneqq \bra{x}[q]\op{\rho} \ket{x'}[q]$.
Using \eqr{eq:ssdbasis}, the state can be expressed in the subsystem basis,
\begin{align} \label{eq:densitymatrixSSD}
    \op{\rho}
    &= \int \hspace{-7pt} \intalpha du du' \,
    \sum_{m,m' \in \mathbb{Z}} \,
    \sum_{\ell,\ell' = 0}^{d-1} \nonumber
    \\
    & \quad \times
    \rho(\ell, \ell', m, m', u,u')
    \ketbra{\ell}{\ell'}[L][L] \otimes \ketbra{m, u}{m', u'}[G][G]
\end{align}
with the matrix elements,
\begin{align}
    \rho (\ell, \ell', & m, m', u,u') \nonumber \\
    &\coloneqq
    \bigl( \bra{\ell}[L] \otimes \bra{m,u}[G] \bigr)
    \op{\rho}
    \bigl( \ket{\ell'}[L] \otimes \ket{m',u'}[G] \bigr)
    \, ,
\end{align}
where we have used the gauge-mode partitioned-position basis, just as in \eqr{eq:babysthirddecomp} (because that is what we use throughout this work).
One could also use the gauge-mode position basis as in \eqr{eq:babysseconddecomp}.
Note that for notational convenience, we have dropped the gauge subscript $G$ everywhere except to label the states. Henceforth, we continue with this convention unless it could cause confusion. As a helpful reminder, whenever an object (such as the matrix element above) is labeled by a logical index $\ell$, $m$ and $u$ within that same expression refer to the gauge mode.

\subsubsection{ Reduced subsystem states: logical and gauge-mode }
\label{sec:partitionedpositionbasis2}

Performing a partial trace of the CV state $\op{\rho}$ over one subsystem---either the logical qudit or the gauge mode---gives the reduced state of the other subsystem. This is useful when the CV state is intended to encode discrete quantum information. In this case, the partial trace over the gauge mode, the \emph{gauge trace}, gives a map from the full CV space to a discrete Hilbert space, after which the tools of standard qubit and qudit quantum information can be applied~\cite{pantaleoni_modular_2020,wan_memory-assisted_2020,tzitrin_progress_2020,hastrup2020cubic}.

The reduced logical state, a $d \times d$ density matrix $\op{\rho}_L$, is found by taking the gauge trace (performed here in the gauge-mode partitioned-position basis),
\begin{align}
    \label{eq:gaugetrace}
    \op{\rho}_L \coloneqq
    \Tr_G [ \op{\rho} ]
    = \sum_{m \in \mathbb{Z}} \intalpha du \,
    \gbra{m,u}
    \op{\rho}
    \gket{m,u}
    \, .
\end{align}
From the standpoint of the reduced logical state, the gauge-mode subsystem acts as an environment, such that any entanglement between the subsystems decoheres $\op{\rho}_L$. Similarly, the reduced gauge-mode state $\op{\rho}_G$ is obtained by a partial trace over the logical qudit subsystem,
\begin{align}
    \label{eq:reducedgaugestate}
    \op{\rho}_G \coloneqq
    \Tr_L [ \op{\rho} ]
    = \sum_{\ell = 0}^{d-1}
    \lbra{\ell}
    \op{\rho}
    \ket{\ell}[L]
    \, .
\end{align}
The reduced gauge state is the state of the virtual mode whose environment is the discrete logical subsystem.

In the following sections, we are interested in encoding discrete quantum information into CV states. For a CV state $\op \rho$ to faithfully encode a given qudit state $\op \sigma$, its reduced logical state $\op{\rho}_L$ must be near $\op \sigma$
(according to some metric).
Ideally, the CV state is a product state over the subsystems,
$
\op \rho = \op{\sigma} \otimes \op{\rho}_{G}
$,
such that the gauge trace yields $\op{\sigma}$. Note that any transformations on the gauge-mode part of the state $\op{\rho}_G$ leave the reduced logical state untouched. More generally, the logical and gauge subsystems are entangled, and as a result the reduced logical state is mixed (as is the reduced gauge state). How well discrete quantum information is encoded into a CV state can be quantified using tools from standard qubit (or qudit) quantum information.  We focus on the \emph{logical fidelity}~\cite{pantaleoni_modular_2020},
\begin{equation} \label{eq:logicalfidelity}
    \mc{F}_\mathcal{L}(\op \rho, \op \sigma ) \coloneqq
    \left(\Tr \left[
    \sqrt{
    \sqrt{ \op \sigma  }
    \Tr_G[\op \rho]
    \sqrt{ \op \sigma  }
        }
    \right] \right)^2
    \, ,
\end{equation}
where $\Tr_G[\op \rho] = \op{\rho}_L$ from \eqr{eq:gaugetrace}. One can define other logical-subsystem metrics as needed.

\subsection{Decomposing single-mode CV operators} \label{sec:singlemodeoperatordecomposition}

Operators on a single CV mode, decomposed in accordance with $\Hcv \simeq \complex^d\otimes\Hcvp$, can be classified into three types. Borrowing terminology from the subsystem-code literature~\cite{poulin_stabilizer_2005}, we categorize them as logical, gauge, and interaction operators.

Logical operators of the form $\op A _L \otimes \id_ G$ (or simply $\op A _L$) act nontrivially only on the logical subsystem in $\complex^d$, whereas gauge operators, $\id _L \otimes \op A _G $ (or simply $\op A _G$), act only on the gauge mode in $\Hcvp$. Note that any purely gauge operations that a CV state undergoes have identical effects under a gauge trace. That is, a tensor-product operation $\op A _L \otimes \op A _G$ is logically equivalent to $\op A _L $---\emph{i.e.},~they act as the same operation on the logical subsystem.

When a subsystem-decomposed CV operator cannot be written as a tensor product between a logical and a gauge operation, it is an \emph{interaction operator}. The effect of an interaction operator can be described by considering its action on tensor-product states. A tensor-product state  $\ket{\psi}[L] \otimes \ket{\phi}[G]$, can be interpreted as a fully faithful representation of the pure qubit state $\ket{\psi}[L]$, since the states coincide upon gauge tracing. Furthermore, the same faithfully encoded state after an unitary operation $\op A _L \otimes \op A _G$ coincides with $\op A _L \ket{\psi}[L]$ after gauge tracing.

Consider, on the other hand, the action on $\ket{\psi}[L] \otimes \ket{\phi}[G]$ of an entangling operation between the logical and gauge subsystems.  In this case, gauge tracing generally results in a mixed logical state due to residual entanglement between the subsystems. This degradation of purity is likely to be unwanted, since it cannot be mitigated using unitary operations on the logical subsystem alone. In other words, we can describe the gauge mode as a (virtual) environment subsystem, and we can identify coupling with the environment via interaction operations as a source of decoherence of the logical states.

In this work, we decompose several operators useful for CV cluster state quantum computing: unitary momentum- and position-shift operators, and a nonunitary ``envelope'' operator defined in Sec.~\ref{sec:finitesqueezevacuum}.

\subsection{ Decomposing arbitrary single-mode shifts } \label{sec:arbshifts}
The decomposition of a single-mode CV operator, described above, can be performed using the resolution of the identity in the subsystem basis, \eqr{eq:subsystembasisidentity}. We apply the subsystem decomposition to shift operators, which are sufficient to reconstruct CV error operations (see Ref.~\cite{terhal_encoding_2016} for the procedure.) More to the point, shifts are useful in two contexts that we are specifically interested in here. First, in CV cluster-state teleportation, active shifts follow each step in the teleportation to correct for random homodyne measurement outcomes~\cite{menicucci_universal_2006}, and second, for quantum computing with GKP  codes (introduced below), specific shifts implement the logical Pauli gates.

General translations in phase space are generated by the displacement operator
$\op{D} (\alpha) \coloneqq
e^{\alpha \op{a}^\dagger - \alpha^* \op{a} }$,
where
$\alpha = \alpha_R + i \alpha_I
$,
with $\alpha_R$ and $\alpha_I$ being the real and imaginary parts of the complex number $\alpha$, respectively. A displacement operator can be separated as
\begin{align}
    \op{D}(\alpha)
        = e^{ i \alpha_R \alpha_I} \op{X}(\sqrt{2} \alpha_R ) \op{Z}(\sqrt{2} \alpha_I)
        \, ,
\end{align}
where shifts in position by $s$ and in momentum by $t$ are, respectively, generated by
    \begin{subequations}
    \begin{align}
        \op{X}(s) & \coloneqq e^{-i s \op{p}} \,, \label{eq:posshift} \\
        \op{Z}(t) & \coloneqq e^{i t \op{q}} \,. \label{eq:momshift}
    \end{align}
    \end{subequations}
We perform a SSD of each of these shift operators below.
The momentum-shift operator decomposes simply, because it is diagonal in $\op{q}$, while the position-shift operator is not and will be decomposed by inspecting its action on a subsystem basis state.

\subsubsection{ Momentum shifts }
\label{subsec:momshifts}

The momentum-shift operator $\op{Z}(t)$, \eqr{eq:momshift}, can be decomposed straightforwardly due to the fact that it is generated by $\op{q}$, which decomposes according to \eqr{eq:qsubsystemdecomp}. Substituting the subsystem decomposition of the $\q$ operator gives
\begin{align}
    \Z\pqty{t}
    = e^{ i t \q }
    = e^{ i t \pqty{ \alpha \l + \alpha d \mGop + \uGop} }
    =
    e^{ i \alpha t \l }
    e^{ i \alpha d t \mGop }
    e^{ i t \uGop }
    \, ,
    \label{eq:zshiftdecomposed}
\end{align}
which is a tensor product between logical and gauge operators; no logical-gauge entangling terms are present.

To understand the purely logical piece, we gain some intuition by considering the case $d=2$. In this case, we get the following representation of the logical part of the operator
\begin{equation}
    \label{eq:zrotation}
    e^{i \theta \l}
    =
    e^{ i \theta / 2 } \op R_{L}^z (\theta)
    \, ,
\end{equation}
where $\op R_{L}^z (\theta) \coloneqq \text{exp} (- i \tfrac{\theta}{2} \Z_{L} )$ is the usual operator that rotates the logical qubit in the Bloch sphere by $+\theta$ about the $Z$ axis. This relation follows from $ \l = \half\pqty*{\id_L - \ope{Z}_L} $. Thus, $\text{exp} (i \theta \l)$ is equivalent to a simple rotation of the logical qubit up to a $\theta$-dependent phase.
In \eqr{eq:zshiftdecomposed}, then, $\theta = \alpha t$.

\subsubsection{ Position shifts }
The position-shift operator $\X(t)$, \eqr{eq:posshift}, is not diagonal in $\op{q}$ and thus takes a more complicated form in the subsystem decomposition. Again note that we drop the labels on gauge eigenvalues (unless necessary to avoid confusion).
We separate a position shift by $t \in \mathbb{R}$ into its integer and fractional parts with respect to bin size $\alpha$ and then further decompose the integer part using a modulo operation with respect to $d$ to get
\begin{align}
    t &= \alpha(dn + k) + v
    \,,
\end{align}
where
    \begin{subequations}
    \begin{align}
        k &\coloneqq \fracpart{I_{\alpha}(t)}{d}
        \, ,
        \\
        n &\coloneqq I_{d}(I_{\alpha}(t) )
        \, ,
        \\
        v &\coloneqq \fracpart{t}{\alpha}
        \, .\,
    \end{align}
    \end{subequations}
This decomposition is the same as was done in \eqr{eq:doubledecomp}.
Then, the position-shift operator separates into two parts,
\begin{equation} \label{eq:shiftdecomp}
    \X\pqty{t} =  \X\pqty*{v} \X\bqty{\alpha \pqty{d n + k}}
    \, ,
\end{equation}
and a shift by a value $v \in [-\frac \alpha 2,\frac \alpha 2)$. The integer-shift portion of \eqr{eq:shiftdecomp} acts on a subsystem-basis eigenstate, \eqr{eq:ssdbasis}, as
\begin{align}
    &\X\bqty{\alpha \pqty{d n + k}}
        \ket{\ell}[L] \otimes \ket{m,u}[G]
         \nonumber \\
    &\ =
    (\X_L)^{k}
    \X_G\bqty{ \alpha n + \alpha I_{d}(\ell+k)}
    \ket{\ell}[L] \otimes
    \ket{m,u}[G]
    \, ,
    \label{eq:Xlargeshift}
\end{align}
where $\op{X}_L$ is the logical-qudit operator in \eqr{eq:quditX}.
Since the values of $\ell$ and $k$ are restricted to the set $\{0,\dots,d-1\}$, $I_{d}(\ell+k)$ can only assume the values $0$ or $1$.
The gauge-mode displacement operator $\X_G(t)$ is defined by its action on gauge-mode, partitioned-position basis states:
\begin{equation}
    \X_G(t) \ket{m,u}_G \coloneqq
        \ket{
        m +
        I_{\alpha}(u+t)
        ,
        \fracpart{u + t}{\alpha}}[G]
    \, .
\end{equation}
The fractional shift portion of \eqr{eq:shiftdecomp} acts on subsystem-basis eigenstates as
    \begin{align}
     &\X\pqty{v} \ket{\ell}[L] \otimes \ket{m,u}[G] \nonumber
     \\
     &\quad =
     \X_L^{I_{\alpha}(u+v)}
     \X_G\bqty{
         \alpha I_{d}\bqty{\ell+I_{\alpha}(u+v)}
         - \alpha I_{\alpha}(u+v)
     }
     \nonumber \\* & \quad \quad \times
     \X_G ( v )
     \ket{ \ell }[L] \otimes \ket{ m , u }[G]
         \label{eq:Xsmallshift}
         \, .
     \end{align}
Since $v$ is small, $I_{\alpha}(u+v)$ can only take on values $\{0, \pm 1\}$. Combining Eqs.~(\ref{eq:Xlargeshift}) and~(\ref{eq:Xsmallshift}), we find the SSD of the position shift operator on a subsystem-basis eigenstate,
{\shrinkeqless
\begin{align}
    &\X\pqty{t} \ket{\ell}[L] \otimes \ket{m,u}[G] \nonumber
    \\*
    &\quad= (\X_L) ^ {k + I_{\alpha}(u+v)}
    \X_G\bqty{ \alpha I_{d}(\ell+k)}
    \X_G\pqty{\alpha n  + v}
    \nonumber \\* & \quad\quad \times
    \X_G\bqty{\alpha
    I_{d}\bqty{\ell + I_{\alpha}(u+v)}  - \alpha I_{\alpha}(u+v)
    }
    \ket{\ell}[L] \otimes \ket{m,u}[G]
    \, .
    \label{eq:decXprod}
\end{align}%
}%
Applying the shift operator on the right-hand side, we can see the action on the subsystem basis state:
\begin{align}
    &\X(t)
    \ket{\ell}[L] \otimes \ket{m,u}[G]
    \nonumber \\
    & \quad =
    \ket{
    \fracpart{
    \ell+k
    + I_{\alpha}(u+v)
    }{d}
    }[L]
    \nonumber \\
    & \qquad \otimes
    \ket{
    n + m +
    I_d \bqty{
    \ell+k + I_{\alpha}(u+v)
    }
    ,
    \fracpart{u + v}{\alpha}
    }[G]
    \, .
    \label{eq:decXbasis}
\end{align}
A more detailed derivation can be found in Appendix~\ref{app:posshiftdec}.

The fact that the SSD is based on a modular decomposition of a real number, \eqr{eq:maps}, is useful for interpreting these equations. A shift by $t$ on a position eigenstate is effectively addition of two real numbers, \emph{i.e.},~$\X(t)\ket{x}[q] = \ket{t + x}[q]$. Since we are representing those real numbers in terms of their logical, gauge-integer ($m$) and and gauge-modular ($u$) registers, the decomposed-$\X(t)$ describes addition between these registers with carrying. A difficulty lies in the fact that the registers themselves are not of the same size, in contrast to binary digit representations, for example.
In this way, a subsystem-basis position shift is analogous to a quantum adder~\cite{quantumadder}, which is a complicated circuit that entangles the registers. Our version is further complicated by the fact that the three subsystems are of different Hilbert-space dimension.

\section{ Decomposing squeezed vacuum and Gottesman-Kitaev-Preskill states }
\label{sec:decsqueezedandGKP}

When $d=2$ (\emph{i.e.},~we are interested in qubits), every pure CV state for a mode has an SSD [using the form in \eqr{eq:babysfirstdecomp}],
    \begin{equation} \label{eq:logicaldecomp}
        \ket{\psi} = \ket{0}[L] \otimes \ket{\bar{\psi}_0}[G] + \ket{1}[L] \otimes \ket{\bar{\psi}_1}[G]
        \, ,
    \end{equation}
where $\ket{\bar{\psi}_0}$ and $\ket{\bar{\psi}_1}$ are states of the gauge mode associated with the logical $\ket{0}[L]$ and $\ket{1}[L]$ states. Note that $\ket{\bar{\psi}_0}$ and $\ket{\bar{\psi}_1}$ are generally unnormalized and not orthogonal. When $\ket{\bar{\psi}_0} \propto \ket{\bar{\psi}_1}$, $\ket \psi$ is a tensor-product state across the subsystems. Such states have no logical-gauge entanglement and pure reduced states in both subsystems.

An important example is the 0-momentum eigenstate, which decomposes simply in the three bases we have considered:
\begin{subequations} \label{eq:0momentumeig}
    \begin{align}
    \label{eq:0momentumdef}
        \ket{0}[p] &= \onrtwopi \int dx\, \ket{x}[q] \\
        & = \onrtwopi \sum_{m \in \mathbb{Z}} \intalpha du\, \ket{m,u}
        \label{eq:0momentumdepp} \\
        &= \ket{+}[L] \otimes \ket{0}_{p,G}
    \, .
    \end{align}
\end{subequations}
The first line gives the position representation and the second the partitioned-position decomposition. The third line gives the SSD, revealing that a 0-momentum eigenstate encodes a logical $\ket{+}[L]$ state and another 0-momentum state in the gauge mode. 0-momentum eigenstates are the building blocks for ideal continuous-variable cluster states and the essential ancillae for ideal CV teleportation.

Another important class of states with tensor-product SSDs are GKP
states, introduced to encode a qubit into a harmonic oscillator~\cite{gottesman_encoding_2001}. GKP encodings have a number of features that make them appealing for quantum computation including Gaussian Clifford operations, Gaussian magic-state preparation~\cite{baragiola_all-gaussian_2019, yamasaki2020cost}, resistance to loss~\cite{albert_performance_2018}, and fault tolerance when used in tandem with canonical CV cluster states~\cite{menicucci_fault-tolerant_2014}. Here, we focus on the structure of GKP states; more detail about GKP codes and their use for quantum computing can be found, \emph{e.g.}, in Refs.~\cite{gottesman_encoding_2001,albert_performance_2018,mensen2020phasespace}.

GKP codes use periodic wave functions to define a two-dimensional subspace into which a qubit is encoded. For a GKP code with position-wave function periodicity $2\alpha$, the orthogonal computational-basis codewords are
\begin{equation} \label{eq:jGKP}
    \ket{j_{\GKP}} = \sum_{m \in \integers} \ket{\alpha (2m + j)}[q] \, ,
\end{equation}
which are used in superposition for an arbitrary GKP state,
\begin{equation} \label{eq:arbGKP}
    \ket{\psi_{\GKP}} = c_0 \ket{0_{\GKP}} + c_1 \ket{1_{\GKP}} \, ,
\end{equation}
with complex amplitudes $c_0$ and $c_1$ that specify a qubit state and satisfy $|c_0|^2 + |c_1|^2 = 1$. The position wave functions for the computational basis states are
\begin{align}
    \psi_{\GKP,j}(x)
    &= \sum_{m\in \mathbb{Z}} \delta(x-\alpha (2m + j)) \\
    &=\Sh_{2\alpha}(x - \alpha j) \label{eq:GKPjwavefunction}
    \, ,
\end{align}
where the $\Sh_T$-function (pronounced ``sha'') is a sum of $\delta$-functions with period $T$,
\begin{equation}
    \Sh_{T}(x) \coloneqq \sum_{n \in \mathbb{Z}} \delta(x-nT)
    \, ,
\end{equation}
also known as a \emph{Dirac comb}.\footnote{Note that in other sources, $\Sh$-functions are defined with scaling factors. For example, Ref.~\cite{mensen2020phasespace}, which includes two of the current paper's authors, scales by the period $\sqrt{T}$.}

It is the periodicity of GKP states that makes their SSD simple. This is no coincidence, as modular bosonic subsystem codes~\cite{pantaleoni_modular_2020} were designed as a generalization of GKP codes.\footnote{A related decomposition based on a continuous set of GKP subspaces was given by Ketterer \emph{et al.}~\cite{ketterer_quantum_2016}. See also Ref.~\cite{fabre_wigner_2020} for a phase-space extension.} The subsystem decomposition proceeds as follows. Each term in the sum in \eqr{eq:jGKP} arises from a periodically placed position eigenstate, which is decomposed using Eqs.~(\ref{eq:ssdbasis})--(\ref{eq:maps}), into
$
\ket{2m + \ell, 0} = \ket{\ell}[L] \otimes \ket{m,0}[G]
$.
This gives
\begin{align} \label{eq:GKPjpartitionedposition}
        \ket{j_{\GKP}} &= \sum_{m \in \integers} \ket{2m + j, 0}
        \\*
        &= \ket{j}[L] \otimes \ket{+_\GKP}[G]
        \, .
\end{align}
From the partitioned-position decomposition in the first line to the SSD in the second, we performed the sum over $m$ to find a $\ket {+_\GKP}$ in the gauge mode (whose position wave function is a Dirac comb spaced by $\alpha$).
Using the SSD of $\ket{j_\GKP}$, we find that arbitrary GKP states, \eqr{eq:arbGKP}, decompose as
\begin{align} \label{eq:decomposedGKPqubits}
    \ket{\psi_{\GKP}}
    = \ket{\psi}[L] \otimes \ket{+_{\GKP}}[G]
    \, ,
\end{align}
where the encoded logical state $\ket{\psi}[L] = c_0 \ket{0}[L] + c_1 \ket{1}[L]$ is the same as the intended qubit state specified at the CV level (given by $c_0$ and $c_1$).

Interestingly, a 0-momentum state and a $\ket{+_\GKP}$ state contain the same logical-subsystem state, $\ket{+}[L]$, and differ only in their gauge-mode state. However, neither 0-momentum states nor ideal GKP states are physical---they both have infinite energy and cannot be normalized. Physical approximations to these states contain some amount of \emph{embedded error}~\cite{gottesman_encoding_2001}, also called finite-squeezing noise~\cite{menicucci_fault-tolerant_2014}, that limits their energy. A consequence is that approximate states are no longer product states in the SSD---their position wave functions are not entirely localized to within the $m$-parity-labeled bins that define the logical subsystem. Below, we consider the SSD and reduced logical state for momentum-squeezed vacuum---which approximate 0-momentum eigenstates---and for one particular approximation to GKP states.

\subsection{ Squeezed vacuum states }
\label{sec:finitesqueezevacuum}

A normalized squeezed vacuum state,
\begin{subequations} \label{eq:sqzmomboth}
\begin{align}
    \ket{0, \sparam }[p]
    &\coloneqq
    \left(\frac{\sparam^2}{ \pi}\right)^{1/4} \int dx \,
    e^{-\frac{\sparam^2}{2} x^2} \ket{x}[q]
    \label{eq:momsqz1}
    \\
    &=
    \sqrt{2\pi}
    \left(\frac{\sparam^2}{ \pi}\right)^{1/4}
    e^{-\frac{\sparam^2}{2} \op{q}^2}
    \ket{0}[p]
    \label{eq:momsqz2}
    \, ,
\end{align}
\end{subequations}
has measured variance $\langle \op{q}^2 \rangle = \frac{1}{2 \sparam^2}$ in the position quadrature and $\langle \op{p}^2 \rangle = \frac{\sparam^2}{2}$ in the momentum quadrature. For values of $\sparam < 1$, the state is squeezed in momentum, which is the parameter regime we are interested in.\footnote{Connections between various representations of squeezed vacuum states and their parametrizations can be found in Ref.~\cite{walshe_2020}.}
Occasionally (especially in figures), we will express $\sparam$ in decibels:
\begin{equation} \label{eq:kappadB}
    \sparam \, (\text{dB}) = -10 \log_{10}{\sparam^2}
    \, .
\end{equation}
The reported value is known as \emph{the squeezing} (reported in decibels), while $\sparam$ itself is called a \emph{squeezing factor}~\cite{alexander_noise_2014}. This expression reports the measured quantum noise (variance) of a quadrature in decibels, with the reference value being the vacuum-noise variance of~$\frac 1 2$.
Equation~\eqref{eq:kappadB} holds for any quantity playing the role of a squeezing factor, including, \emph{e.g.},~$\Delta$ or $\kappa$ from the original definition of an approximate GKP state~\cite{gottesman_encoding_2001}.

In \eqr{eq:momsqz2}, we use Eq.~\eqref{eq:0momentumdef} to represent the state as an \emph{envelope operator}, $\exp\pqty{- \tfrac{1}{2} \sparam^2 \q^2}$, acting on a 0-momentum eigenstate. Since the envelope operator is diagonal in $\op{q}$, it decomposes straightforwardly in the SSD:
\begin{align}
    \exp\pqty{
    - \tfrac{1}{2} \sparam^2 \q^2
        }
      &=
    \envop_L \envop_G \envop_\text{int}
      \label{eq:envelopecs}
      \, ,
\end{align}
where we define mutually commuting logical, gauge, and interaction envelope operators:
\begin{subequations}
\begin{align}
    \envop_L &\coloneqq
      \exp\pqty{ \!
    - \tfrac{1}{2} \sparam^2
          \alpha^2 \l ^{2}
          \! } \, , \\
    \envop_G &\coloneqq
      \exp\bqty{ \!
          - \tfrac{1}{2} \sparam^2
          \pqty{
          2\alpha\mGop + \uGop}^2
      }
      \, , \\
          \envop_\text{int} &\coloneqq
    \exp\bqty{ \!
          - \sparam^2
          \alpha \l \otimes \pqty{
          2\alpha\mGop + \uGop}
      }
      \, .
\end{align}
\end{subequations}
The logical envelope operator $\envop _L$, written alternatively as
\begin{subequations}
\begin{align}
    \envop_L &= \ketbra{0_L}{0_L} + \eta \ketbra{1_L}{1_L} \\
    &=
    \tfrac{1}{2}( 1 + \eta) \id_L    + \tfrac{1}{2} ( 1 - \eta  ) \Z_L
    \, ,
    \label{eq:logenv}
\end{align}
\end{subequations}
serves to decrease the relative amplitude of $\ket{1_L}$ of a logical-subsystem state with weight
    \begin{align}
    \eta
        &\coloneqq \exp\pqty{ - \tfrac{1}{2} \sparam^2 \alpha^2 }
    \end{align}
that depends on the squeezing factor~$\sparam$. The interaction envelope operator $\envop_\text{int}$ generates entanglement between the subsystems and can be written similarly to \eqr{eq:logenv},
\begin{align}
    \envop_{\text{int}}
    &=
    \tfrac{1}{2} \op{I}_L \otimes (\op{I}_G + \op{\eta}_G ) + \tfrac{1}{2} \Z_L \otimes (\op{I}_G - \op{\eta}_G)
    \, ,
\end{align}
with the $\eta$ factor replaced by a gauge-mode operator
\begin{align}
    \op{\eta}_G
    &\coloneqq \exp\bqty{
        - \sparam^2 \alpha \pqty{
        2 \alpha \mGop
        + \uGop }
    }
    \, .
\end{align}
Thus, the interaction envelope operator acts similarly to $\envop_L$, except that the applied logical envelope is conditional on the state of the gauge mode. Finally, the gauge-mode envelope operator $\envop_G$ acts only on the gauge-mode subsystem.

In order to find the subsystem decomposition of a squeezed vacuum state, we use the decomposed envelope operator, \eqr{eq:envelopecs}, in \eqr{eq:momsqz2}. With the SSD of a 0-momentum eigenstate, $\ket{0}[p] = \ket{+}[L]\otimes \ket{0}[p,G]$ [\eqr{eq:envelopecs}], we can then apply each piece of the envelope operator to the appropriate subsystem(s).
Using \eqr{eq:0momentumdepp}, we recognize that the 0-momentum eigenstate in the gauge mode satisfies an important relation:
\begin{align}
    \ket{0}[p,G]
    = \frac{1}{ \sqrt{2\pi} }
    \sum_{m \in \integers} \intalpha du \ket{m,u}[G] \, ,
\end{align}
which allows us to directly apply any operators that act only on the gauge mode, namely
$\envop_G\ket{m,u} = \epsilon_G\pqty{m,u}\ket{m,u}$
and
$\op{\eta}_G\ket{m,u} = \eta_G\pqty{m,u} \ket{m,u}$, where
\begin{subequations}
\begin{align}
    \epsilon_G(m,u)
    & \coloneqq
    \exp[ - \tfrac{1}{2} \sparam^2 (2\alpha m + u)^2] \, , \\
    \eta_G(m,u)
    & \coloneqq
    \exp[ - \sparam^2 \alpha(2\alpha m + u)].
\end{align}
\end{subequations}
With these tools, we obtain the decomposition of a squeezed state from \eqr{eq:envelopecs}:
\begin{subequations}
\begin{align}
\ket{0, \sparam }_p
   &= \left( \frac{\sparam^2}{\pi}\right)^{1/4}
   \envop_L \envop_G \envop_{\text{int}}
    \ket{+}[L]
     \sum_m \intalpha
     du \ket{m,u}[G] \\
    & =
   \ket{0}[L] \otimes \gket{\bar{\psi}_0}
   +
   \ket{1}[L]\otimes \gket{\bar{\psi}_1}
   \label{eq:momsqzdecomp}
    \, ,
\end{align}
\end{subequations}
where $\gket{\bar{\psi}_0}$ and $\gket{\bar{\psi}_1}$ are states of the gauge mode---see \eqr{eq:logicaldecomp}.
The subsystem wave functions for squeezed vacuum,
\eqr{eq:subsystemwavefunction},
\begin{align}
\label{eq:squeezedvacsubsystemeqs}
    \psi_\zeta\pqty{\ell, m,u} &\coloneqq \big( \bra{\ell}[L]\otimes \gbra{m,u} \big) \pket{0, \zeta}\\
    &= \left( \sparam^2 \pi^{-1} \right)^{1/4}  \big[ \eta\,  \eta_G\pqty{m,u} \big]^\ell \epsilon_G\pqty{m,u}
    \, ,
\end{align}
describe the two pieces of the state above according to the logical label $\ell$.
Notice that these subsystem wave functions are not proportional,
\begin{align}
    \psi_\zeta\pqty{1, m,u} = \eta\, \eta_G\pqty{m,u} \psi_\zeta \pqty{0, m,u}
    \, ,
\end{align}
since $\eta_G$ is not a constant function.
This is equivalent to the relation $\ket{\bar{\psi}_0}[G] \not\propto \ket{\bar{\psi}_1}[G]$, meaning that \eqr{eq:momsqzdecomp} is not a tensor-product state between logical and gauge subsystems and therefore does not faithfully encode $\ket{+}$.
However, this faithful encoding is achieved in the limit of high squeezing, $\sparam \rightarrow 0$, where $\eta, \eta_G \rightarrow 1$ giving $\ket{\bar{\psi}_0}[G] = \ket{\bar{\psi}_1}[G]$.

The reduced logical state is found by tracing over the gauge mode [\eqr{eq:gaugetrace}]
    \begin{align} \label{eq:sqzvaclogicalstate}
    \op{\rho}_L
    & = \Tr_G
    \bqty{
    \ket{0, \sparam}[p] \bra{0,\sparam}[p]
    }\\
    & =
    \sum_{m} \intalpha du
    \sum_{\ell, \ell'}
    \psi_{\zeta}(\ell, m,u) \psi_\zeta^*(\ell',m,u)
    \ketbra{\ell}{\ell'}[L][L]
    .
\label{eq:momeigderivation}
\end{align}
The sum over $m$ in the subsystem wave functions, \eqr{eq:squeezedvacsubsystemeqs}, can be rewritten in terms of Jacobi theta functions of the third kind~\cite{bellman_brief_2013}
\begin{align}\label{eq:thetadef}
\vartheta(z,\tau)
\coloneqq
\sum_{m \in \integers}
    \exp\pqty{
    \pi i m^2 \tau + 2\pi i m z
    }
    \, ,
\end{align}
where $z$ is a complex variable, and $\tau$ is a complex number with positive imaginary part. After performing the sum over $m$ in \eqr{eq:momeigderivation}, we obtain the reduced logical density matrix in the computational basis,
{\shrinkeq%
\begin{align}
    &\op{\rho}_{L} \nonumber \\*
    &=\frac{1}{2 \alpha}
    \intalpha du
    \begin{pmatrix}
        \vartheta\pqty{ \frac{u}{ 2 \alpha }
        , \frac{\tau_{\sparam^{-1}}}{2} }
        &
        e^{\frac{-\alpha^2 \sparam^{2}}{4}}
        \vartheta\pqty{ \frac{u}{ 2 \alpha } \!+\! \tfrac{1}{4}
        , \frac{\tau_{\sparam^{-1}}}{2} }
        \\
        e^{\frac{-\alpha^2 \sparam^{2}}{4}}
        \vartheta\pqty{\frac{u}{ 2 \alpha } \!+\! \tfrac{1}{4}
        , \frac{\tau_{\sparam^{-1}}}{2} }
        &
        \vartheta\pqty{\frac{u}{ 2 \alpha } \!+\! \tfrac{1}{2}
        , \frac{\tau_{\sparam^{-1}}}{2} }
    \end{pmatrix}
    \, ,
\end{align}%
}%
with $\tau_{\sparam^{-1}} = 2 \pi i / (2 \alpha \sparam)^2 $, using the general definition for $\tau_\sigma$ given below in \eqr{eq:taufactor}.

\begin{figure}[t!]
    \includegraphics[width=0.45\textwidth]{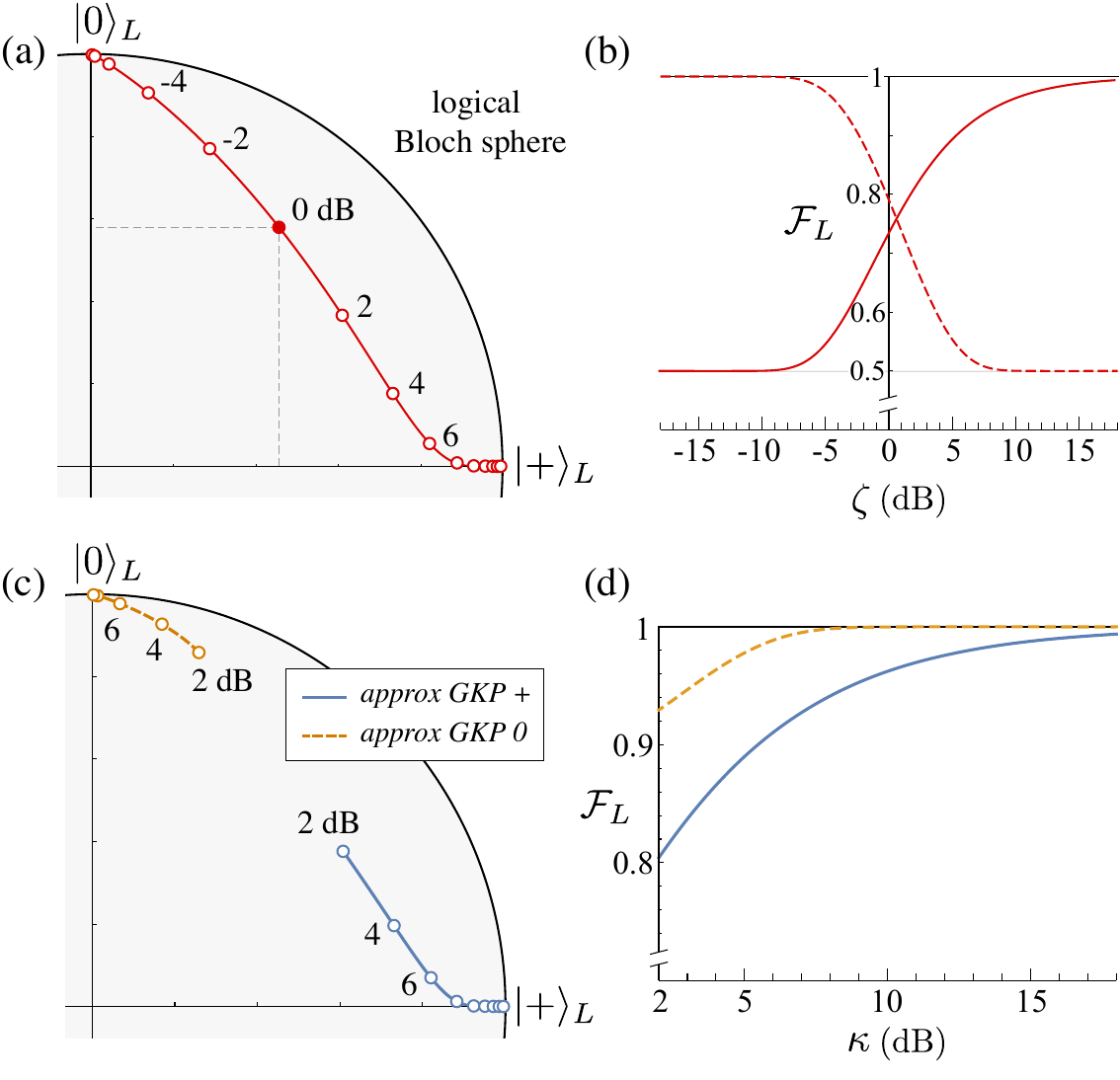}
    \caption{
    Reduced logical state associated with squeezed vacuum states, \eqr{eq:sqzmomboth}, and approximate GKP states, \eqr{eq:GKPstate} for $\alpha = \sqrt{\pi}$.
    (a) Logical Bloch vectors in the $xz$ plane for squeezed states, \eqr{eq:sqzvaclogicalstate}, with squeezing $\sparam$~(dB) [\eqr{eq:kappadB}] between $-18$~dB (highly squeezed in position) and 18~dB (respectively, in momentum). Circles are shown at increments of 2~dB, and 0~dB is the vacuum state. Near the end points, some circles are hidden beneath others.
    (b) Logical fidelity $\mathcal{F}_L$, \eqr{eq:logicalfidelity}, with the qubit states $\ket{+}$ (solid) and $\ket{0}$ (dashed).  The logical state approaches $\ket{+}$ for high momentum squeezing ($\sparam \rightarrow 0$) and $\ket{0}$ for high position squeezing ($\zeta \rightarrow \infty$), which is equivalent to high momentum anti-squeezing.
    (c) Logical Bloch vectors, \eqr{eq:firstgaugetrace}, for approximate GKP states with symmetric embedded error in position and momentum ($\kappa = \Delta)$. Shown are $\ket {+_\GKP}$ ($c_0 = c_1 = \frac{1}{\sqrt{2}}$) and $\ket {0_\GKP}$ ($c_0=1, c_1=0$) states for $\kappa$ in the range 2--18 dB, with circles shown at increments of 2 dB.
    (d) Logical fidelity with intended qubit states $\ket{+}$ (solid) and $\ket{0}$ (dashed). Each logical state approaches its intended qubit state $\ket{+}$ for large $\kappa$.
    }
    \label{fig:approxplus}
\end{figure}

In Fig.~\ref{fig:approxplus}(a), we show the logical Bloch vectors corresponding to $\op{\rho}_L$ for squeezed vacuum states over a large range of squeezing in both position and momentum. In Fig.~\ref{fig:approxplus}(b), we plot the logical fidelity of these states with qubit states $\ket{+}$ and $\ket{0}$. A momentum-squeezed state approximately encodes a $\ket{+}$, while a position squeezed states approximately encodes a $\ket{0}$: the figure quantifies the degree of faithfulness to the logical states with respect to squeezing. The vacuum state (zero squeezing) encodes a mixed logical state in the $xz$ plane, whose Bloch vector depends on the bin size $\alpha$. For large enough $\alpha$, the vacuum position wave function fits almost entirely into the partitioned-position $m=0$ bin, thus encoding a high quality $\ket{0}$. For $\alpha = \sqrt{\pi}$, shown in Fig.~\ref{fig:approxplus}, the vacuum wave function has significant overlap in the neighboring bins ($m = \pm1$), and Fig.~\ref{fig:approxplus}(b) reveals that the  mixed logical state in this case has slightly higher fidelity with $\ket{0}$ than with $\ket{+}$. In summary, the CV states we have seen that encode pure logical qubits are those whose position wave functions are periodic with respect to $2 \alpha$ or only have support in a single partitioned-position bin. For $\alpha = \rpi$, the variance of the vacuum wave function is too large for it to be (even approximately) either one of these.

\subsection{ Approximate GKP states }
\label{sec:approxGKPdecomposition}

There are a number of ways to approximate GKP states; important examples and their relations are analyzed in Matsuura \emph{et al.}~\cite{matsuura2020equivalence}. We construct approximate GKP codewords by first convolving the position wave function of ideal codewords, \eqr{eq:jGKP}, with a normalized Gaussian of standard deviation $\Delta$,
    \begin{align} \label{eq:GaussianNorm}
        G_{\Delta}(x) \coloneqq \frac{1}{\sqrt{2\pi \Delta^2}} e^{- \frac{x^2}{2 \Delta^2}}
        \, ,
    \end{align}
which turns each $\delta$-function into a Gaussian, giving each spike a measured position variance of~$\frac{\Delta^2}{2}$.
Then, applying an overall Gaussian envelope with variance $\frac{1}{\kappa^2}$ damps the spikes far from the origin in the position basis. With this parametrization, each Gaussian spike has a measured momentum variance of~$\frac{\kappa^2}{2}$, which scales with~$\kappa$ in the same way that momentum variance for squeezed vacuum in \eqr{eq:sqzmomboth} scales with~$\sparam$.
We can write these approximate codewords in a compact way in terms of Jacobi theta functions of the third kind, $\vartheta\pqty{x, \tau}$ from \eqr{eq:thetadef}, using the fact that a pulse train of Gaussians with period $T$ is~\cite{pantaleoni_modular_2020}
\begin{equation} \label{eq:pulsetrain}
    \sum_{n \in \integers} G_{\sigma}\pqty{x - n T}
    = \frac{1}{T} \vartheta\pqty{\frac{x}{T}, \frac{2 \pi i\sigma^2}{T^2}}
    \, .
\end{equation}
The (unnormalized) position wave functions for approximate computational-basis codewords, including the broad Gaussian envelope, are~\cite{matsuura2020equivalence}
\begin{align}
    \label{eq:approxGKPwavefunctionposition}
    \bar{\psi}_{\GKP, j}\pqty{x}
    &=
    G_{\kappa^{-1}}(x)
          \vartheta\pqty{\frac{x}
              {2 \alpha} - \frac{j}{2},
          \tau_{\Delta}}
    \, ,
\end{align}
where we used the GKP periodicity $T= 2 \alpha$, $j \in \{0,1\}$ labels the state just as in the ideal case [\eqr{eq:jGKP}], and
    \begin{align} \label{eq:taufactor}
        \tau_\sigma \coloneqq 2 \pi i \pqty{\frac{ \sigma}{2 \alpha}}^2
        =
        \frac{i \pi \sigma^2}{2 \alpha^2}
        \,
    \end{align}
for a given standard deviation~$\sigma$ in the Gaussian pulse train, \eqr{eq:pulsetrain}.
The bar over the wave function in \eqr{eq:approxGKPwavefunctionposition} indicates that it is not normalized. Since approximate GKP wave functions are not orthogonal, superpositions need to be normalized case-by-case; see below. This behavior is also present in even and odd coherent-state superpositions, which are the foundation for bosonic cat codes~\cite{ralph2003computation,lund_fault-tolerant_2008}. It is indeed this non-orthogonality that can cause the encoded logical-qubit state to differ from the intended qubit state~\cite{pantaleoni_modular_2020,tzitrin_progress_2020}.

An arbitrary GKP state $\ket{\psi_\GKP}$ is encoded with coefficients $c_0$ and $c_1$ weighting the approximate codewords, just as in the ideal case in \eqr{eq:arbGKP}, which gives the position wave function~\cite{gottesman_encoding_2001,matsuura2020equivalence}
\begin{align} \label{eq:GKPstate}
    \psi_{\GKP}\pqty{x}
    &= \frac{1}{\sqrt{\mathcal{N}}} \sum_j c_j \bar{\psi}_{\GKP, j}\pqty{x}
    \, ,
\end{align}
with normalization
\begin{align}
    \mathcal{N} &= \int dx\, |\bar{\psi}_{\GKP}(x)|^2
    \\
    &=  \sum_{j, j'} c_j^* c_{j'} \int dx \, [\bar{\psi}_{\GKP, j}\pqty{x}]^* \bar{\psi}_{\GKP, j'}\pqty{x}
    \,.
\end{align}

The SSD of approximate GKP states is found by evaluating the wave function at $ x = u + 2 \alpha m + \alpha \ell $:
\begin{align}
    &\psi_{\GKP}\pqty{\ell, m, u}
     \nonumber\\*
    &= \frac{1}{\sqrt{\mc{N}}}
    G_{\kappa^{-1}}\pqty{
        u + 2\alpha m + \alpha \ell}
        \sum_j c_j
    \vartheta
          \pqty{
          \frac{u}{2 \alpha} - \frac{\ell + j}{2},
          \tau_{\Delta}
          }
    \, ,
    \label{eq:approxGKPwavefunctionpositionDecomposed}
\end{align}
where we used the periodicity of the $\vartheta$-function, $\vartheta\pqty{ z + 1 , \tau } = \vartheta\pqty{ z , \tau }$, to simplify the expression.

\begin{figure*}[t!]
    \centering
    \includegraphics[width=1.0\textwidth]{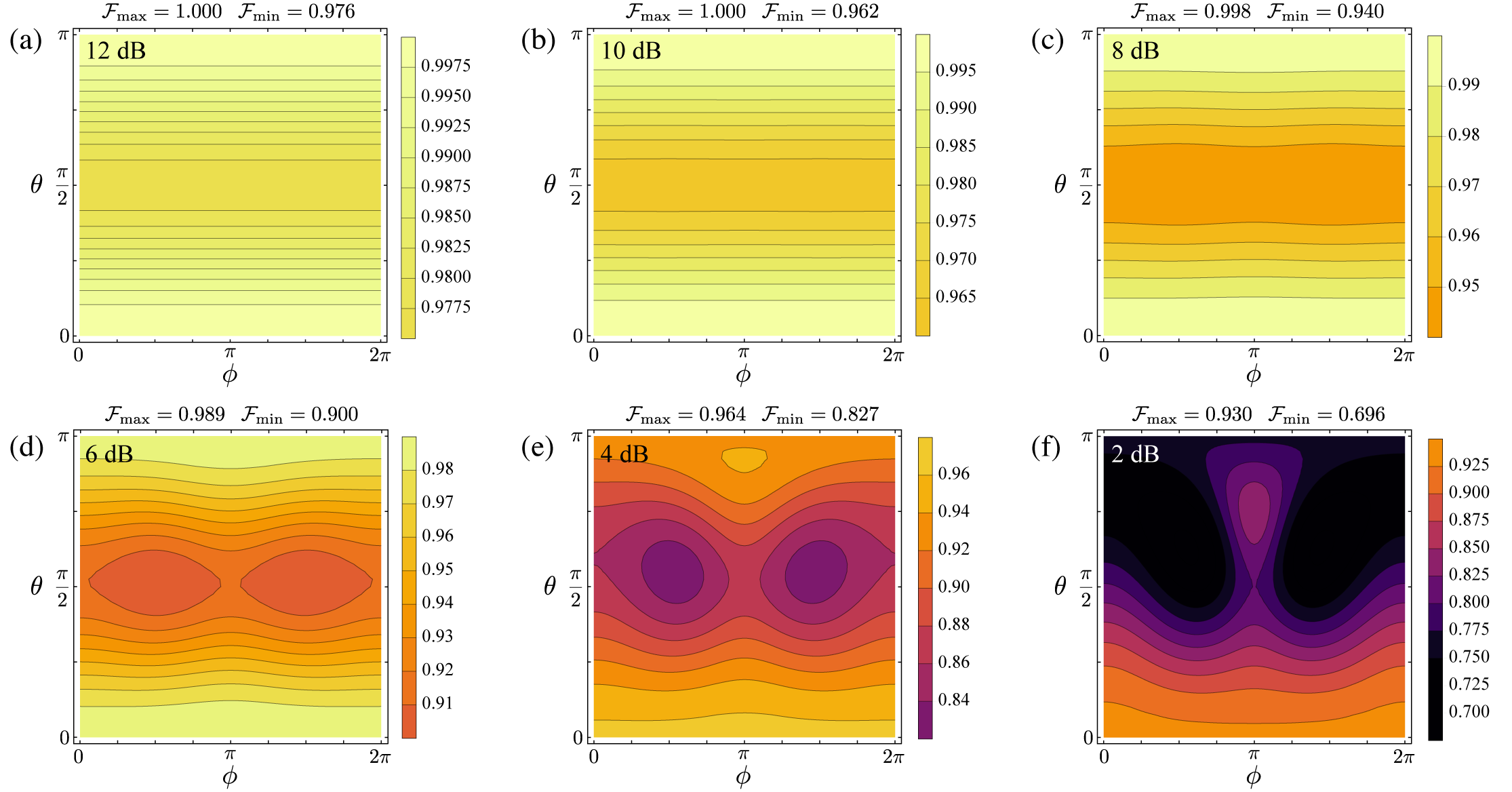}
    \caption{
        Logical fidelity between approximate GKP states, \eqr{eq:GKPstate}, and their intended logical qubit state, with symmetric embedded error in position and momentum ($\kappa = \Delta)$. The approximate GKP states and the intended qubit states are specified by polar and azimuthal angles $(\theta, \phi)$ on the Bloch sphere: $c_0 = \cos\pqty{\theta/2}$ and $c_1 = \sin\pqty{\theta/2} e^{i \phi}$.  Each contour plot shows the full set of logical fidelities for a given $\kappa$ (in dB) that sets the quality of the GKP state; see \eqr{eq:approxGKPwavefunctionposition}. For high-quality GKP states (8--12 dB), the logical fidelities are approximately independent of the azimuthal Bloch-sphere angle~$\phi$.
    }
    \label{fig:approxGKP}
\end{figure*}

We extract the reduced logical state,
    \begin{align} \label{eq:reducedstatematrixelement}
        \op{\rho}_L = \sum_{\ell, \ell'} \rho_L^{\ell\ell'} \ketbra{\ell}{\ell'}[L][L]
        \, ,
    \end{align}
using the gauge trace, \eqr{eq:gaugetrace}, to find the logical-subsystem matrix element
\begin{widetext}
\begin{align}
     \rho_L^{\ell\ell'} =
     \frac{1}{\mathcal{N}}
    e^{-\frac{\kappa^{2}\alpha^{2}}{4}(\ell - \ell')^2 }
        \sum_{j, j'}
    c^{*}_{j} c_{j'}
    \intalpha
    du \,
    \vartheta\pqty{
    \frac{u}{2 \alpha} + \frac{\ell + \ell'}{4},
    \frac{\tau_{\kappa^{-1}}}{2}
    }
    \vartheta\pqty{
    \frac{u}{2 \alpha} + \frac{\ell - j}{2},
    \tau_\Delta
    }
    \vartheta\pqty{
    \frac{u}{2 \alpha} + \frac{\ell' - j'}{2},
    \tau_\Delta
    }
    \label{eq:firstgaugetrace}
    \, .
\end{align}
\end{widetext}
The matrix elements are a sum of four integrals, each of which quantifies the overlap between three Jacobi theta functions, evaluated at different points. These points are determined by the summation indices $j,j'$ and the logical indices $\ell, \ell'$. When the approximate GKP states are high quality---$\Delta, \kappa \ll 1$---one can verify that, in the integration domain, the three Jacobi theta functions only result in a large overlap when evaluated at the same point. In all other cases, the contribution to the sum is small and can be neglected. Finally, note that when the GKP states are ideal ($\kappa = \Delta = 0$), the exponential damping term simplifies to 1,  and each integral (after normalization) in the sum gives $ \delta_{j, \ell} \delta_{j', \ell'} $. In this case, $ \rho_L^{\ell \ell'} = c^*_\ell c_{\ell'} $, indicating that the logical state, $\ket{\psi}[L] = c_0 \ket{0}[L] + c_1 \ket{1}[L]$, is pure and is identical to the intended qubit state.
Note that, contrary to Ref.~\cite{pantaleoni_modular_2020},
no approximations were made here in the derivation of the logical state, \eqr{eq:firstgaugetrace}.

We analyze the qubit quality for approximate GKP states, \eqr{eq:approxGKPwavefunctionposition}, with parameters  $\kappa = \Delta$ by comparing the logical qubit state to the \emph{intended qubit state}, itself specified by the coefficients in the superposition, $\{c_0, c_1\}$. In Figs.~\ref{fig:approxplus}(c) and~\ref{fig:approxplus}(d), we show logical Bloch vectors and logical fidelities for approximate GKP states of quality $\kappa$ intended to encode either a $\ket{+}$ or a $\ket{0}$.

More generally, we would like to know to what degree an approximate GKP state faithfully encodes an arbitrary intended qubit state. In Fig.~\ref{fig:approxGKP}, we show the logical fidelity between approximate GKP states, \eqr{eq:GKPstate}, with $c_0$ and $c_1$ given in polar form and their intended qubit state specified by the same coefficients. For low-quality GKP states (large $\kappa$), the logical fidelities vary significantly depending on the intended qubit state, whereas for high-quality GKP states (small $\kappa$) the fidelity is nearly independent of azimuthal angle $\phi$. In the high-quality regime, the lower bound on the logical fidelity occurs at or near states on the equator of the Bloch sphere ($\theta \approx \frac{\pi}{2}$), \emph{i.e.},~states of the form $\tfrac{1}{\sqrt{2}} \pqty{ \ket{0} + e^{i\phi} \ket{1}}$. On the other hand, the highest logical fidelities occur near the poles, which correspond to computational-basis codewords. The asymmetry in the fidelities of different codewords (most apparent for lower-quality GKP states) is due to the fact that the SSD here is based on the modular decomposition of the position operator. Performing this decomposition on the momentum operator instead reverses the roles of these low- and high-fidelity states.

\section{Teleportation of GKP-encoded quantum information}
\label{sec:measurementsCVCS}

CV cluster states are resources for universal, measurement-based quantum computing~\cite{menicucci_universal_2006}. In cluster-state protocols, desired unitary gates are applied to an input state by sequentially measuring nodes and performing necessary corrections depending on the measurement outcomes. Each measured node teleports the state at that node to the adjacent one with a gate applied that depends on the measurement type.  A common and important type of measurement performed for CVCS quantum computing is homodyne detection, an easy-to-perform Gaussian measurement, which implements arbitrary multimode Gaussian unitaries on the input state~\cite{menicucci_universal_2006}.

When the input state is a GKP-encoded state, Gaussian operations suffice for fault-tolerant universal quantum computation~\cite{menicucci_fault-tolerant_2014}. Furthermore, when parts or all of the cluster state are composed of $\ket{+_\GKP}$  states---\emph{i.e.},~hybrid or GKP cluster states---logical GKP Clifford unitaries are still implemented via homodyne detection. In addition, the GKP nodes perform automatic syndrome extraction for GKP error correction~\cite{walshe_2020}. Error correction is critical for fault tolerance because the physical states that comprise a cluster state---momentum-squeezed states [\eqr{eq:momsqz2}] and approximate GKP states [\eqr{eq:approxGKPwavefunctionposition}]---contain embedded noise depending on their quality. Even in the absence of external noise processes, this embedded noise is transferred to the input state during computation and will accumulate to catastrophic levels if not mitigated.

We focus here on the teleportation gadget~\cite{braunstein_teleportation_1998,menicucci_universal_2006} that forms the foundation of CV cluster-state quantum computing.
This is the measurement-based computation that implements the identity gate (modulo a double Fourier transform) on an arbitrary input state $\ket{\psi}$ after teleporting it through two nodes of a CV cluster state.

Canonical CV cluster states are built from 0-momentum eigenstates coupled together by two-mode controlled-$Z$ gates~\cite{menicucci_universal_2006,gu_quantum_2009},
\begin{align}
    \CZ[g] \coloneqq e^{i g \q_1 \otimes \q_2 }
    \, ,
    \label{eq:CVcontrolledZ}
\end{align}
for weight $g=1$. The ideal teleportation gadget is prepared by attaching an arbitrary input state $\ket{\psi}$ to one mode of a two-mode CV cluster state.
The circuit for this teleportation gadget (which proceeds right-to-left) is
\begin{equation}
\begin{aligned}
    \Qcircuit @C=0.8em @R=1.0em{
        & & & &
        \control \cw[1] & \push{\, \bra{s}[p]}[1]
        & \ctrl{1}
        & \qw
        & \rstick{\hspace{-1.5em}\ket{\psi}[]}
        \\
        & \control \cw[1] & \push{\, \bra{t}[p]}[1]
        & \ctrl{1} &
        \gate{X(-s)}\cwx[-1]
        & \qw
        & \ctrl{-1}
        & \qw
        & \rstick{\hspace{-1.5em}\ket{0}[p]}
        \\
        \lstick{(out)} & \gate{X(-t)} \cwx[-1]
        & \qw
        & \ctrl{-1}
        & \qw
        & \qw
        & \qw
        & \qw
        & \rstick{
        \hspace{-1.5em}\ket{0}[p] \, ,
        }
    }
    \label{eq:idealtel}
\end{aligned}
\end{equation}
where the entangling gates are CV controlled-$Z$ gates [\eqr{eq:CVcontrolledZ}] with weight $g=1$, and the input mode and the first ancilla mode are measured via homodyne detection of the $p$ quadrature with respective outcomes $s$ and $t$. Measurements are represented in the circuit by bras $\bra{\cdot}[p]$, and each outcome is used to perform a position-shift correction $\op{X}(\cdot)$, \eqr{eq:posshift}. This feed-forward procedure is indicated by the double lines (classical lines) connecting measurements and gates.  The ultimate action of the the circuit is to perfectly teleport the input state to the final (unmeasured) mode with two Fourier transforms, one arising from each wire:
\begin{align} \label{eq:idealteleportation}
    \ket{\psi^\text{tel}} = \op{T}_\text{ideal}(s,t) \ket{\psi} = \op{F}^2 \ket{\psi}
    \, .
\end{align}
One way to understand the origin of the Fourier transforms is to recognize that a canonical CV cluster state is the Choi state for $\op F$~\cite{walshe_2020}, so ordinary one-mode teleportation using this state~\cite{gu_quantum_2009,menicucci_universal_2006} applies the corresponding gate. Doing so twice (with corrections), as in \eqr{eq:idealtel}, applies $\op F^2$.

Note that the action of the ideal teleportation circuit is independent of the measurement outcomes because the active position-shift corrections fully undo any conditional effects. This is a feature of teleportation using ideal 0-momentum states as ancillae; it does not hold true in the following section, wherein squeezed-momentum states replace the 0-momentum eigenstates.

\subsection{Noisy CV cluster-state teleportation}
\label{sec:noisyteleportation}

In physical settings, the cluster state is produced from finite-energy momentum-squeezed states, \eqr{eq:sqzmomboth}, with $\sparam < 1$ rather than ideal 0-momentum eigenstates.  In this case, the teleportation is no longer perfect~\cite{menicucci_universal_2006}, with the input state acquiring noise that originates in the finite squeezing. The teleportation circuit for this setting (which proceeds right-to-left) is
\begin{equation}
\begin{aligned}
    \Qcircuit @C=0.8em @R=1.0em{
        & & & &
        \control \cw[1] & \push{\, \bra{s}[p] }[1]
        & \ctrl{1}
        & \qw
        & \rstick{\hspace{-1.5em} \ket{\psi}}
        \\
        & \control \cw[1] & \push{\, \bra{t}[p] }[1]
        & \ctrl{1} &
        \gate{X(-s)}\cwx[-1]
        & \qw
        & \ctrl{-1}
        & \qw
        & \rstick{\hspace{-1.5em} \ket{0, \sparam }_p }
        \\
        \lstick{(out)} & \gate{X(-t)} \cwx[-1]
        & \qw
        & \ctrl{-1}
        & \qw
        & \qw
        & \qw
        & \qw
        & \rstick{
        \hspace{-1.5em} \ket{0, \sparam }_p \, ,
        }
    }
    \label{eq:noisytel}
\end{aligned}
\end{equation}
whose ultimate action is to teleport the input state $\ket{\psi}$ to the final mode with an additional operation $\op{T}_{\sparam}(s,t)$,
\begin{align}   \label{eq:faultykop}
    \ket{\psitelst}
    &\coloneqq
    \op{T}_{\sparam} \pqty{s,t} \ket{\psi} \\
    &=
    G_{\sparam^{-1}} \pqty{\q + t}
    G_{\sparam^{-1}} \pqty{\p + s}
    \F^{2} \ket{\psi}
    \, ,
\end{align}
with a normalized Gaussian function given in \eqr{eq:GaussianNorm}. The noisy teleportation operator $\op{T}_{\sparam} \pqty{s,t}$ is a non-unitary Kraus operator that produces output states conditional on the outcomes $s,t$.
This is because, in contrast to ideal teleportation in \eqr{eq:idealteleportation}, the corrective shifts here do not fully restore the state after teleportation; instead, noisy teleportation gives Gaussian envelopes, in different quadratures, whose centers depend on the measurement outcomes.
Finally, since the envelope operators are not unitary, the output state in \eqr{eq:faultykop} is not normalized. This is indicated by the overbar $\bar{\psi}$. The joint probability of obtaining the outcomes $s$ and $t$,
    \begin{align} \label{eq:teleportationprobability}
        \Pr(s,t)
        =
        \braket{\psitelst}{\psitelst}
    \end{align}
can be used to renormalize the state,
$
\ket{\psitelstnobar}
=
\frac{1}{\sqrt{\Pr(s,t)}} \ket{\psitelst}
$.

We examine the effects of the finite-squeezing envelope operators in \eqr{eq:faultykop} using the position representation of the teleported state,
\begin{align} \label{eq:wavefunctionteleportedstate}
    \psitelstx
    = \braket{x}{\psitelst}[q][]
    = \bra{x}[q] \op{T}_{\sparam}(s,t) \ket{\psi}.
\end{align}
To evaluate this expression, we use the fact that the Fourier operator $\op{F}$ acts on basis states as $\F \ket{x}[q] = \ket{x}[p]$ and $\F^\dagger \ket{x}[q] = \ket{-x}[p]$, which transforms between position and momentum wave functions.
For a state $\ket{\psi}$ with position wave function $\psi(x)$, \eqr{eq:positionwavefunction}, the momentum wave function is
\begin{align}
    &\tilde{\psi}(x)
    \coloneqq \braket{x}{\psi}[p]
    = \bra{x}[q] \F^{\dag} \ket{\psi}
    = \mathcal{F}[\psi](x)
    \, ,
\end{align}
where the Fourier transform of a function $f$ is
    \begin{align}
        \mathcal{F}[f](x) \coloneqq \frac{1}{\sqrt{2\pi}} \int dy \, e^{-ixy} f(y)
        \, .
    \end{align}

Inserting a complete set of momentum eigenstates between the Gaussian envelope operators, we find the position wave function of the teleported state, \eqr{eq:wavefunctionteleportedstate},
\begin{align}     \label{eq:telstate1}
    \psitelstx
    &  =
    \frac{G_{\sparam^{-1}}\pqty{x+t}}{\sqrt{2\pi}}
    \int \!dy \ e^{ - i x y }
    G_{\sparam^{-1}}\pqty{s-y}
    \tilde{\psi}(y)
    \\
    &  =
    G_{\sparam^{-1}}\pqty{x+t}
    \mathcal{F}[ h \cdot \tilde{\psi} ](x)
    \, ,
\end{align}
where $h(y) \coloneqq G_{\sparam^{-1}} \pqty{s-y}$, we used $\bra{y}[p] \F^2 \ket{\psi} = \tilde{\psi}(-y)$, and we made the change of variables $y \rightarrow -y$ going from the first line to the second.
We rewrite \eqr{eq:telstate1} in a more convenient form using the convolution theorem, $
\mathcal{F}[{f \cdot g}] = \mathcal{F}[f] \ast \mathcal{F}[  g]
$, where the convolution of two functions $f$ and $g$ is $[f\ast g] (x) \coloneqq \int dy \, f(y) g(x-y)$.
The Fourier transforms we need are $\mathcal{F}[h](x) = e^{-isx} G_{\sparam}(x)$ and $\mathcal{F}[\tilde{\psi}](x) = \psi(-x)$, so that the position wave function of the teleported state is
\begin{align}
    \psitelstx
    =
    G_{\sparam^{-1}}(x+t)
    \int dy \, e^{-isy} G_{ \sparam }(y) \psi(y-x)
    \, .
    \label{eq:telstatefinal}
\end{align}
This expression is useful in the next section, where we obtain the SSD of the teleported state for different input states.

\subsection{Noisy teleportation of ideal GKP states} \label{sec:telmeas}

Intuitively, it is clear from the noisy teleportation map in \eqr{eq:faultykop} that some information in the input state is lost due to the envelope operators.
We quantify this statement for qubit information encoded using the GKP encoding. The subsystem decomposition is a useful tool for this analysis since with it a logical-subsystem qubit can be associated with any CV state. This logical qubit can then be compared to the intended qubit state using the logical fidelity, \eqr{eq:logicalfidelity}.

We consider an ideal GKP state $\ket{\psi_\GKP}$, \eqr{eq:GKPstate}, sent through the noisy teleportation circuit in \eqr{eq:noisytel}. Because ideal GKP states are not normalizable, we ignore normalization in the following derivation.\footnote{Alternatively, one can consider ideal GKP states as the limit of finite-energy states~\cite{mensen2020phasespace}.}  First, note that the double Fourier transform present in both ideal and noisy teleportation leaves the state invariant, $\op{F}^2 \ket{\psi_\GKP} = \ket{\psi_\GKP}$, due to the parity symmetry of all GKP codes~\cite{grimsmo_quantum_2020}.\footnote{Square-lattice GKP codes ($\alpha = \sqrt{\pi}$) have the additional property that $\op{F}$ acts as a logical Hadamard gate.}
For each computational basis state $\ket{j_{\GKP}}$ with wave function $\Sh_{2\alpha}(x - \alpha j)$, \eqr{eq:GKPjwavefunction}, we obtain the teleported position wave function from \eqr{eq:telstatefinal}:
\begin{align}
    & \psitelGKPjstx
    \nonumber
    \\ &  =
    G_{\sparam^{-1}}(x+t)
    \sum_{m \in \mathbb{Z}}
    e^{ - is \pqty{ x - 2\alpha m - \alpha j } }
    G_{ \sparam }(x - 2 \alpha m - \alpha j )
    \\ &  =
    e^{ - \frac{ s^2 \sparam^2 }{ 2 } }
    G_{\sparam^{-1}}(x+t)
    \vartheta\pqty{
    \frac{ x + i s \sparam^2  }{ 2\alpha }
    - \frac{j}{2} , \tau_{ \sparam }
    }
    \, ,
    \label{eq:noisyteleportedGKPj}
\end{align}
where $\tau_\sigma$ is defined in \eqr{eq:taufactor}.

The noisy teleported wave function resembles that for an approximate GKP state with $ \Delta = \kappa $, \eqr{eq:approxGKPwavefunctionposition}, with two notable differences that depend on the measurement outcomes $s$ and $t$. First, the Gaussian envelope in \eqr{eq:noisyteleportedGKPj} is centered at $t$. Second, the position of the peaks of the theta function is disturbed by $i s \sparam^2$ in a nontrivial way---imaginary shifts affect the periodicity of theta functions by introducing a secondary period that depends on the imaginary part of the translation (see Ref.~\cite{bellman_brief_2013} for an in-depth analysis of theta functions). Note that when using the momentum wave function of the state instead, the outcome-dependent distortion effects swap their roles. Namely, $s$ is associated with a translated Gaussian envelope and $t$ with an imaginary translation of the theta function.

Interestingly, for the outcomes $s = t = 0$, these teleported states are identical to the approximate GKP codewords defined in \eqr{eq:approxGKPwavefunctionposition} (with symmetric noise $\Delta = \kappa = \sparam$ inherited from the ancillae). Other outcomes shift the envelopes and change the logical state, which we investigate below.

For an arbitrary, ideal GKP state $\ket{\psi_\GKP}$, the position wave function for the (unnormalized) teleported state is given by superpositions of \eqr{eq:noisyteleportedGKPj}:
\begin{align}  \label{eq:teleportedGKPwavefunction}
    &
    \psitelGKPstx
    \nonumber \\* & \quad
    =
    G_{\sparam^{-1}}(x+t) \sum_{j \in \{0,1\}} c_j \vartheta\pqty{
    \frac{ x + i s \sparam^{2}  }{ 2\alpha }
    - \frac{j}{2} , \tau_{ \sparam^{-1} } }
    \, .
\end{align}
The amplitudes $c_0$ and $c_1$ parameterize the intended qubit state encoded into the ideal GKP code.
The teleported state,
\begin{equation} \label{eq:teleportedGKPdensitymatrixpositionbasis}
        \op{\bar{\rho}}^\text{tel}_\GKP(s,t)
        \coloneqq \ketbra{\psitelGKPst}{\psitelGKPst}
        \, ,
\end{equation}
can be written in the position basis, \eqr{eq:densitymatrixpositionbasis}, with matrix elements
\begin{equation} \label{eq:positionbasismatrixelementsteleport}
        \bar{\rho}^\text{tel}_\GKP(x,x';s,t) =
        \psitelGKPstx
        \bqty{\psitelGKPstxprime}^{*}
        \, .
\end{equation}
Teleportation introduces noise that corrupts the encoded qubit information, which we quantify below.

\subsubsection{Subsystem decomposition and logical state}

Performing an SSD of the teleported GKP wave function, \eqr{eq:teleportedGKPwavefunction}, gives
\begin{align}     \label{eq:teleportedGKPwavefunctionSSD}
    \bar{\psi}^\text{tel}_\GKP(\ell, m, u; s,t)
    &
    = \nonumber
    G_{\sparam^{-1}}(2 \alpha m + u + \alpha \ell +t)
    \\
    & \quad \times \! \!
    \sum_{j\in \{0,1\}}
    c_j \vartheta\pqty{
    \frac{ u + i s \sparam^2 }{2\alpha}
    + \frac{\ell - j}{2}
    ,
    \tau_{\sparam}
    }
    \, ,
\end{align}
where the periodicity, $\vartheta(z+m,\tau) = \vartheta(z, \tau)$, was used to remove the gauge bin-number $m$ inside the theta function.
The density matrix associated with this state, \eqr{eq:teleportedGKPdensitymatrixpositionbasis}, can be represented in the subsystem basis, \eqr{eq:densitymatrixSSD},
with matrix elements
\begin{align} \label{eq:matrixelementGKPteleportedSSD}
    &\bar{\rho}^\text{tel}_\GKP
        (\ell,\ell',m,m',u,u'; s,t)
    \nonumber \\*
    &\quad
    =\bar{\psi}^\text{tel}_\GKP(\ell, m, u;s,t)
    \big[
    \bar{\psi}^\text{tel}_\GKP(\ell', m', u';s,t)
    \big]^*
    \, .
\end{align}

Performing a gauge trace extracts the logical qubit associated with a
noisy-teleported GKP state. The gauge trace is executed by setting $m=m'$ and $u=u'$ in the subsystem-basis matrix elements, \eqr{eq:matrixelementGKPteleportedSSD}, and then summing or integrating over these variables, as appropriate.
This procedure gives the unnormalized matrix elements for the reduced logical state [see \eqr{eq:reducedstatematrixelement}]:
\begin{align} \label{eq:gaugetraceGKPteleported}
    \bar{\rho}^{\ell\ell'}_L
    & = \sum_{m\in \mathbb{Z}}
    \intalpha du\
    \bar{\rho}^\text{tel}_\GKP (\ell,\ell',m,m,u,u;s,t)
    \, .
\end{align}
Also, note that we do not include teleportation-outcome labels $(s,t)$ on reduced logical states.
We now insert the wave function for the teleported GKP state, \eqr{eq:teleportedGKPwavefunctionSSD}, into the expression.
Only the Gaussian factor of $\bar{\psi}^\text{tel}_\GKP(\ell, m, u)$ depends on gauge bin number $m$, so we perform the sum over $m$ in the gauge-trace expression, giving the factor
\begin{align}
    \nonumber
    \sum_{m \in \mathbb{Z}}
    G_{\sparam^{-1}}\pqty{2 \alpha m + u + \alpha \ell + t}
    G_{\sparam^{-1}}\pqty{2 \alpha m + u + \alpha \ell' + t}
    \\
    =
    \sqrt{ \frac{\pi }{ 4 \alpha^2 \sparam^{2} } }
    e^{
    -\frac{\alpha^2 \sparam^{2}}{4}
    (\ell - \ell')^{2}
    }
    \vartheta\pqty{
    \frac{u + t}{2 \alpha} \!- \!\frac{\ell + \ell'}{4},
    \frac{\tau_{\sparam^{-1}}}{2}
    }
    \, .
\end{align}
The exponential term degrades logical states by damping non-diagonal terms ($\ell \neq \ell'$) of the reduced logical matrix. The strength of the damping depends on the amount of squeezing in the cluster state $\sparam$. Inserting this factor into the gauge-trace expression, \eqr{eq:gaugetraceGKPteleported}, we find the reduced logical state representing the logical information carried by an ideal GKP codeword damaged by noisy teleportation:
\begin{widetext}
\begin{align}
    \rho^{\ell\ell'}_L  =
    \frac{1}{\mathcal{N}}
    e^{
      -\frac{\alpha^2 \sparam^2}{4}
      (\ell - \ell')^{2}
      }
    \sum_{j, j'} c^*_{j} c_{j'}
    \intalpha du\
    \vartheta\pqty{
        \frac{u + t}{2 \alpha} - \frac{\ell + \ell'}{4},
        \frac{\tau_{\sparam^{-1}}}{2}
    }
    \vartheta\pqty{
    \frac{ u + i s \sparam^{2} }{2 \alpha} + \frac{ \ell - j }{2},
    \tau_\sparam
                  }
    \vartheta\pqty{
    \frac{ u + i s \sparam^{2} }{2 \alpha} + \frac{ \ell' - j' }{2},
    \tau_\sparam
                  }
    \, ,
    \label{eq:gkpfaultytel}
\end{align}
\end{widetext}
where $\tau_\sigma$ is defined in \eqr{eq:taufactor}, and we include the normalization $\mathcal{N}$.
For outcomes $s=t=0$, this formula is identical to that for an approximate GKP state with $\Delta = \kappa = \sparam$,  \eqr{eq:firstgaugetrace}. For all other outcomes, the logical state is distorted by the fact that the the envelope operators in \eqr{eq:faultykop} have shifted centers.

\subsubsection{General form for GKP reduced logical states}

We now introduce a compact form for a qubit density matrix that is useful for representing the reduced logical states for ideal GKP, approximate GKP, and teleported GKP states.

First, a multivariate Siegel (or Riemann) theta function defined for $\vec{z} \in \mathbb{C}^N$ is
\begin{align}
    \Theta\pqty{\vec{z}, \boldsymbol{\tau} }
    \coloneqq
    \sum_{ \vec{m} \in \integers^{N}}
    \exp\bqty{
    2 \pi i
        \pqty{
            \tfrac{1}{2}
            \vec{m}^{\tp} \boldsymbol{\tau}  \vec{m}
            + \vec{m}^{\tp}  \vec{z}
            }
    }
    \, ,
    \label{eq:Siegeltheta}
\end{align}
where $\vec{m} \in \integers^N$ and $\boldsymbol{\tau}$ is a complex, symmetric matrix in the Siegel upper half-space~\cite{bellman_brief_2013}: $\boldsymbol{\tau} \in \mathbb{C}^{N \times N}$ with ${\Im[\boldsymbol{\tau} ] > 0}$ ---\emph{i.e.}, $\Im[\boldsymbol{\tau} ]$ is positive definite as a matrix.
When $\boldsymbol{\tau}$ is diagonal with diagonal elements $\tau_{ii}$, this expression is equivalent to a product of Jacobi theta functions of the third kind,
$\Theta(\vec{z},\boldsymbol{\tau}) = \prod_{i=1}^N \vartheta \pqty{z_i,\tau_{ii}}$, a form that we recognize in \eqr{eq:gkpfaultytel}.

The reduced logical states considered in the present work can be described
(up to normalization)
by a $2\times2$ matrix
\begin{align}     \label{eq:parametrizedstatematrix}
    \op \rho_{K,\boldsymbol{\tau}}(\vec{w})
    \coloneqq
    \sum_{\ell, \ell'}
    \rho^{\ell \ell'}_{K,\boldsymbol{\tau}}(\vec{w})
    \ket{\ell}[L] \bra{\ell'}[L]
    \
\end{align}
with $\ell \ell'$-th matrix elements
\begin{align}
    &\rho^{\ell\ell'}_{K,\boldsymbol{\tau}}\pqty{\vec{w}}
    \coloneqq
    e^{-\frac{1}{4}K^{2}\alpha^{2}\pqty{\ell - \ell'}^{2}}
    \nonumber \\* & \qquad \times
    \! \!
    \intalpha du\
    \sum_{j, j'}
    c^{*}_j c_{j'}
    \Theta \! \bqty{
        \! \pqty{
            \threevec{
                u / (2 \alpha) + (\ell + \ell')/4}{
                u / (2 \alpha) + \ell / 2 -j / 2 }{
                u / (2 \alpha) + \ell' / 2 - j' / 2 }
        } \!\! + \!
        \frac{\vec{w}}{2 \alpha}
        , \boldsymbol{\tau}
    }
    \, .
    \label{eq:generalstate}
\end{align}
This is a general form, with the parameters $K$, $\boldsymbol{\tau}$, and $\vec w$ determined case by case. $K\geq 0$ is a positive real parameter, and the matrix $\boldsymbol{\tau}$ lies in the Siegel upper half-space (a fact that is required for the theta sum to converge). Both $K$ and $\boldsymbol{\tau}$ will reflect the width of the spikes in GKP states, while $\vec w$ is 3-dimensional complex vector that will capture the effects of the measurement outcomes from teleportation. [The physical meaning of complex displacements in arguments to theta functions is discussed after \eqr{eq:noisyteleportedGKPj}.]

First, we can write the reduced logical state for an approximate GKP state with amplitudes $c_0$ and $c_1$ using Eqs.~\eqref{eq:parametrizedstatematrix} and \eqref{eq:generalstate}. For $\ket{\psi_{\GKP}} = c_0 \ket{0_{\GKP}} + c_1 \ket{1_{\GKP}} $, we have the gauge trace of an approximate GKP state with position-spike width $\Delta$ and position-envelope width $\kappa^{-1}$ (see \secr{sec:approxGKPdecomposition}):
\begin{align}
    \op{\rho}_L =
    \frac{1}{\mathcal{N}}
    \Tr_G \big[\ket*{\psi^{\kappa,\Delta}_{\GKP}}\bra*{\psi^{\kappa,\Delta}_{\GKP}} \big]
    &=
    \frac{1}{\mathcal{N}}
    \op \rho_{\kappa,\boldsymbol{\tau}}(\vec{0})
\end{align}
with
normalization $\mathcal{N}$ required to satisfy $\Tr[\op{\rho}_L] = 1$, and the $3 \times 3$ matrix
\begin{align}
    \boldsymbol{\tau}
    =
    \diag\pqty{ \tfrac{1}{2}
    \tau_{\kappa^{-1}},
    \tau_\Delta,
    \tau_\Delta
    }
    \label{eq:secondgaugetrace}
    \, .
\end{align}
To get the expression for ideal GKP states, it is sufficient to take the limits of the above for zero-width position spikes, $\Delta \rightarrow 0$, and flat envelope, $\kappa \rightarrow 0$, so that
$
\Tr_G[\ket{\psi_{\GKP}}\bra{\psi_{\GKP}}]
=
\op \rho_{0,\boldsymbol{\tau}_0}\pqty{\vec{0}}
$%
, where
$
\boldsymbol{\tau}_0
=
\diag(+i\infty,i 0^{+},i 0^{+})
$%
---the limits are taken so that the $\boldsymbol{\tau}$ matrix stays in the Siegel upper half-space. Regularizing the resulting matrix yields a normalized logical state~\cite{matsuura2020equivalence, mensen2020phasespace}.

For noisy teleportation of ideal GKP codewords through a cluster state composed of squeezed-momentum states parametrized by $\sparam$, the reduced logical state, whose matrix elements are given in \eqr{eq:gkpfaultytel}, can be written compactly using \eqr{eq:parametrizedstatematrix} as
\begin{align}
    \op{\rho}_L
    =
    \frac{1}{\mathcal{N}}
    \Tr_G \big[\ket*{\psi^{\text{tel}}_{\GKP}}\bra*{\psi^{\text{tel}}_{\GKP}} \big]
    &=
    \frac{1}{\mathcal{N}}
    \op \rho_{
        \kappa , \boldsymbol{\tau}
        }
        \pqty{\vec{w}}
    \, ,
    \label{eq:decomposedlogdensmatideal}
\end{align}
where $\mathcal{N}$ is the normalization,
\begin{align}
    \boldsymbol{\tau} &=
    \diag\pqty{ \tfrac{1}{2}
    \tau_{\sparam^{-1}},
    \tau_\sparam,
    \tau_\sparam
    }
    \, ,
\end{align}
and the teleportation measurement outcomes $s$ and $t$ appear in the vector
\begin{align}
    \vec{w} = (t, is \sparam^2, is \sparam^2)^\tp.
\end{align}
Outcome-dependent translations, including complex ones, are discussed above in the paragraph after \eqr{eq:noisyteleportedGKPj}.

In addition to the cases considered above, noisy teleportation of approximate GKP states can also be described by \eqr{eq:parametrizedstatematrix}, as can the situation where the teleportation measurement outcomes are averaged over. We consider these cases in the following subsection.

\subsection{Noisy teleportation of approximate GKP states} \label{sec:noisyteleportapproxGKP}

In a physical setting, the input GKP states to the teleportation circuit, \eqr{eq:noisytel}, will themselves contain finite-squeezing noise. We consider here approximate GKP states, \eqr{eq:approxGKPwavefunctionposition}, parametrized by small position-spike width $\Delta$ and broad position-envelope width $\kappa^{-1}$ teleported through a cluster state comprised of momentum-squeezed vacuum with narrow momentum wave function variance $\sparam^{2}$, \eqr{eq:momsqz1}. The derivation to obtain the logical density matrix proceeds, just as for ideal GKP states, by applying the noisy teleportation operator, \eqr{eq:faultykop}, to an approximate GKP state as in~\eqref{eq:firstgaugetrace} and subsequently tracing over the gauge mode. A detailed derivation can be found in Appendix~\ref{appendix:noisyteleportation}. The exact form of the logical state is cumbersome, but it can be greatly simplified in the limit of high quality GKP and squeezed states.  Ignoring parameter dependence that is higher order than $\Delta^{2}$, $\kappa^{2}$, and $\sparam^{2}$,  which corresponds roughly to a squeezing of at least $8$~dB, the logical reduced state is approximately
\begin{align}
    \op{\rho}_L &
    =
    \frac{1}{\mathcal{N}}
    \Tr_G \big[
    \op{T}_{\sparam}(s,t)
    \ket*{\psi_{\GKP}^{\kappa,\Delta}}
    \bra*{\psi_{\GKP}^{\kappa,\Delta}}
    \op{T}^{\dag}_{\sparam}(s,t)
    \big]
    \\
    & \approx \label{eq:logicalstatenoisyteleportedapproxGKP}
    \frac{1}{\mathcal{N}}
    \op \rho_{
        \sqrt{ \kappa^2 + \sparam^2 } ,
        \boldsymbol{\tau}
        }
        \pqty{\vec{w}}
    \, ,
\end{align}
where
$\mathcal{N}$ is the normalization,
\begin{align}
    \boldsymbol{\tau}
    & = \diag \pqty{
     \tfrac{1}{2} \tau_{(\kappa^2 + \sparam^{2})^{-1/2} } ,
    \tau_{ \Delta} + \tau_\sparam ,
    \tau_{ \Delta} + \tau_\sparam
    }
    \, ,
\end{align}
and the teleportation measurement outcomes $s$ and $t$ appear in the vector
\begin{align}
    \vec{w} = \left[\sparam^{2}\pqty{\kappa^2 + \sparam^{2} }^{-1} t, is\sparam^{2}, is \sparam^{2} \right]^\tp.
\end{align}
Note that the logical state in \eqr{eq:logicalstatenoisyteleportedapproxGKP} can also describe two situations considered previously. First, it describes a perfectly teleported approximate GKP state  with $\sparam \rightarrow 0$ while keeping $\Delta$ and $\kappa$ finite. Second, it describes noisy teleportation of an ideal GKP state (setting $\Delta = \kappa = 0$ and keeping $\sparam$ finite).

The logical state above contains the information about how an intended qubit state, parametrized by the probability amplitudes $c_0$ and $c_1$, is damaged both by an encoding into approximate GKP states (see \secr{sec:approxGKPdecomposition}) and further by noisy teleportation over a finitely squeezed cluster state. (Recall that for a faithful encoding, the normalized matrix elements satisfy $\rho_L^{\ell\ell'} = c^*_{\ell} c_{\ell'}$, indicating that the logical state $\op{\rho}_L$ is identical to the intended qubit state.) One can quantify the quality of the encoding using the logical fidelity, $\mc{F}_L( \op{\rho}^\text{tel}_\GKP, \op \sigma)$  [\eqr{eq:logicalfidelity}], where $\op{\sigma}$ is the intended qubit state, and $\op{\rho}^\text{tel}_\GKP$ has logical state $\op{\rho}_L$ given by \eqr{eq:logicalstatenoisyteleportedapproxGKP}.

\subsubsection{Averaging over teleportation outcomes} \label{sec:averaged}

We conclude with the case where the teleported GKP state is averaged over measurement outcomes. This models the situation where the measurement outcomes $s$ and $t$ are forgotten after the teleportation protocol~\cite{gu_quantum_2009} and represents a type of worst-case-scenario estimate---one in which we cannot use the outcome dependence to our advantage. This situation is described, mathematically, as follows.

A generic pure state $\ket{\psi}$ after noisy teleportation is
$
\op{T}_{\sparam} \pqty{s,t} \ket{\psi}
$%
, where
$
\op{T}_{\sparam}
$ is the noisy teleportation operator in \eqr{eq:faultykop}. The density operator corresponding to this state is then
\begin{align}
    \op{\bar{\rho}}^\text{tel}(s,t)
    =
    \op{T}_{\sparam} \pqty{s,t}
    \ketbra{\psi}{\psi}[][]
    [\op{T}_{\sparam}\pqty{s,t} ]^\dagger
    \, .
    \label{eq:teldensmat}
\end{align}
Averaging this conditional state over all outcomes $s$ and $t$ gives the average (unconditional) state,
\begin{align} \label{rhoavgoverrhotel}
    \op{\rho}_\text{avg}^\text{tel}
    &\coloneqq \iint ds dt \, \op{\bar{\rho}}^\text{tel}\pqty{s,t} \\
    &= \iint dx dx' \, \rho_\text{avg}^\text{tel}\pqty{x, x'} \ketbra{x}{x'}[q][q]
    \, .
\end{align}
There is no need to include the probability distribution for the outcomes in \eqr{rhoavgoverrhotel}, since each conditional state $\op{\bar{\rho}}^\text{tel}(s,t)$ is the product of the probability distribution of the outcomes $s,t$ and the normalized state~\cite{nielsen2002quantum}.
The average-state matrix elements
$
\rho_\text{avg}^{\text{tel}}(x,x')
$
are found by integrating the matrix elements $
\bar{\rho}^{\text{tel}}(x,x';s,t)
= \psitelstx [\psitelstxprime]^{*}
$, with $\psitelstx$ being the position wave function after noisy teleportation, \eqr{eq:telstatefinal}. This gives
\begin{align}
    \label{eq:avgstatematrixelement}
    &\rho_\text{avg}^{\text{tel}}(x,x')
    = \iint ds dt \, \bar{\rho}^{\text{tel}}(x,x';s,t) \,
    \\& \quad =
    G_{ \sqrt{2} \sparam^{-1} } ( x' - x )
    \int dy\
    G_{  \frac{\sparam}{\sqrt{2}}}(y)
    \psi^{*}( y - x )
    \psi( y - x' )
    \, ,
\end{align}
where $\psi(x)$ is the position wave function of the input state.

\begin{figure}[t]
    \centering
    \includegraphics[ width=0.45\textwidth ]{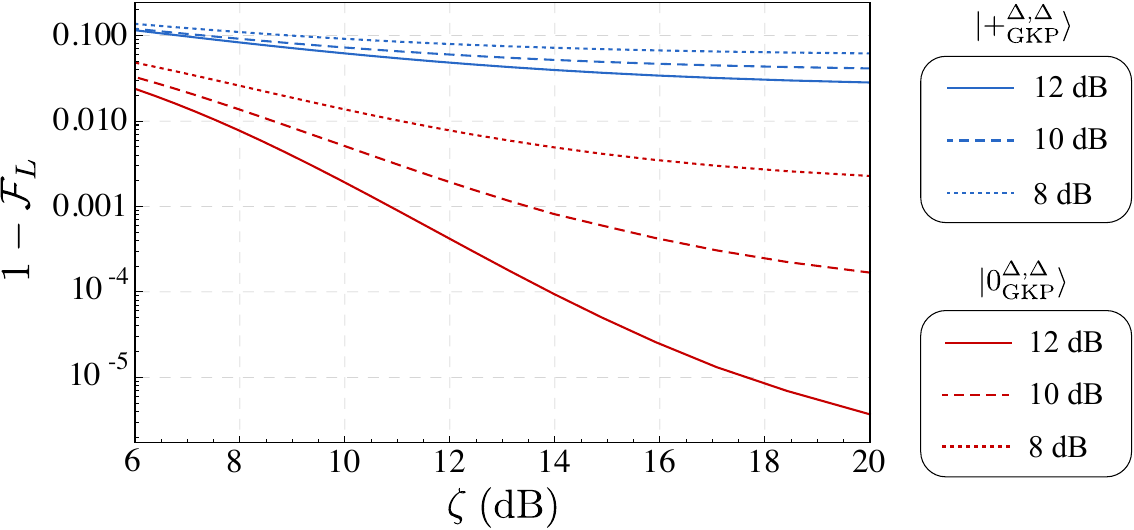}
    \caption{%
        Average-state logical infidelity $1-\mathcal{F}_L$ between the intended qubit state and an approximate GKP state (with symmetric noise $\kappa = \Delta$) after noisy teleportation, averaged over outcomes.
        Each curve describes the infidelity for an approximate GKP state of quality $\Delta$ given momentum-squeezed vacuum of quality $\sparam$ in the teleportation circuit. The blue lines ($\ket{+}$) represent the upper bound to the infidelity one can achieve after teleportation, whereas the red lines ($\ket{0}$) represent the lower bound (cf. Fig~\ref{fig:approxGKP}).
    }
    \label{fig:fidelities}
\end{figure}

We plug the wave function for an input approximate GKP state (setting $\kappa = \Delta$ for simplicity) into \eqr{eq:avgstatematrixelement} to find the matrix elements for an averaged-over-outcomes, teleported, approximate GKP state encoding an intended qubit given by the amplitudes $c_0$ and $c_1$:
\begin{align} \label{eq:avgGKPmatrixelements}
    \rho^\text{tel}_{\GKP, \text{avg}} \pqty{x,x'} &= \nonumber
    G_{\sqrt{2}\sparam^{-1}}\pqty{x - x'}
    G_{\Delta^{-1}}\pqty{x}
    G_{\Delta^{-1}}\pqty{x'}
    \nonumber \\* & \quad \times
    \sum_{j, j'} c^{*}_{j} c_{j'}
    \Theta\pqty{
    \bqty{\twovec
            { \frac{x }{2\alpha} - \frac{j }{2} }
            { \frac{x'}{2\alpha} - \frac{j'}{2} }
         },
    \boldsymbol{\tau}}
     \, .
\end{align}
The 2-dimensional Siegel theta function, \eqr{eq:Siegeltheta}, involves the $2 \times 2$ matrix
\begin{subequations}
\begin{align}
    \boldsymbol{\tau}
    &=
    \pqty{
    \begin{matrix}
        \tau_{\Delta} + \tfrac{1}{2} \tau_{\sparam} && \tfrac{1}{2} \tau_{\sparam} \\
        \tfrac{1}{2} \tau_{\sparam} && \tau_{\Delta} + \tfrac{1}{2} \tau_{\sparam} \\
    \end{matrix}
    }
    \\ &=
    \tau_{\Delta}
    \pqty{
    \begin{matrix}
        1 &&  0 \\
        0 && 1
    \end{matrix}
    }
    +
    \frac{1}{2} \tau_{\sparam}
        \pqty{
    \begin{matrix}
        1 && 1 \\
        1 && 1 \\
    \end{matrix}
    }
    \, .
    \label{eq:matsumform}
\end{align}
\end{subequations}
In the sum form, \eqr{eq:matsumform}, we recognize a contribution from the input GKP state (that depends on $\Delta$) and from the cluster state (that depends on $\zeta$).

We now extract the logical state by performing a gauge trace on $\op{\rho}_\text{GKP,avg}^\text{tel}$, whose matrix elements are given by \eqr{eq:avgGKPmatrixelements}. Assuming high quality input GKP state and momentum-squeezed states, we keep terms up to $\sparam^2$ and $\Delta^2$. Then, the reduced logical state for an averaged-over-outcomes, teleported, approximate GKP state is
\begin{align}
    \op{\rho}_L = \Tr_G[\op{\rho}_\text{GKP,avg}^\text{tel}] &=
    \op{\rho}_{
        \sqrt{\sparam^2 + \Delta^2},
        \boldsymbol{\tau}
    }\pqty{\vec{0}}
     \, ,
    \label{eq:decomposedlogdensmataveraged}
\end{align}
where the 3-dimensional Siegel theta function has matrix
\begin{align}
    \boldsymbol{\tau} &=
    \pqty{\begin{matrix}
            \frac{1}{2}\tau_{\Delta^{-1}} && 0 && 0 \\
            0 && \tau_{\Delta} + \tfrac{1}{2} \tau_{\sparam} && \tfrac{1}{2} \tau_{\sparam} \\
            0 && \tfrac{1}{2} \tau_{\sparam} && \tau_{\Delta} + \tfrac{1}{2} \tau_{\sparam} \\
    \end{matrix}}
    \, .
\end{align}
Note that, unlike for pure teleported GKP states previously considered, $\boldsymbol{\tau}$ is not diagonal.

Using this logical state, we find the average-state logical infidelity with various intended qubit states in Fig~\ref{fig:fidelities}. Approximate $\ket{+_\GKP}$ states perform considerably worse than approximate $\ket{0_\GKP}$ states, even at higher single-spike and envelope quality. This is an artefact of having performed the decomposition in~$q$; see the end of \secr{sec:approxGKPdecomposition}.

\section{Conclusion}

In this work, we have presented a partitioning of the Hilbert space of a bosonic mode into that of a discrete logical subsystem and a gauge-mode subsystem. Such a decomposition endows any state of the mode with logical information. Since this subsystem decomposition~(SSD) is based on a modular decomposition of the position operator, Gottesman-Kitaev-Preskill~(GKP) codewords (themselves periodic in position) are natural carriers of logical information.

The formalism is sufficiently flexible, however, that we may also ask, \emph{What is the logical information carried by states other than GKP states, regardless of whether they were designed for use as a bosonic code state?} An interesting example of such a state is the $0$-momentum eigenstate $\ket{0}[p]$, encoding the very same logical information as $\ket{+_\GKP}$ (\emph{i.e.},~a logical $\ket{+}$ state). Moreover, we use the SSD to find the logical information in finite-energy approximations to these states: approximate GKP states and momentum-squeezed states.  We find that these CV states do not perfectly encode the intended qubit state due to logical-gauge entanglement, although the faithfulness of the encoding (quantified by the logical fidelity) increases with the quality of the CV state.

Our decomposition is, at heart, just a different basis for the Hilbert space of a single mode---albeit one that changes its tensor-product decomposition: ${\Hcv \simeq \mathbb{C}^d \otimes \Hcvp}$. As such, operators can also be inspected in a decomposed fashion. The present work features several examples of operators in the subsystem decomposition. We mention in passing that the Hilbert-space relation can go in the other direction---two physical subsystems can be compiled into a larger, virtual ``supersystem.'' Joint properties of the subsystems can then be analyzed at the level of the supersystem, which may be useful in future work.


In \secr{sec:finitesqueezevacuum}, we decompose the non-unitary ``envelope operator'' used to generate momentum-squeezed states from 0-momentum states and approximate GKP states from ideal ones. The decomposition reveals a purely logical operation that damps towards a logical $\ket{0}$ and a more complex, logical-gauge entangling operator that causes further decoherence. Such decompositions show us how CV noise operators affect encoded information.

As a second example, in \secr{sec:arbshifts}, we decompose unitary momentum- and position-shift operators, which have entirely different behavior since the starting point of our SSD is the position basis. We showed that momentum shifts do not entangle the subsystems, acting instead as a local qubit rotation and separate (but related) gauge-mode action. On the other hand, position shifts typically do entangle them. An important exception to this is a small position shift acting on a high-quality GKP state, which acts trivially on the logical qubit. In this case, the shift acts exclusively on the gauge mode. From this perspective, it becomes clear why GKP error correction for small position shifts can be recovered: They simply leave the logical part of a well-encoded GKP state invariant. The only point of doing the recovery, then, is to return the gauge mode to the middle of its position bin so that it remains robust against further shifts. For small momentum shifts, the story is different since such shifts affect the gauge mode and rotate the qubit by a related amount. In this case, the correction (displacing the mode back) both restores the gauge mode to where it should be and also rotates the qubit back to where it belongs. This asymmetry in $q$ and $p$ explains the dominance of $Z_L$-axis dephasing noise in the SSD picture of high-quality encoded states---a fact that shows up throughout the figures in this work.

As a practical application, in \secr{sec:telmeas}, we employ the SSD to study the effects of teleportation through a finitely squeezed CV cluster states on qubits encoded into approximate GKP states. The reduced logical state and the logical fidelity were found to be powerful tools for characterizing the degree of success of noisy teleportation, as they allow us to track and assess the teleported logical quantum information directly. This study provides a framework for the analysis of CV noise as logical channels on qubits encoded into modes.

In fact, numerous problems are ripe for analysis with the SSD. For instance, it is straightforward to use what we have learned here to study the logical properties of CV measurement-based quantum computing (beyond just teleportation)~\cite{menicucci_universal_2006}---in particular, its
fault-tolerant realization using a GKP-doped CV cluster state~\cite{menicucci_fault-tolerant_2014}. Additional potential applications include analyzing the logical properties of other CV quantum computing schemes involving GKP qubits~\cite{fukui_high-threshold_2018,fukui_high-threshold_2019,bourassa2021blueprint}, as well as concatenation of GKP code with qubit codes~\cite{fukui_analog_2017,hanggli_enhanced_2020,noh2020fault,fukui2021efficient} and the use of GKP states for quantum communication~\cite{fukui2020all,rozpkedek2020quantum}. We leave these to future work.

\begin{acknowledgments}
The authors would like to acknowledge Rafael Alexander, Lucas Mensen, and Victor Albert for useful discussions.
This work was supported by the Australian Research Council Centre of Excellence for Quantum Computation and Communication Technology (Project No. CE170100012).
\end{acknowledgments}

\bibliography{updated_references}

\begin{thebibliography}{62}%
\makeatletter
\providecommand \@ifxundefined [1]{%
 \@ifx{#1\undefined}
}%
\providecommand \@ifnum [1]{%
 \ifnum #1\expandafter \@firstoftwo
 \else \expandafter \@secondoftwo
 \fi
}%
\providecommand \@ifx [1]{%
 \ifx #1\expandafter \@firstoftwo
 \else \expandafter \@secondoftwo
 \fi
}%
\providecommand \natexlab [1]{#1}%
\providecommand \enquote  [1]{``#1''}%
\providecommand \bibnamefont  [1]{#1}%
\providecommand \bibfnamefont [1]{#1}%
\providecommand \citenamefont [1]{#1}%
\providecommand \href@noop [0]{\@secondoftwo}%
\providecommand \href [0]{\begingroup \@sanitize@url \@href}%
\providecommand \@href[1]{\@@startlink{#1}\@@href}%
\providecommand \@@href[1]{\endgroup#1\@@endlink}%
\providecommand \@sanitize@url [0]{\catcode `\\12\catcode `\$12\catcode
  `\&12\catcode `\#12\catcode `\^12\catcode `\_12\catcode `\%12\relax}%
\providecommand \@@startlink[1]{}%
\providecommand \@@endlink[0]{}%
\providecommand \url  [0]{\begingroup\@sanitize@url \@url }%
\providecommand \@url [1]{\endgroup\@href {#1}{\urlprefix }}%
\providecommand \urlprefix  [0]{URL }%
\providecommand \Eprint [0]{\href }%
\providecommand \doibase [0]{https://doi.org/}%
\providecommand \selectlanguage [0]{\@gobble}%
\providecommand \bibinfo  [0]{\@secondoftwo}%
\providecommand \bibfield  [0]{\@secondoftwo}%
\providecommand \translation [1]{[#1]}%
\providecommand \BibitemOpen [0]{}%
\providecommand \bibitemStop [0]{}%
\providecommand \bibitemNoStop [0]{.\EOS\space}%
\providecommand \EOS [0]{\spacefactor3000\relax}%
\providecommand \BibitemShut  [1]{\csname bibitem#1\endcsname}%
\let\auto@bib@innerbib\@empty
\bibitem [{\citenamefont {Raussendorf}\ and\ \citenamefont
  {Briegel}(2001{\natexlab{a}})}]{raussendorf_one-way_2001}%
  \BibitemOpen
  \bibfield  {author} {\bibinfo {author} {\bibfnamefont {R.}~\bibnamefont
  {Raussendorf}}\ and\ \bibinfo {author} {\bibfnamefont {H.~J.}\ \bibnamefont
  {Briegel}},\ }\bibinfo {title} {``A {One}-{Way} {Quantum} {Computer},''\
  }\href {https://link.aps.org/doi/10.1103/PhysRevLett.86.5188} {\bibfield
  {journal} {\bibinfo  {journal} {Physical Review Letters}\ }\textbf {\bibinfo
  {volume} {86}},\ \bibinfo {pages} {5188} (\bibinfo {year}
  {2001}{\natexlab{a}})}\BibitemShut {NoStop}%
\bibitem [{\citenamefont {Raussendorf}\ and\ \citenamefont
  {Briegel}(2001{\natexlab{b}})}]{raussendorf_computational_2001}%
  \BibitemOpen
  \bibfield  {author} {\bibinfo {author} {\bibfnamefont {R.}~\bibnamefont
  {Raussendorf}}\ and\ \bibinfo {author} {\bibfnamefont {H.}~\bibnamefont
  {Briegel}},\ }\href {http://arxiv.org/abs/quant-ph/0108067} {\bibinfo {title}
  {``Computational model underlying the one-way quantum computer,''\ }}
  (\bibinfo {year} {2001}{\natexlab{b}}),\ \bibinfo {note} {arXiv:
  quant-ph/0108067}\BibitemShut {NoStop}%
\bibitem [{\citenamefont {Nielsen}(2004)}]{nielsen_optical_2004}%
  \BibitemOpen
  \bibfield  {author} {\bibinfo {author} {\bibfnamefont {M.~A.}\ \bibnamefont
  {Nielsen}},\ }\bibinfo {title} {``Optical {Quantum} {Computation} {Using}
  {Cluster} {States},''\ }\href
  {https://link.aps.org/doi/10.1103/PhysRevLett.93.040503} {\bibfield
  {journal} {\bibinfo  {journal} {Physical Review Letters}\ }\textbf {\bibinfo
  {volume} {93}},\ \bibinfo {pages} {040503} (\bibinfo {year}
  {2004})}\BibitemShut {NoStop}%
\bibitem [{\citenamefont {Nielsen}(2006)}]{nielsen_cluster-state_2006}%
  \BibitemOpen
  \bibfield  {author} {\bibinfo {author} {\bibfnamefont {M.~A.}\ \bibnamefont
  {Nielsen}},\ }\bibinfo {title} {``Cluster-state quantum computation,''\
  }\href {https://www.sciencedirect.com/science/article/pii/S0034487706800145}
  {\bibfield  {journal} {\bibinfo  {journal} {Reports on Mathematical Physics}\
  }\textbf {\bibinfo {volume} {57}},\ \bibinfo {pages} {147} (\bibinfo {year}
  {2006})}\BibitemShut {NoStop}%
\bibitem [{\citenamefont {Braunstein}(1998)}]{braunstein_error_1998}%
  \BibitemOpen
  \bibfield  {author} {\bibinfo {author} {\bibfnamefont {S.~L.}\ \bibnamefont
  {Braunstein}},\ }in\ \href
  {https://link.springer.com/chapter/10.1007/978-94-015-1258-9_3}
  {{\selectlanguage {english}\emph {\bibinfo {booktitle} {Quantum {Information}
  with {Continuous} {Variables}}}}}\ (\bibinfo  {publisher} {Springer,
  Dordrecht},\ \bibinfo {year} {1998})\ pp.\ \bibinfo {pages}
  {19--29}\BibitemShut {NoStop}%
\bibitem [{\citenamefont {Menicucci}\ \emph {et~al.}(2006)\citenamefont
  {Menicucci}, \citenamefont {van Loock}, \citenamefont {Gu}, \citenamefont
  {Weedbrook}, \citenamefont {Ralph},\ and\ \citenamefont
  {Nielsen}}]{menicucci_universal_2006}%
  \BibitemOpen
  \bibfield  {author} {\bibinfo {author} {\bibfnamefont {N.~C.}\ \bibnamefont
  {Menicucci}}, \bibinfo {author} {\bibfnamefont {P.}~\bibnamefont {van
  Loock}}, \bibinfo {author} {\bibfnamefont {M.}~\bibnamefont {Gu}}, \bibinfo
  {author} {\bibfnamefont {C.}~\bibnamefont {Weedbrook}}, \bibinfo {author}
  {\bibfnamefont {T.~C.}\ \bibnamefont {Ralph}},\ and\ \bibinfo {author}
  {\bibfnamefont {M.~A.}\ \bibnamefont {Nielsen}},\ }\bibinfo {title}
  {``Universal {Quantum} {Computation} with {Continuous}-{Variable} {Cluster}
  {States},''\ }\href {https://link.aps.org/doi/10.1103/PhysRevLett.97.110501}
  {\bibfield  {journal} {\bibinfo  {journal} {Physical Review Letters}\
  }\textbf {\bibinfo {volume} {97}},\ \bibinfo {pages} {110501} (\bibinfo
  {year} {2006})}\BibitemShut {NoStop}%
\bibitem [{\citenamefont {Takeda}\ and\ \citenamefont
  {Furusawa}(2019)}]{takeda_toward_2019}%
  \BibitemOpen
  \bibfield  {author} {\bibinfo {author} {\bibfnamefont {S.}~\bibnamefont
  {Takeda}}\ and\ \bibinfo {author} {\bibfnamefont {A.}~\bibnamefont
  {Furusawa}},\ }\bibinfo {title} {``Toward large-scale fault-tolerant
  universal photonic quantum computing,''\ }\href
  {https://doi.org/10.1063/1.5100160} {\bibfield  {journal} {\bibinfo
  {journal} {APL Photonics}\ }\textbf {\bibinfo {volume} {4}},\ \bibinfo
  {pages} {060902} (\bibinfo {year} {2019})}\BibitemShut {NoStop}%
\bibitem [{\citenamefont {Yoshikawa}\ \emph {et~al.}(2016)\citenamefont
  {Yoshikawa}, \citenamefont {Yokoyama}, \citenamefont {Kaji}, \citenamefont
  {Sornphiphatphong}, \citenamefont {Shiozawa}, \citenamefont {Makino},\ and\
  \citenamefont {Furusawa}}]{yoshikawa_invited_2016}%
  \BibitemOpen
  \bibfield  {author} {\bibinfo {author} {\bibfnamefont {J.}~\bibnamefont
  {Yoshikawa}}, \bibinfo {author} {\bibfnamefont {S.}~\bibnamefont {Yokoyama}},
  \bibinfo {author} {\bibfnamefont {T.}~\bibnamefont {Kaji}}, \bibinfo {author}
  {\bibfnamefont {C.}~\bibnamefont {Sornphiphatphong}}, \bibinfo {author}
  {\bibfnamefont {Y.}~\bibnamefont {Shiozawa}}, \bibinfo {author}
  {\bibfnamefont {K.}~\bibnamefont {Makino}},\ and\ \bibinfo {author}
  {\bibfnamefont {A.}~\bibnamefont {Furusawa}},\ }\bibinfo {title} {``Invited
  {Article}: {Generation} of one-million-mode continuous-variable cluster state
  by unlimited time-domain multiplexing,''\ }\href
  {https://aip-scitation-org.ezproxy.lib.rmit.edu.au/doi/full/10.1063/1.4962732}
  {\bibfield  {journal} {\bibinfo  {journal} {APL Photonics}\ }\textbf
  {\bibinfo {volume} {1}},\ \bibinfo {pages} {060801} (\bibinfo {year}
  {2016})}\BibitemShut {NoStop}%
\bibitem [{\citenamefont {Asavanant}\ \emph {et~al.}(2017)\citenamefont
  {Asavanant}, \citenamefont {Nakashima}, \citenamefont {Shiozawa},
  \citenamefont {Yoshikawa},\ and\ \citenamefont
  {Furusawa}}]{asavanant_generation_2017}%
  \BibitemOpen
  \bibfield  {author} {\bibinfo {author} {\bibfnamefont {W.}~\bibnamefont
  {Asavanant}}, \bibinfo {author} {\bibfnamefont {K.}~\bibnamefont
  {Nakashima}}, \bibinfo {author} {\bibfnamefont {Y.}~\bibnamefont {Shiozawa}},
  \bibinfo {author} {\bibfnamefont {J.}~\bibnamefont {Yoshikawa}},\ and\
  \bibinfo {author} {\bibfnamefont {A.}~\bibnamefont {Furusawa}},\
  }{\selectlanguage {english}\bibinfo {title} {``Generation of highly pure
  {Schrödinger}’s cat states and real-time quadrature measurements via
  optical filtering,''\ }}\href
  {https://www.osapublishing.org/abstract.cfm?URI=oe-25-26-32227} {\bibfield
  {journal} {\bibinfo  {journal} {Optics Express}\ }\textbf {\bibinfo {volume}
  {25}},\ \bibinfo {pages} {32227} (\bibinfo {year} {2017})}\BibitemShut
  {NoStop}%
\bibitem [{\citenamefont {Asavanant}\ \emph {et~al.}(2019)\citenamefont
  {Asavanant}, \citenamefont {Shiozawa}, \citenamefont {Yokoyama},
  \citenamefont {Charoensombutamon}, \citenamefont {Emura}, \citenamefont
  {Alexander}, \citenamefont {Takeda}, \citenamefont {Yoshikawa}, \citenamefont
  {Menicucci}, \citenamefont {Yonezawa},\ and\ \citenamefont
  {Furusawa}}]{asavanant_generation_2019}%
  \BibitemOpen
  \bibfield  {author} {\bibinfo {author} {\bibfnamefont {W.}~\bibnamefont
  {Asavanant}}, \bibinfo {author} {\bibfnamefont {Y.}~\bibnamefont {Shiozawa}},
  \bibinfo {author} {\bibfnamefont {S.}~\bibnamefont {Yokoyama}}, \bibinfo
  {author} {\bibfnamefont {B.}~\bibnamefont {Charoensombutamon}}, \bibinfo
  {author} {\bibfnamefont {H.}~\bibnamefont {Emura}}, \bibinfo {author}
  {\bibfnamefont {R.~N.}\ \bibnamefont {Alexander}}, \bibinfo {author}
  {\bibfnamefont {S.}~\bibnamefont {Takeda}}, \bibinfo {author} {\bibfnamefont
  {J.}~\bibnamefont {Yoshikawa}}, \bibinfo {author} {\bibfnamefont {N.~C.}\
  \bibnamefont {Menicucci}}, \bibinfo {author} {\bibfnamefont {H.}~\bibnamefont
  {Yonezawa}},\ and\ \bibinfo {author} {\bibfnamefont {A.}~\bibnamefont
  {Furusawa}},\ }{\selectlanguage {english}\bibinfo {title} {``Generation of
  time-domain-multiplexed two-dimensional cluster state,''\ }}\href
  {https://science-sciencemag-org.ezproxy.lib.rmit.edu.au/content/366/6463/373}
  {\bibfield  {journal} {\bibinfo  {journal} {Science}\ }\textbf {\bibinfo
  {volume} {366}},\ \bibinfo {pages} {373} (\bibinfo {year}
  {2019})}\BibitemShut {NoStop}%
\bibitem [{\citenamefont {Takeda}\ and\ \citenamefont
  {Furusawa}(2017)}]{takeda_universal_2017}%
  \BibitemOpen
  \bibfield  {author} {\bibinfo {author} {\bibfnamefont {S.}~\bibnamefont
  {Takeda}}\ and\ \bibinfo {author} {\bibfnamefont {A.}~\bibnamefont
  {Furusawa}},\ }\bibinfo {title} {``Universal Quantum Computing with
  Measurement-Induced Continuous-Variable Gate Sequence in a Loop-Based
  Architecture,''\ }\href
  {https://link.aps.org/doi/10.1103/PhysRevLett.119.120504} {\bibfield
  {journal} {\bibinfo  {journal} {Phys. Rev. Lett.}\ }\textbf {\bibinfo
  {volume} {119}},\ \bibinfo {pages} {120504} (\bibinfo {year}
  {2017})}\BibitemShut {NoStop}%
\bibitem [{\citenamefont {Larsen}\ \emph {et~al.}(2019)\citenamefont {Larsen},
  \citenamefont {Guo}, \citenamefont {Breum}, \citenamefont
  {Neergaard-Nielsen},\ and\ \citenamefont
  {Andersen}}]{larsen_deterministic_2019}%
  \BibitemOpen
  \bibfield  {author} {\bibinfo {author} {\bibfnamefont {M.~V.}\ \bibnamefont
  {Larsen}}, \bibinfo {author} {\bibfnamefont {X.}~\bibnamefont {Guo}},
  \bibinfo {author} {\bibfnamefont {C.~R.}\ \bibnamefont {Breum}}, \bibinfo
  {author} {\bibfnamefont {J.~S.}\ \bibnamefont {Neergaard-Nielsen}},\ and\
  \bibinfo {author} {\bibfnamefont {U.~L.}\ \bibnamefont {Andersen}},\
  }{\selectlanguage {english}\bibinfo {title} {``Deterministic generation of a
  two-dimensional cluster state,''\ }}\href
  {https://science-sciencemag-org.ezproxy.lib.rmit.edu.au/content/366/6463/369}
  {\bibfield  {journal} {\bibinfo  {journal} {Science}\ }\textbf {\bibinfo
  {volume} {366}},\ \bibinfo {pages} {369} (\bibinfo {year}
  {2019})}\BibitemShut {NoStop}%
\bibitem [{\citenamefont {Asavanant}\ \emph {et~al.}(2020)\citenamefont
  {Asavanant}, \citenamefont {Charoensombutamon}, \citenamefont {Yokoyama},
  \citenamefont {Ebihara}, \citenamefont {Nakamura}, \citenamefont {Alexander},
  \citenamefont {Endo}, \citenamefont {Yoshikawa}, \citenamefont {Menicucci},
  \citenamefont {Yonezawa},\ and\ \citenamefont
  {Furusawa}}]{asavanant2020onehundred}%
  \BibitemOpen
  \bibfield  {author} {\bibinfo {author} {\bibfnamefont {W.}~\bibnamefont
  {Asavanant}}, \bibinfo {author} {\bibfnamefont {B.}~\bibnamefont
  {Charoensombutamon}}, \bibinfo {author} {\bibfnamefont {S.}~\bibnamefont
  {Yokoyama}}, \bibinfo {author} {\bibfnamefont {T.}~\bibnamefont {Ebihara}},
  \bibinfo {author} {\bibfnamefont {T.}~\bibnamefont {Nakamura}}, \bibinfo
  {author} {\bibfnamefont {R.~N.}\ \bibnamefont {Alexander}}, \bibinfo {author}
  {\bibfnamefont {M.}~\bibnamefont {Endo}}, \bibinfo {author} {\bibfnamefont
  {J.}~\bibnamefont {Yoshikawa}}, \bibinfo {author} {\bibfnamefont {N.~C.}\
  \bibnamefont {Menicucci}}, \bibinfo {author} {\bibfnamefont {H.}~\bibnamefont
  {Yonezawa}},\ and\ \bibinfo {author} {\bibfnamefont {A.}~\bibnamefont
  {Furusawa}},\ }\href@noop {} {\bibinfo {title} {``One-hundred step
  measurement-based quantum computation multiplexed in the time domain with 25
  MHz clock frequency,''\ }} (\bibinfo {year} {2020}),\ \Eprint
  {https://arxiv.org/abs/2006.11537} {arXiv:2006.11537 [quant-ph]} \BibitemShut
  {NoStop}%
\bibitem [{\citenamefont {Larsen}\ \emph {et~al.}(2020)\citenamefont {Larsen},
  \citenamefont {Guo}, \citenamefont {Breum}, \citenamefont
  {Neergaard-Nielsen},\ and\ \citenamefont
  {Andersen}}]{larsen2020deterministic}%
  \BibitemOpen
  \bibfield  {author} {\bibinfo {author} {\bibfnamefont {M.~V.}\ \bibnamefont
  {Larsen}}, \bibinfo {author} {\bibfnamefont {X.}~\bibnamefont {Guo}},
  \bibinfo {author} {\bibfnamefont {C.~R.}\ \bibnamefont {Breum}}, \bibinfo
  {author} {\bibfnamefont {J.~S.}\ \bibnamefont {Neergaard-Nielsen}},\ and\
  \bibinfo {author} {\bibfnamefont {U.~L.}\ \bibnamefont {Andersen}},\
  }\href@noop {} {\bibinfo {title} {``Deterministic multi-mode gates on a
  scalable photonic quantum computing platform,''\ }} (\bibinfo {year}
  {2020}),\ \Eprint {https://arxiv.org/abs/2010.14422} {arXiv:2010.14422
  [quant-ph]} \BibitemShut {NoStop}%
\bibitem [{\citenamefont {Menicucci}(2014)}]{menicucci_fault-tolerant_2014}%
  \BibitemOpen
  \bibfield  {author} {\bibinfo {author} {\bibfnamefont {N.~C.}\ \bibnamefont
  {Menicucci}},\ }{\selectlanguage {english}\bibinfo {title}
  {``Fault-{Tolerant} {Measurement}-{Based} {Quantum} {Computing} with
  {Continuous}-{Variable} {Cluster} {States},''\ }}\href
  {https://link.aps.org/doi/10.1103/PhysRevLett.112.120504} {\bibfield
  {journal} {\bibinfo  {journal} {Physical Review Letters}\ }\textbf {\bibinfo
  {volume} {112}} (\bibinfo {year} {2014})}\BibitemShut {NoStop}%
\bibitem [{\citenamefont {Cochrane}\ \emph {et~al.}(1999)\citenamefont
  {Cochrane}, \citenamefont {Milburn},\ and\ \citenamefont
  {Munro}}]{cochrane_macroscopically_1999}%
  \BibitemOpen
  \bibfield  {author} {\bibinfo {author} {\bibfnamefont {P.~T.}\ \bibnamefont
  {Cochrane}}, \bibinfo {author} {\bibfnamefont {G.~J.}\ \bibnamefont
  {Milburn}},\ and\ \bibinfo {author} {\bibfnamefont {W.~J.}\ \bibnamefont
  {Munro}},\ }\bibinfo {title} {``Macroscopically distinct
  quantum-superposition states as a bosonic code for amplitude damping,''\
  }\href {https://link.aps.org/doi/10.1103/PhysRevA.59.2631} {\bibfield
  {journal} {\bibinfo  {journal} {Physical Review A}\ }\textbf {\bibinfo
  {volume} {59}},\ \bibinfo {pages} {2631} (\bibinfo {year}
  {1999})}\BibitemShut {NoStop}%
\bibitem [{\citenamefont {Gottesman}\ \emph {et~al.}(2001)\citenamefont
  {Gottesman}, \citenamefont {Kitaev},\ and\ \citenamefont
  {Preskill}}]{gottesman_encoding_2001}%
  \BibitemOpen
  \bibfield  {author} {\bibinfo {author} {\bibfnamefont {D.}~\bibnamefont
  {Gottesman}}, \bibinfo {author} {\bibfnamefont {A.}~\bibnamefont {Kitaev}},\
  and\ \bibinfo {author} {\bibfnamefont {J.}~\bibnamefont {Preskill}},\
  }{\selectlanguage {english}\bibinfo {title} {``Encoding a qubit in an
  oscillator,''\ }}\href {https://link.aps.org/doi/10.1103/PhysRevA.64.012310}
  {\bibfield  {journal} {\bibinfo  {journal} {Physical Review A}\ }\textbf
  {\bibinfo {volume} {64}} (\bibinfo {year} {2001})}\BibitemShut {NoStop}%
\bibitem [{\citenamefont {Chuang}\ \emph {et~al.}(1997)\citenamefont {Chuang},
  \citenamefont {Leung},\ and\ \citenamefont {Yamamoto}}]{chuang_bosonic_1997}%
  \BibitemOpen
  \bibfield  {author} {\bibinfo {author} {\bibfnamefont {I.~L.}\ \bibnamefont
  {Chuang}}, \bibinfo {author} {\bibfnamefont {D.~W.}\ \bibnamefont {Leung}},\
  and\ \bibinfo {author} {\bibfnamefont {Y.}~\bibnamefont {Yamamoto}},\
  }\bibinfo {title} {``Bosonic quantum codes for amplitude damping,''\ }\href
  {https://link.aps.org/doi/10.1103/PhysRevA.56.1114} {\bibfield  {journal}
  {\bibinfo  {journal} {Physical Review A}\ }\textbf {\bibinfo {volume} {56}},\
  \bibinfo {pages} {1114} (\bibinfo {year} {1997})}\BibitemShut {NoStop}%
\bibitem [{\citenamefont {Terhal}\ and\ \citenamefont
  {Weigand}(2016)}]{terhal_encoding_2016}%
  \BibitemOpen
  \bibfield  {author} {\bibinfo {author} {\bibfnamefont {B.~M.}\ \bibnamefont
  {Terhal}}\ and\ \bibinfo {author} {\bibfnamefont {D.}~\bibnamefont
  {Weigand}},\ }\bibinfo {title} {``Encoding a qubit into a cavity mode in
  circuit QED using phase estimation,''\ }\href
  {https://link.aps.org/doi/10.1103/PhysRevA.93.012315} {\bibfield  {journal}
  {\bibinfo  {journal} {Phys. Rev. A}\ }\textbf {\bibinfo {volume} {93}},\
  \bibinfo {pages} {012315} (\bibinfo {year} {2016})}\BibitemShut {NoStop}%
\bibitem [{\citenamefont {Albert}\ \emph {et~al.}(2018)\citenamefont {Albert},
  \citenamefont {Noh}, \citenamefont {Duivenvoorden}, \citenamefont {Young},
  \citenamefont {Brierley}, \citenamefont {Reinhold}, \citenamefont {Vuillot},
  \citenamefont {Li}, \citenamefont {Shen}, \citenamefont {Girvin},
  \citenamefont {Terhal},\ and\ \citenamefont
  {Jiang}}]{albert_performance_2018}%
  \BibitemOpen
  \bibfield  {author} {\bibinfo {author} {\bibfnamefont {V.~V.}\ \bibnamefont
  {Albert}}, \bibinfo {author} {\bibfnamefont {K.}~\bibnamefont {Noh}},
  \bibinfo {author} {\bibfnamefont {K.}~\bibnamefont {Duivenvoorden}}, \bibinfo
  {author} {\bibfnamefont {D.~J.}\ \bibnamefont {Young}}, \bibinfo {author}
  {\bibfnamefont {R.~T.}\ \bibnamefont {Brierley}}, \bibinfo {author}
  {\bibfnamefont {P.}~\bibnamefont {Reinhold}}, \bibinfo {author}
  {\bibfnamefont {C.}~\bibnamefont {Vuillot}}, \bibinfo {author} {\bibfnamefont
  {L.}~\bibnamefont {Li}}, \bibinfo {author} {\bibfnamefont {C.}~\bibnamefont
  {Shen}}, \bibinfo {author} {\bibfnamefont {S.~M.}\ \bibnamefont {Girvin}},
  \bibinfo {author} {\bibfnamefont {B.~M.}\ \bibnamefont {Terhal}},\ and\
  \bibinfo {author} {\bibfnamefont {L.}~\bibnamefont {Jiang}},\
  }{\selectlanguage {english}\bibinfo {title} {``Performance and structure of
  single-mode bosonic codes,''\ }}\href
  {https://link.aps.org/doi/10.1103/PhysRevA.97.032346} {\bibfield  {journal}
  {\bibinfo  {journal} {Physical Review A}\ }\textbf {\bibinfo {volume} {97}},\
  \bibinfo {pages} {032346} (\bibinfo {year} {2018})}\BibitemShut {NoStop}%
\bibitem [{\citenamefont {Joshi}\ \emph {et~al.}(2021)\citenamefont {Joshi},
  \citenamefont {Noh},\ and\ \citenamefont {Gao}}]{joshi2020quantum}%
  \BibitemOpen
  \bibfield  {author} {\bibinfo {author} {\bibfnamefont {A.}~\bibnamefont
  {Joshi}}, \bibinfo {author} {\bibfnamefont {K.}~\bibnamefont {Noh}},\ and\
  \bibinfo {author} {\bibfnamefont {Y.~Y.}\ \bibnamefont {Gao}},\ }\bibinfo
  {title} {``Quantum information processing with bosonic qubits in circuit
  {QED},''\ }\href {https://doi.org/10.1088/2058-9565/abe989} {\bibfield
  {journal} {\bibinfo  {journal} {Quantum Science and Technology}\ }\textbf
  {\bibinfo {volume} {6}},\ \bibinfo {pages} {033001} (\bibinfo {year}
  {2021})}\BibitemShut {NoStop}%
\bibitem [{\citenamefont {Cai}\ \emph {et~al.}(2021)\citenamefont {Cai},
  \citenamefont {Ma}, \citenamefont {Wang}, \citenamefont {Zou},\ and\
  \citenamefont {Sun}}]{cai_bosonic_2021}%
  \BibitemOpen
  \bibfield  {author} {\bibinfo {author} {\bibfnamefont {W.}~\bibnamefont
  {Cai}}, \bibinfo {author} {\bibfnamefont {Y.}~\bibnamefont {Ma}}, \bibinfo
  {author} {\bibfnamefont {W.}~\bibnamefont {Wang}}, \bibinfo {author}
  {\bibfnamefont {C.-L.}\ \bibnamefont {Zou}},\ and\ \bibinfo {author}
  {\bibfnamefont {L.}~\bibnamefont {Sun}},\ }\bibinfo {title} {``Bosonic
  quantum error correction codes in superconducting quantum circuits,''\ }\href
  {https://www.sciencedirect.com/science/article/pii/S2667325820300145}
  {\bibfield  {journal} {\bibinfo  {journal} {Fundamental Research}\ }\textbf
  {\bibinfo {volume} {1}},\ \bibinfo {pages} {50} (\bibinfo {year}
  {2021})}\BibitemShut {NoStop}%
\bibitem [{\citenamefont {Wu}\ \emph {et~al.}(2020)\citenamefont {Wu},
  \citenamefont {Alexander}, \citenamefont {Liu},\ and\ \citenamefont
  {Zhang}}]{wu_quantum_2020}%
  \BibitemOpen
  \bibfield  {author} {\bibinfo {author} {\bibfnamefont {B.-H.}\ \bibnamefont
  {Wu}}, \bibinfo {author} {\bibfnamefont {R.~N.}\ \bibnamefont {Alexander}},
  \bibinfo {author} {\bibfnamefont {S.}~\bibnamefont {Liu}},\ and\ \bibinfo
  {author} {\bibfnamefont {Z.}~\bibnamefont {Zhang}},\ }{\selectlanguage
  {english}\bibinfo {title} {``Quantum computing with multidimensional
  continuous-variable cluster states in a scalable photonic platform,''\
  }}\href {https://link.aps.org/doi/10.1103/PhysRevResearch.2.023138}
  {\bibfield  {journal} {\bibinfo  {journal} {Physical Review Research}\
  }\textbf {\bibinfo {volume} {2}},\ \bibinfo {pages} {023138} (\bibinfo {year}
  {2020})}\BibitemShut {NoStop}%
\bibitem [{\citenamefont {Bourassa}\ \emph {et~al.}(2021)\citenamefont
  {Bourassa}, \citenamefont {Alexander}, \citenamefont {Vasmer}, \citenamefont
  {Patil}, \citenamefont {Tzitrin}, \citenamefont {Matsuura}, \citenamefont
  {Su}, \citenamefont {Baragiola}, \citenamefont {Guha}, \citenamefont
  {Dauphinais}, \citenamefont {Sabapathy}, \citenamefont {Menicucci},\ and\
  \citenamefont {Dhand}}]{bourassa2021blueprint}%
  \BibitemOpen
  \bibfield  {author} {\bibinfo {author} {\bibfnamefont {J.~E.}\ \bibnamefont
  {Bourassa}}, \bibinfo {author} {\bibfnamefont {R.~N.}\ \bibnamefont
  {Alexander}}, \bibinfo {author} {\bibfnamefont {M.}~\bibnamefont {Vasmer}},
  \bibinfo {author} {\bibfnamefont {A.}~\bibnamefont {Patil}}, \bibinfo
  {author} {\bibfnamefont {I.}~\bibnamefont {Tzitrin}}, \bibinfo {author}
  {\bibfnamefont {T.}~\bibnamefont {Matsuura}}, \bibinfo {author}
  {\bibfnamefont {D.}~\bibnamefont {Su}}, \bibinfo {author} {\bibfnamefont
  {B.~Q.}\ \bibnamefont {Baragiola}}, \bibinfo {author} {\bibfnamefont
  {S.}~\bibnamefont {Guha}}, \bibinfo {author} {\bibfnamefont {G.}~\bibnamefont
  {Dauphinais}}, \bibinfo {author} {\bibfnamefont {K.~K.}\ \bibnamefont
  {Sabapathy}}, \bibinfo {author} {\bibfnamefont {N.~C.}\ \bibnamefont
  {Menicucci}},\ and\ \bibinfo {author} {\bibfnamefont {I.}~\bibnamefont
  {Dhand}},\ }\bibinfo {title} {``Blueprint for a {S}calable {P}hotonic
  {F}ault-{T}olerant {Q}uantum {C}omputer,''\ }\href
  {https://doi.org/10.22331/q-2021-02-04-392} {\bibfield  {journal} {\bibinfo
  {journal} {{Quantum}}\ }\textbf {\bibinfo {volume} {5}},\ \bibinfo {pages}
  {392} (\bibinfo {year} {2021})}\BibitemShut {NoStop}%
\bibitem [{\citenamefont {Larsen}\ \emph {et~al.}(2021)\citenamefont {Larsen},
  \citenamefont {Chamberland}, \citenamefont {Noh}, \citenamefont
  {Neergaard-Nielsen},\ and\ \citenamefont
  {Andersen}}]{larsen2021faulttolerant}%
  \BibitemOpen
  \bibfield  {author} {\bibinfo {author} {\bibfnamefont {M.~V.}\ \bibnamefont
  {Larsen}}, \bibinfo {author} {\bibfnamefont {C.}~\bibnamefont {Chamberland}},
  \bibinfo {author} {\bibfnamefont {K.}~\bibnamefont {Noh}}, \bibinfo {author}
  {\bibfnamefont {J.~S.}\ \bibnamefont {Neergaard-Nielsen}},\ and\ \bibinfo
  {author} {\bibfnamefont {U.~L.}\ \bibnamefont {Andersen}},\ }\href@noop {}
  {\bibinfo {title} {``A fault-tolerant continuous-variable measurement-based
  quantum computation architecture,''\ }} (\bibinfo {year} {2021}),\ \Eprint
  {https://arxiv.org/abs/2101.03014} {arXiv:2101.03014 [quant-ph]} \BibitemShut
  {NoStop}%
\bibitem [{\citenamefont {Pantaleoni}\ \emph {et~al.}(2020)\citenamefont
  {Pantaleoni}, \citenamefont {Baragiola},\ and\ \citenamefont
  {Menicucci}}]{pantaleoni_modular_2020}%
  \BibitemOpen
  \bibfield  {author} {\bibinfo {author} {\bibfnamefont {G.}~\bibnamefont
  {Pantaleoni}}, \bibinfo {author} {\bibfnamefont {B.~Q.}\ \bibnamefont
  {Baragiola}},\ and\ \bibinfo {author} {\bibfnamefont {N.~C.}\ \bibnamefont
  {Menicucci}},\ }\bibinfo {title} {``Modular {Bosonic} {Subsystem} {Codes},''\
  }\href {https://link.aps.org/doi/10.1103/PhysRevLett.125.040501} {\bibfield
  {journal} {\bibinfo  {journal} {Physical Review Letters}\ }\textbf {\bibinfo
  {volume} {125}},\ \bibinfo {pages} {040501} (\bibinfo {year}
  {2020})}\BibitemShut {NoStop}%
\bibitem [{\citenamefont {Aharonov}\ \emph {et~al.}(1969)\citenamefont
  {Aharonov}, \citenamefont {Pendleton},\ and\ \citenamefont
  {Petersen}}]{aharonov_modular_1969}%
  \BibitemOpen
  \bibfield  {author} {\bibinfo {author} {\bibfnamefont {Y.}~\bibnamefont
  {Aharonov}}, \bibinfo {author} {\bibfnamefont {H.}~\bibnamefont
  {Pendleton}},\ and\ \bibinfo {author} {\bibfnamefont {A.}~\bibnamefont
  {Petersen}},\ }{\selectlanguage {english}\bibinfo {title} {``Modular
  variables in quantum theory,''\ }}\href
  {https://link.springer.com/article/10.1007/BF00670008} {\bibfield  {journal}
  {\bibinfo  {journal} {International Journal of Theoretical Physics}\ }\textbf
  {\bibinfo {volume} {2}},\ \bibinfo {pages} {213} (\bibinfo {year}
  {1969})}\BibitemShut {NoStop}%
\bibitem [{\citenamefont {Zak}(1967)}]{zak_finite_1967}%
  \BibitemOpen
  \bibfield  {author} {\bibinfo {author} {\bibfnamefont {J.}~\bibnamefont
  {Zak}},\ }\bibinfo {title} {``Finite {Translations} in {Solid}-{State}
  {Physics},''\ }\href {https://link.aps.org/doi/10.1103/PhysRevLett.19.1385}
  {\bibfield  {journal} {\bibinfo  {journal} {Physical Review Letters}\
  }\textbf {\bibinfo {volume} {19}},\ \bibinfo {pages} {1385} (\bibinfo {year}
  {1967})}\BibitemShut {NoStop}%
\bibitem [{\citenamefont {Zak}(1968)}]{zak_dynamics_1968}%
  \BibitemOpen
  \bibfield  {author} {\bibinfo {author} {\bibfnamefont {J.}~\bibnamefont
  {Zak}},\ }\bibinfo {title} {``Dynamics of {Electrons} in {Solids} in
  {External} {Fields},''\ }\href
  {https://link.aps.org/doi/10.1103/PhysRev.168.686} {\bibfield  {journal}
  {\bibinfo  {journal} {Physical Review}\ }\textbf {\bibinfo {volume} {168}},\
  \bibinfo {pages} {686} (\bibinfo {year} {1968})}\BibitemShut {NoStop}%
\bibitem [{\citenamefont {Englert}\ \emph {et~al.}(2006)\citenamefont
  {Englert}, \citenamefont {Lee}, \citenamefont {Mann},\ and\ \citenamefont
  {Revzen}}]{englert_periodic_2006}%
  \BibitemOpen
  \bibfield  {author} {\bibinfo {author} {\bibfnamefont {B.-G.}\ \bibnamefont
  {Englert}}, \bibinfo {author} {\bibfnamefont {K.~L.}\ \bibnamefont {Lee}},
  \bibinfo {author} {\bibfnamefont {A.}~\bibnamefont {Mann}},\ and\ \bibinfo
  {author} {\bibfnamefont {M.}~\bibnamefont {Revzen}},\ }{\selectlanguage
  {english}\bibinfo {title} {``Periodic and discrete {Zak} bases,''\ }}\href
  {http://stacks.iop.org/0305-4470/39/i=7/a=011} {\bibfield  {journal}
  {\bibinfo  {journal} {Journal of Physics A: Mathematical and General}\
  }\textbf {\bibinfo {volume} {39}},\ \bibinfo {pages} {1669} (\bibinfo {year}
  {2006})}\BibitemShut {NoStop}%
\bibitem [{\citenamefont {Albert}\ \emph {et~al.}(2017)\citenamefont {Albert},
  \citenamefont {Pascazio},\ and\ \citenamefont
  {Devoret}}]{albert_general_2017}%
  \BibitemOpen
  \bibfield  {author} {\bibinfo {author} {\bibfnamefont {V.~V.}\ \bibnamefont
  {Albert}}, \bibinfo {author} {\bibfnamefont {S.}~\bibnamefont {Pascazio}},\
  and\ \bibinfo {author} {\bibfnamefont {M.~H.}\ \bibnamefont {Devoret}},\
  }{\selectlanguage {english}\bibinfo {title} {``General phase spaces: from
  discrete variables to rotor and continuum limits,''\ }}\href
  {http://stacks.iop.org/1751-8121/50/i=50/a=504002} {\bibfield  {journal}
  {\bibinfo  {journal} {Journal of Physics A: Mathematical and Theoretical}\
  }\textbf {\bibinfo {volume} {50}},\ \bibinfo {pages} {504002} (\bibinfo
  {year} {2017})}\BibitemShut {NoStop}%
\bibitem [{\citenamefont {Albert}\ \emph {et~al.}(2020)\citenamefont {Albert},
  \citenamefont {Covey},\ and\ \citenamefont {Preskill}}]{albert2020robust}%
  \BibitemOpen
  \bibfield  {author} {\bibinfo {author} {\bibfnamefont {V.~V.}\ \bibnamefont
  {Albert}}, \bibinfo {author} {\bibfnamefont {J.~P.}\ \bibnamefont {Covey}},\
  and\ \bibinfo {author} {\bibfnamefont {J.}~\bibnamefont {Preskill}},\
  }\bibinfo {title} {``Robust Encoding of a Qubit in a Molecule,''\ }\href
  {https://link.aps.org/doi/10.1103/PhysRevX.10.031050} {\bibfield  {journal}
  {\bibinfo  {journal} {Phys. Rev. X}\ }\textbf {\bibinfo {volume} {10}},\
  \bibinfo {pages} {031050} (\bibinfo {year} {2020})}\BibitemShut {NoStop}%
\bibitem [{\citenamefont {Janssen}(1988)}]{janssen_zak_1988}%
  \BibitemOpen
  \bibfield  {author} {\bibinfo {author} {\bibfnamefont {A.}~\bibnamefont
  {Janssen}},\ }\bibinfo {title} {``The {Zak} transform: {A} signal transform
  for sampled time-continuous signals,''\ }\href@noop {} {\bibfield  {journal}
  {\bibinfo  {journal} {Philips J. Res}\ }\textbf {\bibinfo {volume} {43}},\
  \bibinfo {pages} {23} (\bibinfo {year} {1988})}\BibitemShut {NoStop}%
\bibitem [{\citenamefont {Raynal}\ \emph {et~al.}(2012)\citenamefont {Raynal},
  \citenamefont {Kalev}, \citenamefont {Suzuki},\ and\ \citenamefont
  {Englert}}]{raynal_encoding_2012}%
  \BibitemOpen
  \bibfield  {author} {\bibinfo {author} {\bibfnamefont {P.}~\bibnamefont
  {Raynal}}, \bibinfo {author} {\bibfnamefont {A.}~\bibnamefont {Kalev}},
  \bibinfo {author} {\bibfnamefont {J.}~\bibnamefont {Suzuki}},\ and\ \bibinfo
  {author} {\bibfnamefont {B.-G.}\ \bibnamefont {Englert}},\ }\bibinfo {title}
  {``Encoding many qubits in a rotor,''\ }\href
  {https://aip-scitation-org.ezproxy.lib.rmit.edu.au/doi/abs/10.1063/1.4746063}
  {\bibfield  {journal} {\bibinfo  {journal} {AIP Conference Proceedings}\
  }\textbf {\bibinfo {volume} {1469}},\ \bibinfo {pages} {63} (\bibinfo {year}
  {2012})}\BibitemShut {NoStop}%
\bibitem [{\citenamefont {Wan}\ \emph {et~al.}(2020)\citenamefont {Wan},
  \citenamefont {Neville},\ and\ \citenamefont
  {Kolthammer}}]{wan_memory-assisted_2020}%
  \BibitemOpen
  \bibfield  {author} {\bibinfo {author} {\bibfnamefont {K.~H.}\ \bibnamefont
  {Wan}}, \bibinfo {author} {\bibfnamefont {A.}~\bibnamefont {Neville}},\ and\
  \bibinfo {author} {\bibfnamefont {S.}~\bibnamefont {Kolthammer}},\ }\bibinfo
  {title} {``Memory-assisted decoder for approximate Gottesman-Kitaev-Preskill
  codes,''\ }\href {https://link.aps.org/doi/10.1103/PhysRevResearch.2.043280}
  {\bibfield  {journal} {\bibinfo  {journal} {Phys. Rev. Research}\ }\textbf
  {\bibinfo {volume} {2}},\ \bibinfo {pages} {043280} (\bibinfo {year}
  {2020})}\BibitemShut {NoStop}%
\bibitem [{\citenamefont {Tzitrin}\ \emph {et~al.}(2020)\citenamefont
  {Tzitrin}, \citenamefont {Bourassa}, \citenamefont {Menicucci},\ and\
  \citenamefont {Sabapathy}}]{tzitrin_progress_2020}%
  \BibitemOpen
  \bibfield  {author} {\bibinfo {author} {\bibfnamefont {I.}~\bibnamefont
  {Tzitrin}}, \bibinfo {author} {\bibfnamefont {J.~E.}\ \bibnamefont
  {Bourassa}}, \bibinfo {author} {\bibfnamefont {N.~C.}\ \bibnamefont
  {Menicucci}},\ and\ \bibinfo {author} {\bibfnamefont {K.~K.}\ \bibnamefont
  {Sabapathy}},\ }\bibinfo {title} {``Progress towards practical qubit
  computation using approximate {Gottesman}-{Kitaev}-{Preskill} codes,''\
  }\href {https://link.aps.org/doi/10.1103/PhysRevA.101.032315} {\bibfield
  {journal} {\bibinfo  {journal} {Physical Review A}\ }\textbf {\bibinfo
  {volume} {101}},\ \bibinfo {pages} {032315} (\bibinfo {year}
  {2020})}\BibitemShut {NoStop}%
\bibitem [{\citenamefont {Hastrup}\ \emph {et~al.}(2020)\citenamefont
  {Hastrup}, \citenamefont {Larsen}, \citenamefont {Neergaard-Nielsen},
  \citenamefont {Menicucci},\ and\ \citenamefont
  {Andersen}}]{hastrup2020cubic}%
  \BibitemOpen
  \bibfield  {author} {\bibinfo {author} {\bibfnamefont {J.}~\bibnamefont
  {Hastrup}}, \bibinfo {author} {\bibfnamefont {M.~V.}\ \bibnamefont {Larsen}},
  \bibinfo {author} {\bibfnamefont {J.~S.}\ \bibnamefont {Neergaard-Nielsen}},
  \bibinfo {author} {\bibfnamefont {N.~C.}\ \bibnamefont {Menicucci}},\ and\
  \bibinfo {author} {\bibfnamefont {U.~L.}\ \bibnamefont {Andersen}},\
  }\href@noop {} {\bibinfo {title} {``Cubic phase gates are not suitable for
  non-Clifford operations on GKP states,''\ }} (\bibinfo {year} {2020}),\
  \Eprint {https://arxiv.org/abs/2009.05309} {arXiv:2009.05309 [quant-ph]}
  \BibitemShut {NoStop}%
\bibitem [{\citenamefont {Poulin}(2005)}]{poulin_stabilizer_2005}%
  \BibitemOpen
  \bibfield  {author} {\bibinfo {author} {\bibfnamefont {D.}~\bibnamefont
  {Poulin}},\ }\bibinfo {title} {``Stabilizer {Formalism} for {Operator}
  {Quantum} {Error} {Correction},''\ }\href
  {https://link.aps.org/doi/10.1103/PhysRevLett.95.230504} {\bibfield
  {journal} {\bibinfo  {journal} {Physical Review Letters}\ }\textbf {\bibinfo
  {volume} {95}},\ \bibinfo {pages} {230504} (\bibinfo {year}
  {2005})}\BibitemShut {NoStop}%
\bibitem [{\citenamefont {Vedral}\ \emph {et~al.}(1996)\citenamefont {Vedral},
  \citenamefont {Barenco},\ and\ \citenamefont {Ekert}}]{quantumadder}%
  \BibitemOpen
  \bibfield  {author} {\bibinfo {author} {\bibfnamefont {V.}~\bibnamefont
  {Vedral}}, \bibinfo {author} {\bibfnamefont {A.}~\bibnamefont {Barenco}},\
  and\ \bibinfo {author} {\bibfnamefont {A.}~\bibnamefont {Ekert}},\ }\bibinfo
  {title} {``Quantum networks for elementary arithmetic operations,''\ }\href
  {https://link.aps.org/doi/10.1103/PhysRevA.54.147} {\bibfield  {journal}
  {\bibinfo  {journal} {Phys. Rev. A}\ }\textbf {\bibinfo {volume} {54}},\
  \bibinfo {pages} {147} (\bibinfo {year} {1996})}\BibitemShut {NoStop}%
\bibitem [{\citenamefont {Baragiola}\ \emph {et~al.}(2019)\citenamefont
  {Baragiola}, \citenamefont {Pantaleoni}, \citenamefont {Alexander},
  \citenamefont {Karanjai},\ and\ \citenamefont
  {Menicucci}}]{baragiola_all-gaussian_2019}%
  \BibitemOpen
  \bibfield  {author} {\bibinfo {author} {\bibfnamefont {B.~Q.}\ \bibnamefont
  {Baragiola}}, \bibinfo {author} {\bibfnamefont {G.}~\bibnamefont
  {Pantaleoni}}, \bibinfo {author} {\bibfnamefont {R.~N.}\ \bibnamefont
  {Alexander}}, \bibinfo {author} {\bibfnamefont {A.}~\bibnamefont
  {Karanjai}},\ and\ \bibinfo {author} {\bibfnamefont {N.~C.}\ \bibnamefont
  {Menicucci}},\ }\bibinfo {title} {``All-{Gaussian} {Universality} and {Fault}
  {Tolerance} with the {Gottesman}-{Kitaev}-{Preskill} {Code},''\ }\href
  {https://link.aps.org/doi/10.1103/PhysRevLett.123.200502} {\bibfield
  {journal} {\bibinfo  {journal} {Physical Review Letters}\ }\textbf {\bibinfo
  {volume} {123}},\ \bibinfo {pages} {200502} (\bibinfo {year}
  {2019})}\BibitemShut {NoStop}%
\bibitem [{\citenamefont {Yamasaki}\ \emph {et~al.}(2020)\citenamefont
  {Yamasaki}, \citenamefont {Matsuura},\ and\ \citenamefont
  {Koashi}}]{yamasaki2020cost}%
  \BibitemOpen
  \bibfield  {author} {\bibinfo {author} {\bibfnamefont {H.}~\bibnamefont
  {Yamasaki}}, \bibinfo {author} {\bibfnamefont {T.}~\bibnamefont {Matsuura}},\
  and\ \bibinfo {author} {\bibfnamefont {M.}~\bibnamefont {Koashi}},\ }\bibinfo
  {title} {``Cost-reduced all-Gaussian universality with the
  Gottesman-Kitaev-Preskill code: Resource-theoretic approach to cost
  analysis,''\ }\href
  {https://link.aps.org/doi/10.1103/PhysRevResearch.2.023270} {\bibfield
  {journal} {\bibinfo  {journal} {Phys. Rev. Research}\ }\textbf {\bibinfo
  {volume} {2}},\ \bibinfo {pages} {023270} (\bibinfo {year}
  {2020})}\BibitemShut {NoStop}%
\bibitem [{\citenamefont {Mensen}\ \emph {et~al.}(2020)\citenamefont {Mensen},
  \citenamefont {Baragiola},\ and\ \citenamefont
  {Menicucci}}]{mensen2020phasespace}%
  \BibitemOpen
  \bibfield  {author} {\bibinfo {author} {\bibfnamefont {L.~J.}\ \bibnamefont
  {Mensen}}, \bibinfo {author} {\bibfnamefont {B.~Q.}\ \bibnamefont
  {Baragiola}},\ and\ \bibinfo {author} {\bibfnamefont {N.~C.}\ \bibnamefont
  {Menicucci}},\ }\href@noop {} {\bibinfo {title} {``Phase-space methods for
  representing, manipulating, and correcting Gottesman-Kitaev-Preskill
  qubits,''\ }} (\bibinfo {year} {2020}),\ \Eprint
  {https://arxiv.org/abs/2012.12488} {arXiv:2012.12488 [quant-ph]} \BibitemShut
  {NoStop}%
\bibitem [{\citenamefont {Ketterer}\ \emph {et~al.}(2016)\citenamefont
  {Ketterer}, \citenamefont {Keller}, \citenamefont {Walborn}, \citenamefont
  {Coudreau},\ and\ \citenamefont {Milman}}]{ketterer_quantum_2016}%
  \BibitemOpen
  \bibfield  {author} {\bibinfo {author} {\bibfnamefont {A.}~\bibnamefont
  {Ketterer}}, \bibinfo {author} {\bibfnamefont {A.}~\bibnamefont {Keller}},
  \bibinfo {author} {\bibfnamefont {S.~P.}\ \bibnamefont {Walborn}}, \bibinfo
  {author} {\bibfnamefont {T.}~\bibnamefont {Coudreau}},\ and\ \bibinfo
  {author} {\bibfnamefont {P.}~\bibnamefont {Milman}},\ }\bibinfo {title}
  {``Quantum information processing in phase space: {A} modular variables
  approach,''\ }\href {https://link.aps.org/doi/10.1103/PhysRevA.94.022325}
  {\bibfield  {journal} {\bibinfo  {journal} {Physical Review A}\ }\textbf
  {\bibinfo {volume} {94}},\ \bibinfo {pages} {022325} (\bibinfo {year}
  {2016})}\BibitemShut {NoStop}%
\bibitem [{\citenamefont {Fabre}\ \emph {et~al.}(2020)\citenamefont {Fabre},
  \citenamefont {Keller},\ and\ \citenamefont {Milman}}]{fabre_wigner_2020}%
  \BibitemOpen
  \bibfield  {author} {\bibinfo {author} {\bibfnamefont {N.}~\bibnamefont
  {Fabre}}, \bibinfo {author} {\bibfnamefont {A.}~\bibnamefont {Keller}},\ and\
  \bibinfo {author} {\bibfnamefont {P.}~\bibnamefont {Milman}},\ }\bibinfo
  {title} {``Wigner distribution on a double-cylinder phase space for studying
  quantum error-correction protocols,''\ }\href
  {https://link.aps.org/doi/10.1103/PhysRevA.102.022411} {\bibfield  {journal}
  {\bibinfo  {journal} {Phys. Rev. A}\ }\textbf {\bibinfo {volume} {102}},\
  \bibinfo {pages} {022411} (\bibinfo {year} {2020})}\BibitemShut {NoStop}%
\bibitem [{\citenamefont {Walshe}\ \emph {et~al.}(2020)\citenamefont {Walshe},
  \citenamefont {Baragiola}, \citenamefont {Alexander},\ and\ \citenamefont
  {Menicucci}}]{walshe_2020}%
  \BibitemOpen
  \bibfield  {author} {\bibinfo {author} {\bibfnamefont {B.~W.}\ \bibnamefont
  {Walshe}}, \bibinfo {author} {\bibfnamefont {B.~Q.}\ \bibnamefont
  {Baragiola}}, \bibinfo {author} {\bibfnamefont {R.~N.}\ \bibnamefont
  {Alexander}},\ and\ \bibinfo {author} {\bibfnamefont {N.~C.}\ \bibnamefont
  {Menicucci}},\ }\bibinfo {title} {``Continuous-variable gate teleportation
  and bosonic-code error correction,''\ }\href
  {https://link.aps.org/doi/10.1103/PhysRevA.102.062411} {\bibfield  {journal}
  {\bibinfo  {journal} {Phys. Rev. A}\ }\textbf {\bibinfo {volume} {102}},\
  \bibinfo {pages} {062411} (\bibinfo {year} {2020})}\BibitemShut {NoStop}%
\bibitem [{\citenamefont {Alexander}\ \emph {et~al.}(2014)\citenamefont
  {Alexander}, \citenamefont {Armstrong}, \citenamefont {Ukai},\ and\
  \citenamefont {Menicucci}}]{alexander_noise_2014}%
  \BibitemOpen
  \bibfield  {author} {\bibinfo {author} {\bibfnamefont {R.~N.}\ \bibnamefont
  {Alexander}}, \bibinfo {author} {\bibfnamefont {S.~C.}\ \bibnamefont
  {Armstrong}}, \bibinfo {author} {\bibfnamefont {R.}~\bibnamefont {Ukai}},\
  and\ \bibinfo {author} {\bibfnamefont {N.~C.}\ \bibnamefont {Menicucci}},\
  }\bibinfo {title} {``Noise analysis of single-mode {Gaussian} operations
  using continuous-variable cluster states,''\ }\href
  {https://link.aps.org/doi/10.1103/PhysRevA.90.062324} {\bibfield  {journal}
  {\bibinfo  {journal} {Physical Review A}\ }\textbf {\bibinfo {volume} {90}},\
  \bibinfo {pages} {062324} (\bibinfo {year} {2014})}\BibitemShut {NoStop}%
\bibitem [{\citenamefont {Bellman}(2013)}]{bellman_brief_2013}%
  \BibitemOpen
  \bibfield  {author} {\bibinfo {author} {\bibfnamefont {R.}~\bibnamefont
  {Bellman}},\ }\href@noop {} {{\selectlanguage {english}\emph {\bibinfo
  {title} {A {Brief} {Introduction} to {Theta} {Functions}}}}}\ (\bibinfo
  {publisher} {Courier Corporation},\ \bibinfo {year} {2013})\BibitemShut
  {NoStop}%
\bibitem [{\citenamefont {Matsuura}\ \emph {et~al.}(2020)\citenamefont
  {Matsuura}, \citenamefont {Yamasaki},\ and\ \citenamefont
  {Koashi}}]{matsuura2020equivalence}%
  \BibitemOpen
  \bibfield  {author} {\bibinfo {author} {\bibfnamefont {T.}~\bibnamefont
  {Matsuura}}, \bibinfo {author} {\bibfnamefont {H.}~\bibnamefont {Yamasaki}},\
  and\ \bibinfo {author} {\bibfnamefont {M.}~\bibnamefont {Koashi}},\ }\bibinfo
  {title} {``Equivalence of approximate Gottesman-Kitaev-Preskill codes,''\
  }\href {https://link.aps.org/doi/10.1103/PhysRevA.102.032408} {\bibfield
  {journal} {\bibinfo  {journal} {Phys. Rev. A}\ }\textbf {\bibinfo {volume}
  {102}},\ \bibinfo {pages} {032408} (\bibinfo {year} {2020})}\BibitemShut
  {NoStop}%
\bibitem [{\citenamefont {Ralph}\ \emph {et~al.}(2003)\citenamefont {Ralph},
  \citenamefont {Gilchrist}, \citenamefont {Milburn}, \citenamefont {Munro},\
  and\ \citenamefont {Glancy}}]{ralph2003computation}%
  \BibitemOpen
  \bibfield  {author} {\bibinfo {author} {\bibfnamefont {T.~C.}\ \bibnamefont
  {Ralph}}, \bibinfo {author} {\bibfnamefont {A.}~\bibnamefont {Gilchrist}},
  \bibinfo {author} {\bibfnamefont {G.~J.}\ \bibnamefont {Milburn}}, \bibinfo
  {author} {\bibfnamefont {W.~J.}\ \bibnamefont {Munro}},\ and\ \bibinfo
  {author} {\bibfnamefont {S.}~\bibnamefont {Glancy}},\ }\bibinfo {title}
  {``Quantum computation with optical coherent states,''\ }\href
  {https://link.aps.org/doi/10.1103/PhysRevA.68.042319} {\bibfield  {journal}
  {\bibinfo  {journal} {Phys. Rev. A}\ }\textbf {\bibinfo {volume} {68}},\
  \bibinfo {pages} {042319} (\bibinfo {year} {2003})}\BibitemShut {NoStop}%
\bibitem [{\citenamefont {Lund}\ \emph {et~al.}(2008)\citenamefont {Lund},
  \citenamefont {Ralph},\ and\ \citenamefont
  {Haselgrove}}]{lund_fault-tolerant_2008}%
  \BibitemOpen
  \bibfield  {author} {\bibinfo {author} {\bibfnamefont {A.~P.}\ \bibnamefont
  {Lund}}, \bibinfo {author} {\bibfnamefont {T.~C.}\ \bibnamefont {Ralph}},\
  and\ \bibinfo {author} {\bibfnamefont {H.~L.}\ \bibnamefont {Haselgrove}},\
  }\bibinfo {title} {``Fault-{Tolerant} {Linear} {Optical} {Quantum}
  {Computing} with {Small}-{Amplitude} {Coherent} {States},''\ }\href
  {https://link.aps.org/doi/10.1103/PhysRevLett.100.030503} {\bibfield
  {journal} {\bibinfo  {journal} {Physical Review Letters}\ }\textbf {\bibinfo
  {volume} {100}},\ \bibinfo {pages} {030503} (\bibinfo {year}
  {2008})}\BibitemShut {NoStop}%
\bibitem [{\citenamefont {Braunstein}\ and\ \citenamefont
  {Kimble}(1998)}]{braunstein_teleportation_1998}%
  \BibitemOpen
  \bibfield  {author} {\bibinfo {author} {\bibfnamefont {S.~L.}\ \bibnamefont
  {Braunstein}}\ and\ \bibinfo {author} {\bibfnamefont {H.~J.}\ \bibnamefont
  {Kimble}},\ }\bibinfo {title} {``Teleportation of Continuous Quantum
  Variables,''\ }\href {https://link.aps.org/doi/10.1103/PhysRevLett.80.869}
  {\bibfield  {journal} {\bibinfo  {journal} {Phys. Rev. Lett.}\ }\textbf
  {\bibinfo {volume} {80}},\ \bibinfo {pages} {869} (\bibinfo {year}
  {1998})}\BibitemShut {NoStop}%
\bibitem [{\citenamefont {Gu}\ \emph {et~al.}(2009)\citenamefont {Gu},
  \citenamefont {Weedbrook}, \citenamefont {Menicucci}, \citenamefont {Ralph},\
  and\ \citenamefont {van Loock}}]{gu_quantum_2009}%
  \BibitemOpen
  \bibfield  {author} {\bibinfo {author} {\bibfnamefont {M.}~\bibnamefont
  {Gu}}, \bibinfo {author} {\bibfnamefont {C.}~\bibnamefont {Weedbrook}},
  \bibinfo {author} {\bibfnamefont {N.~C.}\ \bibnamefont {Menicucci}}, \bibinfo
  {author} {\bibfnamefont {T.~C.}\ \bibnamefont {Ralph}},\ and\ \bibinfo
  {author} {\bibfnamefont {P.}~\bibnamefont {van Loock}},\ }\bibinfo {title}
  {``Quantum computing with continuous-variable clusters,''\ }\href
  {https://link.aps.org/doi/10.1103/PhysRevA.79.062318} {\bibfield  {journal}
  {\bibinfo  {journal} {Physical Review A}\ }\textbf {\bibinfo {volume} {79}},\
  \bibinfo {pages} {062318} (\bibinfo {year} {2009})}\BibitemShut {NoStop}%
\bibitem [{\citenamefont {Grimsmo}\ \emph {et~al.}(2020)\citenamefont
  {Grimsmo}, \citenamefont {Combes},\ and\ \citenamefont
  {Baragiola}}]{grimsmo_quantum_2020}%
  \BibitemOpen
  \bibfield  {author} {\bibinfo {author} {\bibfnamefont {A.~L.}\ \bibnamefont
  {Grimsmo}}, \bibinfo {author} {\bibfnamefont {J.}~\bibnamefont {Combes}},\
  and\ \bibinfo {author} {\bibfnamefont {B.~Q.}\ \bibnamefont {Baragiola}},\
  }\bibinfo {title} {``Quantum {Computing} with {Rotation}-{Symmetric}
  {Bosonic} {Codes},''\ }\href
  {https://link.aps.org/doi/10.1103/PhysRevX.10.011058} {\bibfield  {journal}
  {\bibinfo  {journal} {Physical Review X}\ }\textbf {\bibinfo {volume} {10}},\
  \bibinfo {pages} {011058} (\bibinfo {year} {2020})}\BibitemShut {NoStop}%
\bibitem [{\citenamefont {Nielsen}\ and\ \citenamefont
  {Chuang}(2002)}]{nielsen2002quantum}%
  \BibitemOpen
  \bibfield  {author} {\bibinfo {author} {\bibfnamefont {M.~A.}\ \bibnamefont
  {Nielsen}}\ and\ \bibinfo {author} {\bibfnamefont {I.}~\bibnamefont
  {Chuang}},\ }\href@noop {} {\bibinfo {title} {``Quantum computation and
  quantum information,''\ }} (\bibinfo {year} {2002})\BibitemShut {NoStop}%
\bibitem [{\citenamefont {Fukui}\ \emph {et~al.}(2018)\citenamefont {Fukui},
  \citenamefont {Tomita}, \citenamefont {Okamoto},\ and\ \citenamefont
  {Fujii}}]{fukui_high-threshold_2018}%
  \BibitemOpen
  \bibfield  {author} {\bibinfo {author} {\bibfnamefont {K.}~\bibnamefont
  {Fukui}}, \bibinfo {author} {\bibfnamefont {A.}~\bibnamefont {Tomita}},
  \bibinfo {author} {\bibfnamefont {A.}~\bibnamefont {Okamoto}},\ and\ \bibinfo
  {author} {\bibfnamefont {K.}~\bibnamefont {Fujii}},\ }\bibinfo {title}
  {``High-Threshold Fault-Tolerant Quantum Computation with Analog Quantum
  Error Correction,''\ }\href
  {https://link.aps.org/doi/10.1103/PhysRevX.8.021054} {\bibfield  {journal}
  {\bibinfo  {journal} {Phys. Rev. X}\ }\textbf {\bibinfo {volume} {8}},\
  \bibinfo {pages} {021054} (\bibinfo {year} {2018})}\BibitemShut {NoStop}%
\bibitem [{\citenamefont {{Fukui}}(2019)}]{fukui_high-threshold_2019}%
  \BibitemOpen
  \bibfield  {author} {\bibinfo {author} {\bibfnamefont {K.}~\bibnamefont
  {{Fukui}}},\ }\href@noop {} {\bibinfo {title} {``{High-threshold
  fault-tolerant quantum computation with the GKP qubit and realistically noisy
  devices},''\ }} (\bibinfo {year} {2019}),\ \Eprint
  {https://arxiv.org/abs/1906.09767} {arXiv:1906.09767 [quant-ph]} \BibitemShut
  {NoStop}%
\bibitem [{\citenamefont {Fukui}\ \emph {et~al.}(2017)\citenamefont {Fukui},
  \citenamefont {Tomita},\ and\ \citenamefont {Okamoto}}]{fukui_analog_2017}%
  \BibitemOpen
  \bibfield  {author} {\bibinfo {author} {\bibfnamefont {K.}~\bibnamefont
  {Fukui}}, \bibinfo {author} {\bibfnamefont {A.}~\bibnamefont {Tomita}},\ and\
  \bibinfo {author} {\bibfnamefont {A.}~\bibnamefont {Okamoto}},\ }\bibinfo
  {title} {``Analog {Quantum} {Error} {Correction} with {Encoding} a {Qubit}
  into an {Oscillator},''\ }\href
  {https://link.aps.org/doi/10.1103/PhysRevLett.119.180507} {\bibfield
  {journal} {\bibinfo  {journal} {Physical Review Letters}\ }\textbf {\bibinfo
  {volume} {119}},\ \bibinfo {pages} {180507} (\bibinfo {year}
  {2017})}\BibitemShut {NoStop}%
\bibitem [{\citenamefont {H\"anggli}\ \emph {et~al.}(2020)\citenamefont
  {H\"anggli}, \citenamefont {Heinze},\ and\ \citenamefont
  {K\"onig}}]{hanggli_enhanced_2020}%
  \BibitemOpen
  \bibfield  {author} {\bibinfo {author} {\bibfnamefont {L.}~\bibnamefont
  {H\"anggli}}, \bibinfo {author} {\bibfnamefont {M.}~\bibnamefont {Heinze}},\
  and\ \bibinfo {author} {\bibfnamefont {R.}~\bibnamefont {K\"onig}},\
  }\bibinfo {title} {``Enhanced noise resilience of the
  surface--Gottesman-Kitaev-Preskill code via designed bias,''\ }\href
  {https://link.aps.org/doi/10.1103/PhysRevA.102.052408} {\bibfield  {journal}
  {\bibinfo  {journal} {Phys. Rev. A}\ }\textbf {\bibinfo {volume} {102}},\
  \bibinfo {pages} {052408} (\bibinfo {year} {2020})}\BibitemShut {NoStop}%
\bibitem [{\citenamefont {Noh}\ and\ \citenamefont
  {Chamberland}(2020)}]{noh2020fault}%
  \BibitemOpen
  \bibfield  {author} {\bibinfo {author} {\bibfnamefont {K.}~\bibnamefont
  {Noh}}\ and\ \bibinfo {author} {\bibfnamefont {C.}~\bibnamefont
  {Chamberland}},\ }\bibinfo {title} {``Fault-tolerant bosonic quantum error
  correction with the surface--Gottesman-Kitaev-Preskill code,''\ }\href
  {https://link.aps.org/doi/10.1103/PhysRevA.101.012316} {\bibfield  {journal}
  {\bibinfo  {journal} {Phys. Rev. A}\ }\textbf {\bibinfo {volume} {101}},\
  \bibinfo {pages} {012316} (\bibinfo {year} {2020})}\BibitemShut {NoStop}%
\bibitem [{\citenamefont {Fukui}\ and\ \citenamefont
  {Menicucci}(2021)}]{fukui2021efficient}%
  \BibitemOpen
  \bibfield  {author} {\bibinfo {author} {\bibfnamefont {K.}~\bibnamefont
  {Fukui}}\ and\ \bibinfo {author} {\bibfnamefont {N.~C.}\ \bibnamefont
  {Menicucci}},\ }\href@noop {} {\bibinfo {title} {``An efficient,
  concatenated, bosonic code for additive Gaussian noise,''\ }} (\bibinfo
  {year} {2021}),\ \Eprint {https://arxiv.org/abs/2102.01374} {arXiv:2102.01374
  [quant-ph]} \BibitemShut {NoStop}%
\bibitem [{\citenamefont {Fukui}\ \emph {et~al.}(2020)\citenamefont {Fukui},
  \citenamefont {Alexander},\ and\ \citenamefont {van Loock}}]{fukui2020all}%
  \BibitemOpen
  \bibfield  {author} {\bibinfo {author} {\bibfnamefont {K.}~\bibnamefont
  {Fukui}}, \bibinfo {author} {\bibfnamefont {R.~N.}\ \bibnamefont
  {Alexander}},\ and\ \bibinfo {author} {\bibfnamefont {P.}~\bibnamefont {van
  Loock}},\ }\href@noop {} {\bibinfo {title} {``All-Optical Long-Distance
  Quantum Communication with Gottesman-Kitaev-Preskill qubits,''\ }} (\bibinfo
  {year} {2020}),\ \Eprint {https://arxiv.org/abs/2011.14876} {arXiv:2011.14876
  [quant-ph]} \BibitemShut {NoStop}%
\bibitem [{\citenamefont {Rozpędek}\ \emph {et~al.}(2020)\citenamefont
  {Rozpędek}, \citenamefont {Noh}, \citenamefont {Xu}, \citenamefont {Guha},\
  and\ \citenamefont {Jiang}}]{rozpkedek2020quantum}%
  \BibitemOpen
  \bibfield  {author} {\bibinfo {author} {\bibfnamefont {F.}~\bibnamefont
  {Rozpędek}}, \bibinfo {author} {\bibfnamefont {K.}~\bibnamefont {Noh}},
  \bibinfo {author} {\bibfnamefont {Q.}~\bibnamefont {Xu}}, \bibinfo {author}
  {\bibfnamefont {S.}~\bibnamefont {Guha}},\ and\ \bibinfo {author}
  {\bibfnamefont {L.}~\bibnamefont {Jiang}},\ }\href@noop {} {\bibinfo {title}
  {``Quantum repeaters based on concatenated bosonic and discrete-variable
  quantum codes,''\ }} (\bibinfo {year} {2020}),\ \Eprint
  {https://arxiv.org/abs/2011.15076} {arXiv:2011.15076 [quant-ph]} \BibitemShut
  {NoStop}%
\end{thebibliography}%
\bibliographystyle{apsrev4-2_title}

\appendix

\section{Decomposition of position shifts}\label{app:posshiftdec}

We decompose the position shift operator $\X(t)$ in the subsystem basis, first reiterating the definitions from the main text for convenience. A real parameter $t$ is decomposed in terms of two integers and one real number, \emph{i.e.},%
\begin{subequations}%
\begin{align}
t &= \alpha(dn + k) + v
\, ,
\\
v &= \fracpart{t}{\alpha}
\, ,
\\
n &= I_{d}(I_{\alpha}(t) )
\, ,
\\
k &= \fracpart{I_{\alpha}(t)}{d}
\, ,
\end{align}
\end{subequations}
using the notation and conventions of Sec.~\ref{sec:partitionedpositionbasis2}. The gauge displacement operator $\X_G(t)$ is defined through its action on the gauge basis as
\begin{align}
    \X_G(t) \ket{m,u}_G &\coloneqq
        \ket{
        m +
        \closestint{\frac {u + t}{\alpha} }{}
        ,
        \fracpart{u + t}{\alpha}}[G]
        \\ &=
        \ket{
        m +
        I_{\alpha}(u+t)
        ,
        \fracpart{u + t}{\alpha}}[G]
    \, .
\end{align}
\begin{widetext}%
We first write the action of the translation with on a position eigenstate
$
\ket{x}[q]
   =
\ket{\alpha \ell + d \alpha m + u }[q]
$:
\begin{align}
    \X (t) \ket{x}[q]
    &=
    \X(\alpha(dn + k) + v)
    \ket{\alpha \ell + d \alpha m + u }[q]
    \\ &=
    \ket{\alpha d (n + m) + \alpha(\ell+k) + u + v}[q]
    \\ &=
    \ket{\alpha d (n + m)
    + \alpha(\ell+k)
    + \alpha I_{\alpha}(u+v)
    + \fracpart{u + v}{\alpha}
    }[q]
    \, .
\end{align}
Then, we write the state in the partitioned-position basis, $\ket{m,u}$:
\begin{align}
    \X (t)
    \ket{\ell}[L]\otimes \ket{m,u}[G] \nonumber
    &=
    \ket{
    I_{\alpha}\bqty{
    \alpha d (n + m)
    + \alpha(\ell+k)
    + \alpha I_{\alpha}(u+v)
    }
    ,
    \fracpart{u + v}{\alpha}
    }
     \\ &=
    \ket{
    d (n + m)
    + \ell+k
    + I_{\alpha}(u+v)
    ,
    \fracpart{u + v}{\alpha}
    } \label{eq:middlegroundstate}
    \, .
\end{align}
We can simplify the integer in the first entry of the state by writing it as the sum of its fractional part and closest integer with respect to the spacing $d$,
\begin{align}
    d (n + m)
    + \ell+k
    + I_{\alpha}(u+v)
    &=
    \closestint{
    d (n + m)
    + \ell+k
    + I_{\alpha}(u+v)
    }{d}
    +
    \fracpart{
    d (n + m)
    + \ell+k
    + I_{\alpha}(u+v)
    }{d}
    \\
    &=
    d (n + m) +
    \closestint{
    \ell+k
    + I_{\alpha}(u+v)
    }{d}
    +
    \fracpart{
    \ell+k
    + I_{\alpha}(u+v)
    }{d}
    \\
    &=
    d (n + m) +
    d  I_d \big[
    \ell+k + I_{\alpha}(u+v)
    \big]
    +
    \fracpart{
    \ell+k
    + I_{\alpha}(u+v)
    }{d}
    \, ,
\end{align}
which makes the SSD simpler to perform. Namely, the state in \eqr{eq:middlegroundstate} becomes%
{\shrinkeq%
\begin{align}
    \ket{
    d (n + m)
    + \ell+k
    + I_{\alpha}(u+v)
    ,
    \fracpart{u + v}{\alpha}
    }
    &=
    \ket{
    d (n + m) +
    d  I_d \big[
    \ell+k + I_{\alpha}(u+v)
    \big]
    +
    \fracpart{
    \ell+k
    + I_{\alpha}(u+v)
    }{d}
    ,
    \fracpart{u + v}{\alpha}
    }
     \\
    &=
    \ket{
    \fracpart{
    \ell+k
    + I_{\alpha}(u+v)
    }{d}
    }[L] \otimes
    \ket{
    n + m +
    I_d \big[
    \ell+k + I_{\alpha}(u+v)
    \big]
    ,
    \fracpart{u + v}{\alpha}
    }[G]
    \ .
    \label{eq:decXonabasis}
\end{align}
}
The final line is the expression of the shifted state in the subsystem basis.
Note that the expression
$I_{d}[\ell+k+I_{\alpha}(u+v)]$
is a simple boundary-check function that takes values $\{0,\pm 1\}$, depending on the magnitude of the shifts and the initial state.

While Eq.~\eqref{eq:decXonabasis} is enough to specify the SSD of the position-shift operator, one may wish to represent the action of this operator as the product of a number of logical, gauge, and interaction terms acting on the basis state $\ket{\ell}[L] \otimes \ket{m,u}[G]$. We report the product form in the main text in \eqr{eq:decXprod}.

\section{Noisy teleportation of approximate GKP states} \label{appendix:noisyteleportation}
We present the derivation for an approximate GKP state undergoing noisy teleportation. We consider the general situation with no approximations on the GKP and squeezed-state parameters $\Delta$, $\kappa$, $\sparam$. This is more general than the situation considered in \secr{sec:noisyteleportapproxGKP} since we cannot use simple integrals of Dirac combs, and we can no longer use the periodicity of the theta function to simplify the derivation (it will become clear why in a moment). Nevertheless we will see that these complications are accounted for by multivariate Siegel theta functions~\cite{bellman_brief_2013}.
The derivation is conceptually identical to that for an ideal GKP state, where the starting point is again the Kraus operator for noisy teleportation in \eqr{eq:faultykop}. An approximate GKP state, parametrized by $\Delta$ and $\kappa$ in \eqr{eq:approxGKPwavefunctionposition}, is sent through noisy teleportation using squeezed states parametrized by $\sparam$. Its output position wave function $\bar{\psi}_\GKP^\text{tel}(x) = \bra{x}[q] \op{T}_{\sparam}(s,t) \ket{\psi_{\GKP}}$, obtained by integrating
\eqref{eq:telstatefinal}, is
\begin{align} \label{eq:teleportedstatemegageneral}
    \bar{\psi}_\GKP^\text{tel}(x)
    &=
    G_{\sparam^{-1}}\pqty{x - t}
    G_{K_2\kappa^{-1}}
        \pqty{x - i s \sparam^2}
    \sum_{j} c_j
    \vartheta\bqty{
        \frac{ x - i s \sparam^2 }
            { 2 \alpha K_2^2 }
            - \frac{j}{2},
        \frac{ i \pi }{ 2 \alpha^2 K_2^2 }
        \pqty{\sparam^2 + \Delta^2 K_2^2}
                 }
    \, ,
\end{align}
where we defined the auxiliary parameter ${K_2^2 \coloneqq 1 + \sparam^2 \kappa^2}$.
The integral is obtained  by writing the Jacobi theta functions in their infinite-sum form, \eqr{eq:pulsetrain}, performing the resulting Gaussian integrals, and re-summing the result to obtain another theta function.
After teleportation, the periodicity of the wave function is no longer $2\alpha$, as one can verify by inspecting the first argument of the theta function and eliminating integer numbers. Modified periodicity was studied in the context of approximate GKP states in Ref.~\cite{matsuura2020equivalence}.

We now compute the gauge trace of the state in with the wave function in \eqr{eq:teleportedstatemegageneral}. In doing so, we realize that the sum over the gauge variable $m$ can be written in terms of Siegel theta function, \eqr{eq:Siegeltheta}. For a GKP state $\ket{\psi_{\GKP}} = c_0\ket{0_{\GKP}} + c_1 \ket{1_{\GKP}}$, after noisy teleportation, the $\ell \ell'$-th element of the reduced density matrix is
\begin{align} \label{eq:teleportedGKPmegagenerallogical}
    \rho_L^{\ell \ell'}
    &= \frac{1}{\mathcal{N}}
    e^{- \alpha^2 \frac{ \kappa^2 + \sparam^2 K_2^2 }
                    { 4 K_2^2 }(\ell' - \ell)^2
    + i s \frac{ \alpha \kappa^2 \sparam^2 }{ K_2^2 } (\ell' - \ell)
    }
    \sum_{j, j'} c_j^* c_{j'}
    \intalpha du\
    \Theta\bqty{
        \pqty{\begin{matrix}
              \frac{u}{2 \alpha}
            + \frac{(\ell' + \ell)}{4}
            - t \frac{ K_2^2 \sparam^2 }{ 2 \alpha K_1^2 }
            \\
             t \frac{\sparam^2}{ 2 \alpha K_1^2 }
            - i s \frac{\sparam^2}{ 2 \alpha K_2^2 }
            + \frac{\ell' - \ell}{4 K_2^2}
            - \frac{j'}{2}
            \\
              t \frac{\sparam^2}{ 2 \alpha K_1^2 }
            + i s \frac{\sparam^2}{ 2 \alpha K_2^2 }
            - \frac{\ell' - \ell}{4 K_2^2}
            - \frac{j}{2}
        \end{matrix}}
        , \boldsymbol{\tau}
    }
    \, ,
\end{align}
where $\mathcal{N}$ is the normalization, we have defined $K_1 ^2 \coloneqq \sparam^2 + \kappa^2 + \sparam^4 \kappa^2$, and the $3\times 3$ matrix $\boldsymbol{\tau}$  parameterizing the Siegel theta function is
\begin{align}
    \boldsymbol{\tau} &=
    \frac{ i \pi }{ 4 \alpha^2 K_1^2 K_2^2 }
    \pqty{\begin{matrix}
        K_2^4 &&
        -K_2^2 &&
        -K_2^2 \\
        -K_2^2 &&
        1 + 2 K_1^2 (K_2^2 \Delta^2 + \sparam^2 ) &&
        1 \\
        -K_2^2 &&
        1 &&
        1 + 2 K_1^2 (K_2^2 \Delta^2 + \sparam^2 ) \\
    \end{matrix}}
    \, .
\end{align}
 The matrix $\boldsymbol{\tau}$  is complicated because it features terms that are high order in the quality parameters $\Delta$, $\kappa$, and $\sparam$ (as opposed to the high-squeezing limit below).

\subsection{High-quality, high-squeezing limit}
In the limit where input approximate GKP state is high quality and the momentum-squeezed states are highly squeezed, we ignore factors of order larger than $\sparam^{2}$, $\Delta^2$, and $\kappa^2$. In this limit, the auxiliary parameters simplify to
$
K_1 ^2 \rightarrow \sparam^2 + \kappa^2
$
,
$
K_2^{2} \rightarrow 1
$, thus the teleported wave function in \eqr{eq:teleportedstatemegageneral} is
\begin{align}
    \bar{\psi}_{\GKP}^\text{tel}(x) &\approx
    G_{\sparam^{-1}}\pqty{x - t}
    G_{\kappa^{-1}}
        \pqty{x - i s \sparam^2}
    \sum_{j} c_j
    \vartheta\bqty{
        \frac{ x - i s \sparam^2 }
            { 2 \alpha }
            - \frac{j}{2},
            \tau_{\sparam} + \tau_\Delta
                  }
    \, ,
\end{align}
where we used the definition of $\tau_\sigma$ in \eqr{eq:taufactor}. Note that the periodicity of the wave function is restored to $2\alpha$. The expression for the logical matrix elements, \eqr{eq:teleportedGKPmegagenerallogical}, simplifies considerably:
\begin{align}
    \rho^{\ell \ell'}_{L}
    \approx
    \frac{1}{\mathcal{N}}
    e^{-\frac{\alpha^2(\kappa^2 + \sparam^2)}{4}( \ell' - \ell )^2}
    \sum_{j, j'} c_j^* c_{j'}
    \intalpha du \,
    \Theta\bqty{
        \pqty{\begin{matrix}
            \frac{ u - t \frac{\sparam^2}
            { \sparam^2 + \kappa^2 } }{ 2 \alpha }
            + \frac{\ell' + \ell}{4}
            \\
            \frac{ u - i s \sparam^2 }{ 2 \alpha } + \frac{\ell' - j'}{2}
            \\
            \frac{ u - i s \sparam^2 }{ 2 \alpha } + \frac{\ell - j}{2}
        \end{matrix}}
        ,
        \pqty{\begin{matrix}
        \tfrac{1}{2}\tau_{(\kappa^2 + \zeta^2)^{-1/2}}
                && 0 && 0 \\
               0 && \tau_\Delta + \tau_\sparam && 0 \\
               0 && 0 && \tau_\Delta + \tau_\sparam \\
        \end{matrix}}
    }
    \, .
\end{align}
Note that $\boldsymbol{\tau}$ becomes diagonal in this limit, making it a product of three Jacobi theta functions. (We maintain the Siegel-theta form for compactness.) In the limit of infinite squeezing in the CV cluster state (\emph{i.e.},~$\zeta \to 0^+$), this expression reduces to \eqr{eq:firstgaugetrace}.

\end{widetext}

\end{document}